\documentclass[mnsc]{informs3mod}

\OneAndAHalfSpacedXI

\usepackage{natbib}
 \bibpunct[, ]{(}{)}{,}{a}{}{,}%
 \def\bibfont{\small}%
\usepackage[utf8]{inputenc}
\usepackage{graphicx}
\usepackage{amsmath}
\usepackage{amssymb}
\usepackage{xcolor}
\usepackage{float}
\usepackage[shortlabels]{enumitem}
\usepackage{hyperref}
\usepackage[normalem]{ulem}
\usepackage{float}
\usepackage{bbm}

\newcommand{\mcal}[1]{\ensuremath{\mathcal{#1}}}
\renewcommand{\v}[1]{\ensuremath{\boldsymbol{#1}}}
\renewcommand{\b}[1]{\ensuremath{\overline{#1}}}

\renewcommand{\P}{\ensuremath{\mathbb{P}}}

\newcommand{\E}{\ensuremath{\mathbb{E}}}
\newcommand{\Var}{\mathrm{Var}}
\newcommand{\RMSE}{\mathrm{RMSE}}
\newcommand{\GTE}{\ensuremath{\mathsf{GTE}}}
\newcommand{\LR}{\ensuremath{\mathsf{LR}}}
\newcommand{\CR}{\ensuremath{\mathsf{CR}}}
\newcommand{\cluster}{\ensuremath{\mathsf{Cluster}}}
\newcommand{\TSR}{\ensuremath{\mathsf{TSR}}}
\newcommand{\TSRN}{\ensuremath{\mathsf{TSRN}}}
\newcommand{\TSRI}{\ensuremath{\mathsf{TSRI}}}
\newcommand{\TSRIo}{\ensuremath{\mathsf{TSRI\text{-}1}}}
\newcommand{\TSRIt}{\ensuremath{\mathsf{TSRI\text{-}2}}}
\newcommand{\TSRIk}{\ensuremath{\mathsf{TSRI\text{-}k}}}
\newcommand{\GC}{\ensuremath{\mathsf{GC}}}
\newcommand{\GT}{\ensuremath{\mathsf{GT}}}

\definecolor{cobalt}{rgb}{0.0, 0.28, 0.67}

\ifdefined\DRAFT
\newcommand{\rj}[1]{{\color{blue} [R: #1]}}

\newcommand{\delrj}[1]{{\color{blue} \sout{#1}}}
\newcommand{\gw}[1]{{\color{red} [G: #1]}}

\newcommand{\delgw}[1]{{\color{red} \sout{#1}}}
\newcommand{\hl}[1]{{\color{teal} [H: #1]}}

\newcommand{\delhl}[1]{{\color{teal} \sout{#1}}}

\newcommand{\ECdelete}[1]{{\color{red} [EC-]: #1}}

\else
\newcommand{\rj}[1]{}

\newcommand{\delrj}[1]{}
\newcommand{\gw}[1]{}

\newcommand{\delgw}[1]{}
\newcommand{\hl}[1]{}

\newcommand{\delhl}[1]{}

\newcommand{\ECdelete}[1]{}

\fi

\TheoremsNumberedThrough     
\ECRepeatTheorems

\EquationsNumberedThrough    

\MANUSCRIPTNO{MS-RMA-20-02481.R1}

\begin{document}


\RUNAUTHOR{Johari, Li, Liskovich, and Weintraub}

\RUNTITLE{Experimental Design in Two-Sided Platforms}

\TITLE{Experimental Design in Two-Sided Platforms:\\ An Analysis of Bias}

\ARTICLEAUTHORS{%
\AUTHOR{Ramesh Johari}
\AFF{Management Science and Engineering, Stanford University,
\EMAIL{rjohari@stanford.edu}} 
\AUTHOR{Hannah Li}
\AFF{Management Science and Engineering, Stanford University,
\EMAIL{hannahli@stanford.edu}}

\AUTHOR{Inessa Liskovich}
\AFF{Airbnb Inc.,
\EMAIL{inessa.liskovich@airbnb.com}}

\AUTHOR{Gabriel Y. Weintraub}
\AFF{Graduate School of Business, Stanford University,
\EMAIL{gweintra@stanford.edu}}
} 

\ABSTRACT{%
We develop an analytical framework to study experimental design in two-sided marketplaces. Many of these experiments exhibit {\em interference}, where an intervention applied to one market participant influences the behavior of another participant. This interference leads to biased estimates of the treatment effect of the intervention. We develop a stochastic market model and associated mean field limit to capture dynamics in such experiments, and use our model to investigate how the performance of different designs and estimators is affected by marketplace interference effects. Platforms typically use two common experimental designs: demand-side ``customer" randomization ($\CR$) and supply-side ``listing" randomization ($\LR$), along with their associated estimators. We show that good experimental design depends on market balance: in highly demand-constrained markets, $\CR$ is unbiased, while $\LR$ is biased; conversely, in highly supply-constrained markets, $\LR$ is unbiased, while $\CR$ is biased. We also introduce and study a novel experimental design based on {\em two-sided randomization} ($\TSR$) where both customers {\em and} listings are randomized to treatment and control.  We show that appropriate choices of $\TSR$ designs can be unbiased in {\em both} extremes of market balance, while yielding relatively low  bias in intermediate regimes of market balance.
}


\maketitle
%


\section{Introduction}
\label{sec:intro}

We develop a framework to study experiments (also known as {\em A/B tests}) that two-sided platform operators routinely employ to improve the platform. Platforms use experiments to test many types of interventions that affect interactions between participants in the market; examples include features that change the process by which buyers search for sellers or interventions that alter the information the platform shares with buyers. The goal of the experiment is to introduce the intervention to some fraction of the market and use the resulting outcomes to estimate the effect if the intervention were introduced to the entire market. Platforms rely on these estimated effect sizes to make decisions about whether or not to launch the intervention to the entire market.

However, in marketplace experiments, these estimates are often biased due to interference between market participants. Market participants interact and compete with each other and, as a result, the treatment assigned to one individual may influence the behavior of another individual. These interactions violate the Stable Unit Treatment Value Assumption (SUTVA) (\cite{ImbensRubin15}) that guarantees unbiased estimates of the treatment effect. Previous work has shown that the resulting bias can be quite large, and at times as large as the treatment effect itself (\cite{Blake14, Fradkin2015SearchFA, holtz2020reducing}). In this work, we model the platform competition dynamics, investigate how they influence the performance of different canonical experimental designs, and also introduce novel designs that can yield improved performance.

We are particularly motivated by marketplaces where customers do not purchase goods, but rather {\em book} them for some amount of time.
This covers a broad array of platforms, including freelancing (e.g., Upwork), lodging (e.g., Airbnb and Booking.com), and many services (tutoring, dogwalking, child care, etc.). While we explicitly model such a platform, the model we describe also captures features of a platform where goods are bought and supply must be replenished for future demand.

Our model consists of a fixed number of {\em listings}; {\em customers} arrive sequentially over (continuous) time.   For example, in an online labor platform, a freelancer offering work is a listing, and a client looking to hire a freelancer is a customer. On a lodging site, listings include hotel rooms or private rooms and customers are travelers wanting to book. Naturally, an arriving customer can only book {\em available} listings (i.e., those not currently booked).  The customer forms a consideration set from the set of available listings and then, according to a choice model, chooses which listing to book from this set (including an outside option).  
Once a listing is booked, it is occupied and becomes unavailable until the end of its occupancy time.

We focus on {\em interventions} by the platform that change the parameters governing the choice probability of the customer, such as those described above; we refer to the new choice parameters as the {\em treatment} model and the baseline as the {\em control} model.\footnote{The same modeling framework that we employ in this paper can be used to consider interventions that change other parameters, such as customer arrival rates or the time that listings remain occupied when booked; such application is outside the scope of our current work.}  
We assume the platform wants to use an experiment to assess the difference between the rate at which bookings would occur if all choices were made according to the treatment parameters %
and the corresponding rate if all choices were made according to the control parameters %
.  This difference is the {\em global treatment effect} or $\GTE$.  In particular, we assume that the quantity of interest is the {\em steady-state} (or long-run) $\GTE$, i.e., the long-run average difference in rental rates.\footnote{Our framework can also be used to evaluate other metrics of interest based on experimental outcomes; for simplicity we focus on rate of booking in this work.}

Most platforms employ one of two simple designs for testing such an intervention: either {\em customer-side randomization} (what we call the $\CR$ design) or {\em listing-side randomization} (what we call the $\LR$ design).  In the $\CR$ design, customers are randomized to treatment or control.  All customers in treatment make choices according to the treatment choice model and all customers in control make choices according to the control choice model.  In the $\LR$ design, listings are randomized to treatment or control, and the utility of a listing is then determined by its treatment condition.
As a result, in the $\LR$ design, in general each arriving customer will consider {\em some} listings in the treatment condition and {\em some} listings in the control condition.  As an example, suppose the platform decides to test an intervention that shows badges for certain listings.  In the $\CR$ design, all treatment customers see the badges and no control customers see the badges.  In the $\LR$ design, all customers see the badges on treated listings, and do not see them on control listings.

Each of these designs are associated with natural estimators.  In the $\CR$ design, the platform measures the difference in the rate of bookings between treatment and control customers; %
this is what we call the {\em naive $\CR$ estimator}.  In the $\LR$ design, the platform measures the difference in the rate at which treatment and control listings are booked;
this is what we call the {\em naive $\LR$ estimator}.

To develop some intuition for the potential biases, 
first consider an idealized {\em static} setting where listings are instantly replenished upon being booked; in other words, every arriving customer sees the  full set of original listings as available.  
As a result, in the $\CR$ design there is no  interference between treatment and control customers, and consequently the $\CR$ estimator is unbiased for the true $\GTE$.  On the other hand, in the $\LR$ design, every arriving customer considers {\em both} treatment and control listings when choosing whether to book, creating a linkage across listings through customer choice.  In other words, in the $\LR$ design there is interference between treatment and control, and in general the $\LR$ estimator will be biased for the true $\GTE$.%

Now return to a dynamic model where the inventory of listings is limited, and listings remain unavailable for some time after being booked.  In this case, observe that on top of the preceding discussion, there is a dynamic linkage between customers: the set of listings available for consideration by a customer is dependent on the listings considered and booked by previously arriving customers.  This dynamic effect introduces a new form of bias into estimation and is distinctly unique to our work.  In particular, because of this dynamic bias, in general the naive $\CR$ estimator will be biased as well.

Our paper develops a dynamic model of two-sided markets with inventory dynamics, and uses this framework to compare and contrast both the designs and estimators above.  We also introduce and study a novel class of more general designs based on {\em two-sided randomization} (of which the two examples above are special cases).  In more detail, our contributions and the organization of the paper are as follows. %

\noindent {\bf Benchmark model and formal mean field limit}.  Our first main contribution is to develop a general, flexible theoretical model to capture the dynamics described above.  In Section \ref{sec:model}, we present a model that yields a continuous-time Markov chain in which the state at any given time is the number of currently available listings of each type.  In Section \ref{sec:meanfield}, we propose a formal mean field analog of this continuous-time Markov chain, by considering a limit where the number of listings in the system and the arrival rate of customers both grow to infinity.  Scaling by the number of listings yields a continuum mass of listings in the limit.  In the mean field model, the state at a given time is the mass of available listings and this mass evolves via a system of ODEs.  Using a Lyapunov argument, we show that this system is globally asymptotically stable and give a succinct characterization of the resulting asymptotic steady state of the system as the solution to an optimization problem.  We formally establish that the mean field limit arises as the fluid limit of the corresponding finite market model, as market size grows; in other words, the mean field model is a good approximation to large markets.  The mean field model allows us to tractably analyze different estimators, as well as to study their biases in the large market regime.

\noindent {\bf Designs and estimators: Two-sided, customer-side, and listing-side randomization.}
In Section \ref{sec:experiments}, we introduce a more general form of experimental design, called {\em two-sided randomization} ($\TSR$); an analogous idea was independently proposed recently by \cite{bajari2019double} (see also Section \ref{sec:related}). %
In a $\TSR$ design,  both customers {\em and} listings are randomized to treatment and control.  However, the intervention is only applied when a treatment customer considers a treatment listing; otherwise, if the customer is in control or the  listing is in control, the intervention is not seen by the customer.  (In the example above, a customer would see the badge on a listing only if the customer were treated {\em and} the listing were treated.)  Notably, the $\CR$ and $\LR$ designs are special cases of $\TSR$. %
We also define natural naive estimators for each design.

\noindent {\bf Analysis of bias: The role of market balance.}  In Section \ref{sec:bias} we study the bias of the different designs and estimators proposed.  Our main theoretical results characterize how the bias depends on the relative volumes of supply and demand in the market.  In particular, in the highly {\em demand-constrained} regime (where customers arrive slowly and/or listings replenish quickly): %
{\em the naive $\CR$ estimator becomes unbiased, while the naive $\LR$ estimator is biased.}  On the other hand, in the highly {\em supply-constrained} regime (where customers arrive rapidly and/or listings replenish slowly)
we find that in fact {\em the naive $\LR$ estimator becomes unbiased, while the naive $\CR$ estimator is biased.} These findings suggest that platforms can potentially reduce bias by choosing the type of experiment based on knowledge of market balance.  %

Given the findings that $\CR$ and $\LR$ experiments offer benefits in different extremes, it is natural to ask whether good performance can be achieved in moderately balanced markets by ``interpolating'' between the naive $\CR$ and $\LR$ estimators. We define a naive $\TSR$ estimator that achieves this interpolation and has low bias in both market extremes, but still has large bias for moderate market balance. We then define more sophisticated $\TSR$ estimators that explicitly aim to correct for interference in regimes of moderate market balance. These latter estimators exhibit substantially improved performance in simulations. Appendix \ref{app:simulations} shows that these estimators perform well across a wide range of market parameters.

\noindent{\bf Insights from simulations.}
In Section \ref{sec:variance} we turn to simulations in the finite system to study the variance of the estimators. The simulations corroborate the
theoretical findings that $\TSR$ offers benefits with respect to bias, albeit at the cost of moderate increases in variance. Among the $\TSR$ estimators that we study, we find that those with smaller reductions in bias have smaller increases in variance, while those with larger reductions in bias have larger increases in variance, thus revealing a {\em tradeoff} between bias and variance for the $\TSR$ estimators.

In Section \ref{sec:cluster}, we compare the $\TSR$ approach with cluster-randomized experiments, an existing approach that platforms utilize to reduce bias.
The simulations suggest that while both approaches can reduce bias when the market is tightly clustered, $\TSR$ estimators can reduce bias in highly interconnected markets where cluster randomized experiments cannot.

\vspace{0.1in}

Taken together, our work sheds light on what experimental designs and associated estimators should be used by two-sided platforms depending on market conditions, to alleviate the biases from interference that arise in such contexts.
We view our work as a starting point towards a comprehensive framework for experimental design in two-sided platforms; we discuss some directions for future work in Section \ref{sec:conclusion}.

\section{Related work}
\label{sec:related}

{\bf SUTVA.} The types of interference described in these experiments are violations of the Stable Unit Treatment Value Assumption (SUTVA) in causal inference  \citep{ImbensRubin15}.  SUTVA
requires that the (potential outcome) observation on one unit should be unaffected by the particular assignment of treatments to the other units.t
A large number of recent works have investigated experiment design in the presence of interference, particularly in the context of markets and social networks. 

\noindent {\bf Interference in marketplaces.} Biases from interference can be large: \cite{Blake14} empirically show in an auction experiment that the presence of interference among bidders caused the estimate of the treatment effect to be wrong by a factor of two. \cite{Fradkin_2019} finds through simulations that a marketplace experiment changing search and recommendation algorithms can overestimate the true effect by 50 percent. %
More recent work by \cite{holtz2020reducing} randomizes clusters of similar listings to treatment or control, and finds the bias due to interference can be almost 1/3 of the treatment effect.  Interestingly, \cite{holtz2020reducing} also finds weak empirical evidence that the extent of interference depends on market balance; our paper provides strong theoretical grounding for such a claim.

Inspired by the goal of reducing such bias, other work has developed approaches to bias characterization and reduction both theoretically (e.g., \cite{Basse16} in the context of auctions with budgets), as well as via simulation (e.g., \cite{Holtz18} who explores the performance of $\LR$ designs).  Our work complements this line, by developing a mathematical framework for the study of estimation bias in dynamic platforms.  Key to our analysis is the use of a mean field model to model both transient and steady-state behavior of experiments.  A related approach is taken in \cite{Wager19}, where a mean field analysis is used to study equilibrium effects of an experimental intervention where treatment is incrementally applied in a marketplace (e.g., through small pricing changes).

\noindent {\bf Interference in social networks.}  A bulk of the literature in experimental design with interference considers an interference that arises through some underlying social network: e.g., \cite{Manski13} studies the identification of treatment responses under interference; \cite{Ugander13} introduces a graph cluster based randomization scheme and analyzes the bias and variance of the design; and many other papers, including  \cite{Athey18, Basse19, Saveski17} focus on estimating the spillover effects created by interference. In particular, \cite{pouget2019variance} and \cite{zigler2018bipartite} consider interference on a bipartite network, which is closer to a two-sided marketplace setting. In general, this line of work considers a fixed interference pattern (social network) over time. Our work is distinct because the interference caused by supply and demand competition is endogenous to the experiment and dynamically evolving over time.

\noindent {\bf Other experimental designs.}  In practice, platforms currently mitigate the effects of interference through either clustering techniques that change the unit of observation to reduce spillovers among them (e.g., \cite{chamandy16}), similar to some of the works mentioned above (e.g., \cite{Holtz18, Ugander13}); or by {\em switchback testing} \citep{sneider19}, in which the treatment is turned on and off over time.  Both cause a substantial increase in estimation variance due to a reduction in effective sample size and thus the naive $\CR$ and $\LR$ designs remain popular workhorses in the platform experimentation toolkit. In addition to these broad classes of experiments, other work has also introduced modified experiment designs for specific types of interventions, such as \cite{hathuc2020counterfactual} for ranking experiments. 

\noindent {\bf Two-sided randomization.} Finally, a closely related paper is \cite{bajari2019double}.  Independently of our own work, there the authors propose a more general multiple randomized design of which $\TSR$ is a special case.  They focus on a static model and provide an elegant and complete statistical analysis under a local interference assumption.  By contrast, we focus on a dynamic platform model with market-wide interference patterns and focus on a mean field analysis of bias.

\section{A Markov chain model of platform dynamics}
\label{sec:model}

In this section, we first introduce the basic dynamic platform model that we study in this paper with a finite number $N$ of listings.  In the next section, we describe a formal mean field limit of the model inspired by the regime where $N \to \infty$. This mean field limit model then serves as the framework within which we study the bias  of different experimental designs and associated estimators in the remainder of the paper.  

We consider a two-sided platform where we refer to the supply side as {\em listings} and the demand side as {\em customers}.     %
Customers arrive over time and at the time of arrival, the customer forms a consideration set from the set of available listings in the market and then decides whether to book one of them. If the customer books, then the selected listing 
is occupied for a random length of time during which it is unavailable for other customers. At the end of this booking, the listing again becomes available for use for other customers. %

The formal details of our model are as follows.

\textbf{Time.}  The system evolves in continuous time $t \geq 0$.

\textbf{Listings.} The system consists of a fixed number $N$ of listings.  We refer to ``the $N$'th system" as the instantiation of our model with $N$ listings present.  %
We use a superscript ``$N$" to denote quantities in the $N$'th system where appropriate.  

We allow for heterogeneity in the listings. Each listing $\ell$ has a {\em type} $\theta_\ell \in \Theta$, where $\Theta$ is a finite set (the {\em listing type space}).  Note that in general, the type may encode both observable and unobservable covariates; in particular, our analysis does not presume that the platform is completely informed about the type of each listing. For example, in a lodging site $\theta_\ell$ may encode observed characteristics of a house such as the number of bedrooms, but also characteristics that are unobserved by the platform because they may be difficult or impossible to measure.
Let $m^{(N)}(\theta)$ denote the total number of listings of type $\theta$ in the $N$'th system.  For each $\theta \in \Theta$, we assume that $\lim_{N \to \infty} m^{(N)}(\theta)/N = \rho(\theta) > 0$.  Note that $\sum_\theta \rho(\theta) = 1$.

\textbf{State description.}  At each time $t$, each listing $\ell$ can be either {\em available} or {\em occupied} (i.e., occupied by a customer who previously booked it).  The system state at time $t$ in the $N$'th system is described by $\v{\sigma}_t^{(N)} = (\sigma_t^{(N)}(\theta))$, where $\sigma_t^{(N)}(\theta)$ denotes the number of listings of type $\theta$ available in the system at time $t$. Let $S_t^{(N)} = \sum_\theta \sigma_t^{(N)}(\theta)$ be the total number of listings available at time $t$.  In our subsequent development, we develop a model that makes $\v{\sigma}_t^{(N)}$ a continuous-time Markov process.

\textbf{Customers.} Customers arrive to the platform sequentially and decide whether to book, and if so, which listing to book.  Each customer $j$ has a {\em type} $\gamma_j \in \Gamma$, where $\Gamma$ is a finite set (the {\em customer type space}) that represents customer heterogeneity. As with listings, the type may encode both observable and unobservable covariates, and again, our analysis does not presume that the platform is completely informed about the type of each customer. %
Customers of type $\gamma$ arrive according to a Poisson process of rate $\lambda_\gamma^{(N)}$; these processes are independent across types. Let $\lambda^{(N)} = \sum_\gamma \lambda_\gamma^{(N)}$ be the total arrival rate of customers.   Let $T_j$ denote the arrival time of the $j$'th customer.  

We assume that $\lim_{N \to \infty} \lambda^{(N)}/ N = \lambda > 0$, i.e., the arrival rate of customers grows proportionally with the number of listings when we take the large market limit. Further, we assume that for each $\gamma \in \Gamma$, we have $\lim_{N \to \infty} \lambda_\gamma^{(N)}/\lambda^{(N)} = \phi_\gamma > 0$.  Note that $\sum_\gamma \phi_\gamma = 1$.

\textbf{Consideration sets}. In practice, when customers arrive to a platform, they typically form a {\em consideration set} of possible listings to book; the initial formation of the consideration set may depend on various aspects of the search and recommendation algorithms employed by the platform.   To simplify the model, we capture this process by assuming that on arrival, each listing of type $\theta$ that is available at time $t$ is included in the arriving customer's consideration set independently with probability $\alpha_\gamma(\theta) > 0$ for a customer of type $\gamma$.  For example, $\alpha_\gamma(\theta)$ can capture the possibility that the platform's search ranking is more likely to highlight available listings of type $\theta$ that are more attractive for a customer of type $\gamma$, making these listings more likely to be part of the customer's consideration set; this effect is made clear via our choice model presented below.
After the consideration set is formed, a choice model is then applied to the consideration set to determine whether a booking (if any) is made.  %

Formally, the customer choice process unfolds as follows.  Suppose that customer $j$ arrives at time $T_j$.  For each listing $\ell$, let $C_{j\ell} = 0$ if the listing is unavailable at $T_j$.  Otherwise, if listing $\ell$ is available, then let $C_{j\ell} = 1$ with probability $\alpha_{\gamma_j}(\theta_\ell)$, and let $C_{j\ell} = 0$ with probability $1-\alpha_{\gamma_j}(\theta_\ell)$, independently of all other randomness.  Then the consideration set of customer $j$ is $\{ \ell : C_{j\ell} = 1\}$.

Our theoretical results in this paper are developed with this model of consideration set formation.  Other models of consideration set formation are also reasonable, however.  As one example, customers might sample a consideration set of a fixed size, regardless of total number of listings available. We explore such a consideration set model through simulations in Appendix \ref{app:simulations} and show that similar insights hold.

\textbf{Customer choice}.  Customers choose at most one listing to book and can choose not to book at all.  We assume that customers have a {\em utility} for each listing that depends on both customer and listing types: a type $\gamma$ customer has utility $v_\gamma(\theta) > 0$ for a type $\theta$ listing. %
Let $q_{j\ell}$ denote the probability that  arriving customer $j$ of type $\gamma_j$ books listing $\ell$ of type $\theta_\ell$.

In this paper we assume that customers make choices according to the well-known {\em multinomial logit choice model}.  In particular, given the realization of $\v{C}_j$, we have:
\begin{equation}
\label{eq:mnl}
q_{j\ell} = \frac{ C_{j\ell} v_{\gamma_j}(\theta_\ell)}{\epsilon_{\gamma_j}^{(N)} + \sum_{\ell' = 1}^N C_{j\ell'} v_{\gamma_j}(\theta_{\ell'})}.
\end{equation}
Here $\epsilon_\gamma^{(N)} > 0$ is the value of the {\em outside option} for type $\gamma$ customers in the $N$'th system.  The probability that customer $j$ does not book any listing at all grows with $\epsilon_\gamma^{(N)}$.  We let the outside option scale with $N$; this is motivated by the observation that in practical settings, the probability a customer does not book should remain bounded away from zero even for very large systems.  In particular, we assume that $\lim_{N \to \infty} \epsilon_\gamma^{(N)}/N = \epsilon_\gamma > 0$.  

We note that this specification of choice model, although it relies on the multinomial logit model, can be quite flexible because we allow for arbitrary heterogeneity of listings and customers. 

For later reference, we define:
\begin{equation}
\label{eq:qjtheta}
    q_j(\theta) = \E\left[ \sum_{\ell : \theta_\ell = \theta} q_{j\ell}\right],
\end{equation}
where the expectation is over the randomness in $\v{C}_j$.  With this definition, $q_j(\theta)$ is the probability that customer $j$ books an available listing of type $\theta$, where the probability is computed prior to realization of the consideration set.

\textbf{Dynamics: A continuous-time Markov chain.}  
The system evolves as follows.  Initially all listings are available.\footnote{As the system we study is irreducible and we analyze its steady state behavior, it would not matter if we chose a different initial condition.}  Every time a customer arrives, the choice process described above unfolds.  An occupied listing remains occupied, independent of all other randomness, for an exponential time that is allowed to depend on the type of the listing.\footnote{An even more general model might allow the occupancy time to depend on {\em both} listing type {\em and} the type of the customer who made the booking; such a generalization remains an interesting open direction.} More formally, let $\tau>0$ and for each type $\theta$ define $\nu(\theta)$ such that, once booked, a listing of this type will remain occupied for an exponential time with parameter $\tau \nu(\theta)$.  We overload notation and define $\tau(\theta) = \tau \nu(\theta)$.  Once this time expires, the listing returns to being available.

When fixing $\v{\nu} = (\nu(\theta), \theta \in \Theta)$ and all system parameters except for $\tau$, increasing $\tau$ will make the system less supply constrained and decreasing $\tau$ will make the system more supply constrained, while preserving the relative occupancy times of each listing type.

Our preceding specification turns $\v{\sigma}_t^{(N)}$ into a continuous-time Markov process on a finite state space $\mcal{S}^{(N)} = \{ \v{\sigma} : 0 \leq \sigma(\theta) \leq m^{(N)}(\theta),\  \forall \theta \}$.  We now describe the transition rates of this Markov process.  For a state $\v{\sigma} \in \mcal{S}^{(N)}$, $\sigma(\theta)$ represents the number of available listings of type $\theta$.

There are only two types of transitions possible: either (i) a listing that is currently occupied becomes available, or (ii) a customer arrives, and books a listing that is currently available.  (If a customer arrives but does not book anything, the state of the system is unchanged.)  Let $\v{e}_\theta$ denote the unit basis vector in the direction $\theta$, i.e., $e_\theta(\theta) = 1$, and $e_\theta(\theta') = 0$ for $\theta' \neq \theta$.  The rate of the first type of transition is:
 \begin{equation}
 \label{eq:R1}
  R(\v{\sigma}, \v{\sigma} + \v{e}_\theta) = (m^{(N)}(\theta) - \sigma(\theta)) \tau(\theta),
 \end{equation}
since there are $m^{(N)}(\theta) - \sigma(\theta)$ booked listings of type $\theta$, and each remains occupied for an exponential time with mean $1/\tau(\theta)$, independently of all other randomness.  

The second type of transition requires some more steps to formulate.  In principle, our choice model suggests that the identity of both the arriving guest and individual listings affect system dynamics; however, our state description only tracks the aggregate number of listings of each type available at each time $t$.  The key here is that our entire specification depends on customers {\em only} through their type, and depends on listings {\em only} through their type.  

Formally, suppose a customer $j$ of type $\gamma_j = \gamma$ arrives to find the system in state $\v{\sigma}$.  For each $\theta$ let $D_\gamma(\theta | \v{\sigma})$ be a Binomial$(\sigma(\theta), \alpha_\gamma(\theta))$ random variable, independently across $\theta$.   Recall that for each available listing $\ell$, each $C_{j\ell}$ is a Bernoulli$(\alpha_\gamma(\theta_l))$ random variable.  Recalling $q_j(\theta)$ as defined in \eqref{eq:qjtheta}, it is straightforward to check that:
\begin{equation}
\label{eq:anonymizedchoice}
 q_j(\theta) = r_\gamma(\theta | \v{\sigma}) \triangleq \E \left[\frac{ D_\gamma(\theta|\v{\sigma}) v_{\gamma}(\theta)}{\epsilon_{\gamma}^{(N)} + \sum_{\theta'} D_\gamma(\theta'|\v{\sigma}) v_{\gamma}(\theta')}\right].
\end{equation}
In other words, the probability an arriving customer of type $\gamma$ books a listing of type $\theta$ when the state is $\v{\sigma}$ is given by $r_\gamma(\theta | \v{\sigma})$; and this probability depends on the past history {\em only} through the state $\v{\sigma}$ (ensuring the Markov property holds).  

With this definition at hand, for states $\v{\sigma}$ with $\sigma(\theta) > 0$, the rate of the second type of transition is:
\begin{equation}
\label{eq:R2}
 R(\v{\sigma}, \v{\sigma} - \v{e}_\theta) = \sum_\gamma \lambda_\gamma^{(N)}  r_\gamma(\theta | \v{\sigma}).
\end{equation}

Note that the resulting Markov chain is irreducible, since customers have positive probability of sampling into, and booking from, their consideration set, and every listing in the consideration set has positive probability of being booked.  %

\textbf{Steady state.} 
Since the Markov process defined above is irreducible on a finite state space, there is a unique steady state distribution $\pi^{(N)}$ on $\mcal{S}^{(N)}$ for the process.  %

\section{A mean field model of platform dynamics}
\label{sec:meanfield}

The continuous-time Markov process described in the preceding section is challenging to analyze directly because the customers' choices and consideration sets induce complex dynamics.  Instead, to make progress we consider a formal {\em mean field} limit motivated by the regime where $N \to \infty$, in which the evolution of the system becomes deterministic.  We first present a formal mean field analogue of the Markov process introduced in the previous section and provide intuition for its derivation.  We then formally prove that the sequence of Markov processes converges to this mean field model as $N \rightarrow \infty$. The mean field model provides tractable expressions in the large market regime for  the different estimators we consider, allowing us to study and compare their bias.

The mean field model we study consists of a continuum unit mass of listings.  The total mass of listings of type $\theta$ in the system is $\rho(\theta) > 0$ (recall that $\sum_\theta \rho(\theta) = 1$).  We represent the state at time $t$ by $\v{s}_t = (s(\theta), \theta \in \Theta)$; $s_t(\theta)$ represents the mass of listings of type $\theta$ available at time $t$.  The state space for this model is:
\begin{equation}
    \label{eq:S}
 \mcal{S} = \{ \v{s} : 0 \leq s(\theta) \leq \rho(\theta) \}.
\end{equation}

We first present the intuition behind our mean field model.  Consider a state $\v{s} \in \mcal{S}$ with $s(\theta) > 0$ for all $\theta$.  We view this state as analogous to a state $\v{\sigma} \approx N \v{s}$ in the $N$'th system.  We consider the system dynamics defined by \eqref{eq:R1}-\eqref{eq:R2}.  Note that the rate at which occupied listings of type $\theta$ become available is $(m^{(N)}(\theta) - \sigma(\theta)) \tau(\theta)$, from \eqref{eq:R1}.  If we divide by $N$, then this rate becomes $(\rho(\theta) - s(\theta))\tau(\theta)$ as $N \to \infty$.  On the other hand, note that for large $N$, if $D_\gamma(\theta|\v{\sigma})$ is Binomial$(\sigma(\theta), \alpha_\gamma(\theta))$, then $D_\gamma(\theta|\v{\sigma})/N$ concentrates on $\alpha_\gamma(\theta) s(\theta)$.  Thus the choice probability $r_\gamma(\theta | \v{\sigma})$ is approximately:
\begin{equation}
\label{eq:meanfieldchoice}
p_\gamma( \theta | \v{s}) \triangleq \frac{\alpha_\gamma(\theta) v_\gamma(\theta) s(\theta) }{\epsilon_\gamma + \sum_{\theta'} \alpha_\gamma(\theta') v_\gamma(\theta') s(\theta')}.
\end{equation}
(Here we use the fact that $\epsilon_\gamma^{(N)}/N \to \epsilon_\gamma$ as $N \to \infty$.) This is the mean field multinomial logit choice model for our system.  In the finite model, the rate at which listings of type $\theta$ become occupied is $\sum_\gamma \lambda_\gamma^{(N)}  r_\gamma(\theta | \v{\sigma})$, from \eqref{eq:R2}.  If we divide by $N$, this rate becomes $\lambda \sum_\gamma \phi_\gamma  p_\gamma(\theta | \v{s})$ as $N \to \infty$.  

Inspired by the preceding observations, we define the following system of differential equations for the evolution of $\v{s}_t$:
\begin{equation}
\label{eq:ODE}
\frac{d}{dt} s_t(\theta) = (\rho(\theta) - s_t(\theta))\tau(\theta) - \lambda \sum_\gamma \phi_\gamma  p_\gamma(\theta | \v{s}_t),\ \ \theta \in \Theta.
\end{equation}

This is our formal mean field model. In the remainder of this section, we first show that this system has a unique solution for any initial condition. Then we characterize the behavior of the system. By constructing an appropriate Lyapunov function, we show that the mean field model has a unique limit point to which all trajectories converge (regardless of initial condition). This limit point is the unique steady state of the mean field limit. Finally, we prove that the sequence of Markov processes indeed converges to this mean field model (in an appropriate sense).  Hence, the mean field model provides a close approximation to the evolution of large finite markets.

\subsection{Existence and uniqueness of mean field trajectory}

First, we show the straightforward result that the system of ODEs defined in \eqref{eq:ODE} possesses a unique solution.  This follows by an elementary application of the Picard-Lindel\"{o}f theorem from the theory of differential equations.  The proof is in Appendix \ref{app:proofs}.

\begin{proposition}
\label{prop:ODE_trajectory}
Fix an initial state $\hat{\v{s}} \in \mcal{S}$.  The system \eqref{eq:ODE} has a unique solution $\{\v{s}_t:t\geq 0\}$ satisfying $0 \leq s_t(\theta) \leq \rho(\theta)$ and for all $t$ and $\theta$, and $\v{s}_0 = \hat{\v{s}}$. 
\end{proposition}

\subsection{Existence and uniqueness of mean field steady state}

Now we characterize the behavior of the mean field limit. We show that the system of ODEs in \eqref{eq:ODE} has a unique limit point, to which all trajectories converge regardless of the initial condition.  We refer to this as the {\em steady state} of the mean field system.  We prove the result via the use of a convex optimization problem; the objective function of this problem is a Lyapunov function for the mean field dynamics that guarantees global asymptotic stability of the steady state.

Formally, we have the following result.  The proof is in Appendix \ref{app:proofs}.

\begin{theorem}
\label{thm:ODE_ss}
There exists a unique steady state $\v{s}^* \in \mcal{S}$ for \eqref{eq:ODE}, i.e., a unique vector $\v{s}^* \in \mcal{S}$ solving the following system of equations:
\begin{equation}
\label{eq:FOC}
 (\rho(\theta) - s^*(\theta))\tau(\theta) = \lambda \sum_\gamma \phi_\gamma  p_\gamma(\theta | \v{s}^*),\ \ \theta \in \Theta.
\end{equation}
This limit point has the property that $0 < s^*(\theta) < \rho(\theta)$ for all $\theta$, i.e., it is in the interior of $\mcal{S}$.  Further, this limit point is globally asymptotically stable, i.e., all trajectories of $\eqref{eq:ODE}$ converge to $\v{s}^*$ as $t \to \infty$, for any initial condition $\v{s}_0 \in \mcal{S}$.  

The limit point $\v{s}^*$ is the unique solution to the following optimization problem:
\begin{align}
\text{minimize}  \ \ \ & W(\v{s}) \triangleq \sum_\gamma \left( \lambda_\gamma \log\left(\epsilon_\gamma + \sum_\theta \alpha_\gamma(\theta) v_\gamma(\theta) s(\theta) \right) \right) \nonumber \\
& \hspace{5em}- \tau(\theta) \sum_\theta \rho(\theta) \log s(\theta) + \tau(\theta) \sum_\theta s(\theta)  \label{eq:ss_objective} \\ 
\text{subject to} \ \ \ &0 \leq s(\theta) \leq \rho(\theta), \ \ \theta \in \Theta. \label{eq:ss_constraint}
\end{align}
\end{theorem}

The function $W$ appearing in the proposition statement is {\em not} convex; our proof proceeds by first noting that it suffices to restrict attention to $\v{s}$ such that $s(\theta) > 0$ for all $\theta$, then making the transformation $y(\theta) = \log (s(\theta))$.  The objective function redefined in terms of these transformed variables {\em is} strictly convex, and this allows us to establish the desired result.

\subsection{Convergence to the mean field limit}
Finally, we formally describe the sense in which our system converges to the system in \eqref{eq:ODE}. We first move from analyzing the number of listings available in the $N$'th system to analyzing the proportion of listings available. To this end, define the normalized process $\v{Y}_t^{(N)}$ where
\begin{equation*}
    Y_t^{(N)}(\theta) = \sigma_t^{(N)}(\theta)/N , \ \ \theta \in \Theta.
\end{equation*}
Note that under this definition, $\v{Y}_t^{(N)}$ is also a continuous time Markov process with dynamics induced by the dynamics of $\v{\sigma}_t^{(N)}$; in particular, the chain $\v{Y}_t^{(N)}$ has the same transition rates as $\v{\sigma}_t^{(N)}$, but increments are of size $1/N$. The following theorem establishes the convergence of $Y_t^{(N)}$ to the solution of the ODE described in \eqref{eq:ODE} as $N \rightarrow \infty$.

\begin{theorem}
\label{thm:mf_convergence}
Assume that $\epsilon_\gamma^{(N)} / N \rightarrow \epsilon_\gamma$ for all $\gamma$ and $\lambda_\gamma^{(N)} / N \rightarrow \lambda_\gamma$ for all $\gamma$ as $N\rightarrow \infty$.  Fix $\hat{\v{s}} \in \mcal{S}$ and assume that $\v{Y}_0^{(N)}$ is deterministic, with $Y_0^{(N)}(\theta)  \rightarrow \hat{s}(\theta)$ for all $\theta$. %
Let $\v{s}_t$ denote the unique solution to the system defined in \eqref{eq:ODE}, with initial condition $\v{s}_0 = \hat{\v{s}}$.

Then for all $\delta>0$ and for all times $u>0$,
\begin{equation}
    \P\left[\sup_{0 \leq t \leq u} \| \v{Y}_t^{(N)} - \v{s}_t \| > \delta \right] = O\left(\frac{1}{N}\right).
\end{equation}
\end{theorem}

The proof for this result relies on an application of Kurtz's Theorem for the convergence of pure jump Markov processes; full details are in Appendix \ref{app:proofs}.
We note that this result holds for any sequence of initial conditions $\v{Y}_0^{(N)}$, as long as the proportion of available listings at time $t=0$ converges to a constant vector $\hat{\v{s}}$ as $N \rightarrow \infty$; further, the vector $\hat{\v{s}}$ can be any (feasible) initial state in the mean field model.  %

We now utilize the mean field model to study experimental designs and interference.

\section{Experiments: Designs and estimators}
\label{sec:experiments}

In this section, we leverage the framework developed in the previous section to undertake a study of experimental designs a platform might employ to test interventions in the marketplace.  For simplicity, we focus on interventions that change the {\em choice probability} of one or more types of customers for one or more types of listings, and we assume the platform is interested in estimating the resulting rate at which bookings take place.  However, we believe the same approach we employ here can be applied to study other types of interventions and platform objectives as well.  %

Formally, the platform's goal is to design experiments with associated estimators to assess the performance of the intervention (the {\em treatment}), relative to the status quo (the {\em control}).  In particular, the platform is interested in determining the steady-state rate of booking when the entire market is in the treatment condition (i.e., {\em global treatment}), compared to the steady-state rate of booking when the entire market is in the control condition (i.e., {\em global control}).  We refer to the difference of these two rates as the {\em global treatment effect} ($\GTE$). We focus on these steady-state quantities as a platform is typically interested in the long-run effect of an intervention. 

Two types of canonical experimental designs are typically employed in practice: {\em listing-side randomization} (denoted $\LR$) and {\em customer-side randomization} (denoted $\CR$).  In the former design, listings are randomized to treatment or control; in the latter design, customers are randomized to treatment or control.  Each design also has an associated natural ``naive" estimator of booking rates, that is, a (scaled) difference in means estimators for the two groups.  As we discuss, these estimators will typically be biased, due to interference effects.

The $\LR$ and $\CR$ designs are special cases of a novel, more general {\em two-sided randomization} ($\TSR$) design that we introduce in this work, where {\em both} listings and customers are randomized to treatment and control simultaneously. As we discuss, this type of experiment can be combined with design and analysis techniques to reduce bias. On the design side, $\TSR$ designs allow us to construct experiments that interpolate between $\LR$ and $\CR$ designs in such a way that bias is reduced. On the analysis side,  $\TSR$ designs allow us to observe different competition effects, that we can use to heuristically debias our estimators. ($\TSR$ designs were also independently introduced and studied in recent work by \cite{bajari2019double}; see Section \ref{sec:related} for discussion.)  In the next subsection we develop the relevant formalism for these designs; we then subsequently define natural ``naive" estimators that are commonly used for the $\LR$ and $\CR$ designs, as well as an estimator for a $\TSR$ design.  In the remainder of the paper we study the bias of these different designs and estimators under different market conditions.

\subsection{Experimental design}
Since $\CR$ and $\LR$ are special cases of a $\TSR$ design, we first describe how to embed $\TSR$ experimental designs into our model, and then subsequently describe $\CR$ and $\LR$ designs in our model.

\textbf{Treatment condition.}  We consider a binary treatment: every customer and listing in the market will either be in {\em treatment} or {\em control}.   (Generalization of our model to more than two treatment conditions is relatively straightforward.)  We model the treatment condition by expanding the set of customer and listing types.  For every customer type $\gamma \in \Gamma$, we create two new customer types $(\gamma, 0), (\gamma, 1)$; and for every listing type $\theta \in \Theta$, we create a two new listing types $(\theta, 0), (\theta, 1)$.  The types $(\gamma, 0), (\theta, 0)$ are {\em control types}; the types $(\gamma, 1), (\theta, 1)$ are {\em treatment types}.

\textbf{Two-sided randomization.} In this design, randomization takes place on both sides of the market simultaneously. We assume that a fraction $a_C$ of customers are randomized to treatment, and a fraction $1 -a_C$ to control, independently; and we assume that a fraction $a_L$ of listings are randomized to treatment, and a fraction $1 - a_L$ to control, independently.  %

\textbf{Treatment as a choice probability shift.} Examples of interventions that platforms may wish to test include the introduction of higher quality photos for a hotel listing on a lodging site, showing previous job completion rates of a freelancer on an online labor market, or reducing the friction for an item in the checkout flow. 
Such interventions change the choice probability of listings by customers either through the consideration probabilities or perceived utility for a listing.  In particular, we continue to assume the multinomial logit choice model, and we assume that for a type $\gamma$ customer and a type $\theta$ listing that have been given the intervention, the utility becomes $\tilde{v}_\gamma(\theta) > 0$
and the probability of inclusion in the consideration set becomes $\tilde{\alpha}_\gamma(\theta) > 0$.  Since we focus on changes in choice probabilities, we assume that the holding time parameter of a listing of type $\theta$ is $\nu(\theta)$, regardless of whether it is assigned to treatment or control.\footnote{Our current work allows us to relatively easily incorporate $\nu$ depending on treatment condition of the listing, and as such we can extend our results to study $\LR$ designs where $\nu$ varies with treatment condition.  In general, however, when customers are also randomized to treatment or control, the holding time parameter of a listing should also depend on the treatment condition of the customer who booked that listing.  Adapting our framework to incorporate this possibility remains an interesting direction for future work.}

In the $\TSR$ designs that we consider, a key feature is that the intervention is applied only when {\em a treated customer interacts with a treated listing}.  For example, when an online labor marketplace decides to show previous job completion rates of a freelancer as an intervention,
only treated customers can see these rates, and they only see them when they consider treated freelancers.  We model this by redefining quantities in the experiment as follows:
\begin{align}
 v_{\gamma,0}(\theta, 0) = v_{\gamma, 1}(\theta, 0) = v_{\gamma,0}(\theta, 1) = v_\gamma(\theta);&\quad v_{\gamma,1}(\theta, 1) = \tilde{v}_\gamma(\theta);\label{eq:v_expt}\\
 \alpha_{\gamma,0}(\theta, 0) = \alpha_{\gamma, 1}(\theta, 0) = \alpha_{\gamma,0}(\theta, 1) = \alpha_\gamma(\theta);&\quad 
 \alpha_{\gamma,1}(\theta, 1) = \tilde{\alpha}_\gamma(\theta);
  \label{eq:alpha_expt}\\
\epsilon_{\gamma,0} = \epsilon_{\gamma,1} &= \epsilon_\gamma; \label{eq:epsilon_expt}\\
\nu(\theta,0) = \nu(\theta,1) &= \nu(\theta).\label{eq:nu_expt}
\end{align}
This definition is a natural way to incorporate randomization on each side of the market.  However, we remark here that it is not necessarily canonical; for example, an alternate design would be one where the intervention is applied when {\em either} the customer has been treated {\em or} the listing has been treated.  Even more generally, the design might randomize whether the intervention is applied, based on the treatment condition of the customer and the listing.  In all likelihood, the relative advantages of these designs would depend not only on the bias they yield in any resulting estimators, but also in the variance characteristics of those estimators.  We leave further study and comparison of these designs to future work. %

\textbf{Customer-side and listing-side randomization.}  Two canonical special cases of the $\TSR$ design are as follows.  When $a_L = 1$, all listings are in the treatment condition; in this case, randomization only takes place on the customer side of the market.  This is the customer-side randomization ($\CR$) design.  When $a_C = 1$, all customers are in the treatment condition and randomization only takes place on the listing side of the market.  This is the listing-side randomization ($\LR$) design.

\textbf{System dynamics.}  With the specification above, it is straightforward to adapt our mean field system of ODEs, cf.~\eqref{eq:ODE}, and the associated choice model \eqref{eq:meanfieldchoice}, to this setting.  The key changes are as follows:
\begin{enumerate}
\item The mass of control (resp., treatment) listings of type $(\theta, 0)$ (resp., $(\theta, 1)$) becomes $(1-a_L)\rho(\theta)$ (resp., $a_L \rho(\theta)$).  In other words, abusing notation, we define $\rho(\theta, 0) = (1-a_L) \rho(\theta)$, and $\rho(\theta, 1) = a_L\rho(\theta)$.
\item The arrival rate of control (resp., treatment) customers of type $(\gamma, 0)$ (resp., $(\gamma, 1)$) becomes $(1-a_C)\lambda \phi_\gamma$ (resp., $a_C \lambda \phi_\gamma$).  Thus we define $\phi_{\gamma,0} = (1-a_C)\phi_\gamma$, and $\phi_{\gamma,1} = a_C\phi_\gamma$.
\item The choice probabilities are defined as in \eqref{eq:meanfieldchoice}, with the relevant quantities defined in \eqref{eq:v_expt}-\eqref{eq:epsilon_expt}.
\end{enumerate}

Using Proposition \ref{prop:ODE_trajectory} and Theorems \ref{thm:ODE_ss}, we know that there exists a unique solution to the resulting system of ODEs; and that there exists a unique limit point to which all trajectories converge, regardless of initial condition.  This limit point is the steady state for a given experimental design.  For a $\TSR$ experiment with treatment customer fraction $a_C$, and treatment listing fraction $a_L$, we use the notation $\v{s}_t(a_C, a_L) = (s_t(\theta, j) | a_C, a_L), \theta \in \Theta, j \in \{0,1\})$ to denote the ODE trajectory, and we use $\v{s}^*(a_C, a_L) = (s^*(\theta, j) | a_C, a_L), \theta \in \Theta, j \in \{0,1\})$ to denote the steady state.

\textbf{Rate of booking}.  In our subsequent development, it will be useful to have a shorthand notation for the rate at which bookings of listings of treatment condition $j \in \{0,1\}$ are made by customers of treatment condition $i \in \{0,1\}$, in the interval $[0,T]$.  In particular, we define:
\begin{equation}
\label{eq:QijT}
Q_{ij}(T | a_C, a_L) = \frac{\lambda}{T} \int_0^T \sum_\theta \sum_\gamma \phi_{\gamma,i} p_{\gamma,i}(\theta, j | \v{s}_t(a_C, a_L))\; dt.
\end{equation}
Since $\v{s}_t(a_C, a_L)$ is globally asymptotically stable, bounded, and converges to $\v{s}^*(a_C, a_L)$ as $t \to \infty$, we have:
\begin{equation}
\label{eq:Qijinfty}
 Q_{ij}(\infty | a_C, a_L) \triangleq \lim_{T \to \infty} Q_{ij}(T | a_C, a_L) = \lambda \sum_\theta \sum_\gamma \phi_{\gamma,i} p_{\gamma,i}(\theta, j | \v{s}^*(a_C, a_L)).
\end{equation}

\textbf{Global treatment effect.}  Recall we assume the {\em steady-state rate of booking} is the quantity of interest to the platform.  In particular, the platform is interested in the change in this rate from the {\em global control} condition ($a_C = 0, a_L = 0$) to the {\em global treatment} condition ($a_C = 1, a_L = 1$).  

In the global control condition, the steady state rate at which customers book is: $Q^{\GC} = Q_{00}(\infty | 0, 0)$, and in the global treatment condition, the steady state rate at which customers book is $Q^{\GT} = Q_{11}(\infty | 1,1)$.
Under these definitions, the global treatment effect is $\GTE = Q^{\GT} - Q^{\GC}$.  

We remark that the rate of booking decisions made by arriving customers will change over time, even if the market parameters are constant over time (including the arrival rates of different customer types, as well as the utilities that customers have for each listing type).  This transient change in booking rates is driven by changes in the state $\v{s}_t$; in general, such fluctuations will lead the transient rate of booking to differ from the steady-state rate, for all values of $a_C$ and $a_L$ (including global treatment and global control).  
In this work, we focus on the steady state quantities to capture, informally, the long run change in behavior due to an intervention.
\footnote{Note that in two-sided markets, certain types of interventions will also cause long-run {\em economic} equilibration due to strategic responses on the part of market participants; for example, if prices are lowered during an experiment, this may affect entry decisions of both buyers and sellers, and thus the long-run market equilibrium.  While our model allows the choice probabilities to change due to treatment, a more complete analysis of long run economic equilibration due to interventions remains a direction for future work.}

\subsection{Estimators: Transient and steady state}

The goal of the platform is to use the experiment to estimate $\GTE$.  In this section we consider estimators the platform might use to estimate this quantity.  We first consider the $\CR$ and $\LR$ designs, and we define ``naive'' estimators that the platform might use to estimate the global treatment effect.  These designs and estimators are those most commonly used in practice.  We define these estimators during the transient phase of the experiment %
and then define the associated steady-state versions of these estimators.  
Finally, we combine these estimation approaches in a natural heuristic that can be employed for any general $\TSR$ design.

{\bf Estimators for the $\CR$ design.}
We start by considering the $\CR$ design, i.e., where $a_L = 1$ and $a_C \in (0,1)$.  A simple naive estimate of the rate of booking is to measure the rate at which bookings are made in a given interval of time by control customers, and compare this to the analogous rate for treatment customers.  Formally, suppose the platform runs the experiment for the interval  $t \in [0, T]$, with a fraction $a_C$ of customers in treatment.   
The rate at which  customers of treatment condition $i \in \{0,1\}$ book in this period is $Q_{i1}(T | a_C, 1)$.
The {\em naive $\CR$ estimator} is the difference between treatment and control rates, where we correct 
for differences in the size of the control and treatment groups, by scaling with the respective masses:%
\begin{equation}
\label{eq:naive_CR}
 \widehat{\GTE}^{\CR}(T | a_C) = \frac{Q_{11}(T | a_C,1)}{a_C} - \frac{Q_{01}(T | a_C,1)}{1-a_C}. 
 \end{equation}
We let $\widehat{\GTE}^{\CR}(\infty | a_C) = Q_{11}(\infty | a_C,1)/a_C - Q_{01}(\infty | a_C,1)/(1-a_C)$ denote the steady-state naive $\CR$ estimator.

{\bf Estimators for the $\LR$ design.}
Analogously, we can define a naive estimator for the $\LR$ design, i.e., where $a_C = 1$ and $a_L \in (0,1)$.  Formally, suppose the platform runs the experiment for the interval  $t \in [0, T]$, with fraction $a_L$ of listings in treatment.   
The rate at which listings with treatment condition $j \in \{0,1\}$ are booked in this period is $Q_{1j}(T | 1, a_L)$.  The {\em naive $\LR$ estimator} is the difference between treatment and control rates, scaled by the mass of listings in each group:
\begin{equation}
\label{eq:naive_LR}
 \widehat{\GTE}^{\LR}(T | a_L) = \frac{Q_{11}(T | 1,a_L)}{a_L} - \frac{Q_{10}(T | 1,a_L)}{1-a_L}. 
\end{equation}
We let $\widehat{\GTE}^{\LR}(\infty | a_L) = Q_{11}(\infty | 1,a_L)/a_L - Q_{10}(\infty | 1,a_L)/(1-a_L)$ denote the corresponding steady-state naive $\LR$ estimator.  %

{\bf Estimators for the $\TSR$ design.}  As with the $\LR$ and $\CR$ designs, it is possible to design a natural naive estimator for the $\TSR$ design as well.  In particular, we have the following definition of the {\em naive $\TSR$ estimator}:
\begin{equation}
\label{eq:naive_TSR_est}
 \widehat{\GTE}^{\TSRN}(T | a_C, a_L) = \frac{Q_{11}(T | a_C,a_L)}{a_Ca_L} - \frac{Q_{01}(T | a_C,a_L) + Q_{10}(T | a_C, a_L) + Q_{00}(T | a_C, a_L)}{1 - a_C a_L}.
\end{equation}
To interpret this estimator, observe that the first term is the normalized rate at which treatment customers booked treatment listings in the experiment; we normalize this by $a_C a_L$, since a mass $a_C$ of customers are in treatment, and a mass $a_L$ of listings are in treatment.  This first term estimates the global treatment rate of booking.  The sum $Q_{01}(T | a_C,a_L) + Q_{10}(T | a_C, a_L) + Q_{00}(T | a_C, a_L)$ is the total rate at which control bookings took place: either because the customer was in the control group, or because the listing was in the control group, or both.  (Recall that in the $\TSR$ design, the intervention is only seen when treatment customers interact with treatment listings.)  This is normalized by the complementary mass, $1 - a_C a_L$.  This second term estimates the global control rate of booking.  As before, we can define a steady-state version of this estimator as $\widehat{\GTE}^{\TSRN}(\infty | a_C, a_L)$, with the steady-state versions of the respective quantities on the right hand side of \eqref{eq:naive_TSR_est}.

It is straightforward to check that as $a_L \to 1$, we have $\widehat{\GTE}^{\TSRN}(T | a_C, a_L) \to \widehat{\GTE}^{\CR}(T | a_C)$, the naive $\CR$ estimator.  Similarly, as $a_C \to 1$, we have $\widehat{\GTE}^{\TSRN}(T | a_C, a_L) \to \widehat{\GTE}^{\LR}(T | a_L)$, the naive $\LR$ estimator.  In this sense, the naive $\TSR$ estimator naturally ``interpolates" between the naive $\LR$ and $\CR$ estimators.  In the next section, we exploit this interpolation to choose $a_C$ and $a_L$ as a function of market conditions in such a way as to reduce bias.  

More generally, the $\TSR$ design also contains much more information about competition in the marketplace, and the resulting interference effects, than either the $\CR$ or $\LR$ designs.  Inspired by this observation, together with the idea of interpolating between the naive $\CR$ estimator and the naive $\LR$ estimator, in Section \ref{ssec:TSR_est} we explore alternative, more sophisticated $\TSR$ estimators that heuristically debias interference due to competition effects.  As we show, these estimators can offer substantial bias reduction over the naive $\TSRN$ estimator above.

\section{Analysis of bias}
\label{sec:bias}

We now utilize the framework defined to analyze the bias of two common experiment types, $\LR$ experiments and $\CR$ experiments. 
Recall from Section \ref{sec:intro} that, in a setting where listings are immediately replenished, all customers see the full set of original listings as available. There is no competition between customers but there is still competition between listings, and so, intuitively, we expect $\CR$ to be unbiased and $\LR$ to be biased. Meanwhile, in a setting where listings remain unavailable for some amount of time, the resulting dynamic linkage across customers creates a bias in $\CR$ as well. Now, consider the extreme where the market is highly supply-constrained: most customers who arrive see no available listings, but some customers arrive just as a booked listing becomes available and see a single available listing. Such customers compare the listing against the outside option, but, since no other listings are available, do not compare listings against each other. In this regime, there is no competition across listings but there is competition between customers, and so we expect $\LR$ to be unbiased and $\CR$ to be biased.

In this section, we formalize this intuition about the behavior of the estimators in the extremes of market balance. We establish two key theoretical results: in the limit of a highly supply-constrained market (where $\lambda/\tau \to \infty$), the naive $\LR$ estimator becomes an unbiased estimator of the $\GTE$, while the naive $\CR$ estimator is biased.  On the other hand, in the limit of a highly demand-constrained market (where $\lambda/\tau \to 0$), the naive $\CR$ estimator becomes an unbiased estimator of the GTE, while the naive $\LR$ estimator is biased.  In other words, each of the two naive designs is respectively unbiased in the limits of extreme market imbalance. 
At the same time, we find empirically that neither estimator performs well in the region of moderate market balance.

Inspired by this finding, we consider $\TSR$ and associated estimators that naturally interpolate between the two naive designs depending on market balance. %
Given the findings above, we show that a simple approach to adjusting $a_C$ and $a_L$ as a function of market balance yields performance that balances between the naive $\LR$ estimator and the naive $\CR$ estimator.  Nevertheless, we show there is significant room for improvement, by adjusting for the types of interference that arise using observations from the $\TSR$ experiment.  In particular, we propose a heuristic for a novel interpolating estimator for the $\TSR$ design that aims to correct these biases, and yields surprisingly good numerical performance.

\subsection{Theory: Stead-state bias of $\CR$ and $\LR$ in unbalanced markets}
\label{ssec:ss_theory}

In this subsection, we study theoretically the bias of the steady-state naive $\CR$ and $\LR$ estimators  in the limits where the market is extremely unbalanced (either demand-constrained or supply-constrained).  The key tool we employ is a characterization of the asymptotic behavior of $Q_{ij}(\infty | a_C, a_L)$ as defined in \eqref{eq:Qijinfty} in the limits where $\lambda/\tau \to 0$ and $\lambda/\tau \to \infty$.  We use this characterization in turn to quantify the asymptotic bias of the naive estimators relative to the $\GTE$. We derive these results in the next two subsections and provide a simple example in Section \ref{ssec:Qij_homogeneous} to illustrate the effects.

\subsubsection{Highly demand-constrained markets.}

We start by considering the behavior of naive estimators in the limit where $\lambda/\tau \to 0$.  We start with the following proposition that characterizes behavior of $Q_{ij}(\infty|a_C,a_L)$ as $\lambda/\tau \to 0$.  The proof is in Appendix \ref{app:proofs}.
\begin{proposition}
\label{prop:Qij_demand_constrained}
Fix all system parameters except $\lambda$ and $\tau$, and consider a sequence of systems in which $\lambda/\tau \to 0$.  Then along this sequence,
\begin{equation}
\label{eq:Qij_demand_constrained}
\frac{1}{\lambda} Q_{ij}(\infty | a_C, a_L) \to \sum_{\theta} \sum_{\gamma} \phi_{\gamma,i} p_{\gamma,i}( \theta,j | \v{\rho} ).
\end{equation}
\end{proposition}
The expression on the right hand side depends on both $a_C$ and $a_L$ through $\phi_{\gamma,i}$ and $\v{\rho}$ respectively.  In particular, we recall that $\phi_{\gamma,1} = a_C \phi_\gamma$, $\phi_{\gamma,0} = (1-a_C) \phi_\gamma$, and $\rho(\theta,1) = a_L \rho(\theta)$, $\rho(\theta,0) = (1-a_L) \rho(\theta)$.  In our subsequent discussion in this regime, to emphasize the dependence of $\v{\rho}$ on $a_L$ below, we will write $\v{\rho}(a_L) = (\rho(\theta,j|a_L), \theta \in \Theta, j = 0,1)$.  With this definition, we have  $\rho(\theta,1|a_L) = a_L \rho(\theta)$, $\rho(\theta,0|a_L) = (1-a_L) \rho(\theta)$.

This proposition allows us to characterize the bias of both $\CR$ and $\LR$ estimators in the demand constrained limit. 
Note that Proposition \ref{prop:Qij_demand_constrained} shows that, in this limit, the (scaled) rate of booking behaves {\em as if} the available listings of type $(\theta,j)$ was {\em exactly} $\rho(\theta,j|a_L)$ for every $\theta$ and treatment condition $j = 0,1$.  That is, it is as if every arriving customer sees the entire mass of listings as being available, and so bookings are immediately replenished. This observation drives our first main result, that in the demand constrained limit the $\CR$ estimator is unbiased and $\LR$ estimator is biased.

\begin{theorem}
\label{thm:demand_constrained}
Consider a sequence of systems where $\lambda/\tau \to 0$.  Then for all $a_C$ such that $0 < a_C < 1$, $\widehat{\GTE}^{\CR}(\infty|a_C)/\lambda - \GTE/\lambda \to 0$.  However, for $0 < a_L < 1$, generically over parameter values\footnote{Here "generically" means for all parameter values, except possibly for a set of parameter values of Lebesgue measure zero.} we have
$\lim\ \widehat{\GTE}^{\LR}(\infty|a_C)/\lambda - \GTE/\lambda \neq 0$.
\end{theorem}

The full proof can be found in Appendix \ref{app:proofs}. The key insight is that as the market becomes more demand constrained, there is a weakening of the competition between arriving customers, which leads to less interference in a $\CR$ experiment. In the limit, the $\CR$ estimator becomes unbiased. On the other hand, in an $\LR$ experiment there is a positive mass of control \textit{and} treatment listings available in steady state, leading to competition between listings and bias in the $\LR$ estimator.

\subsubsection{Heavily supply-constrained markets.}

We now characterize the behavior of naive estimators in the limit where $\lambda/\tau \to \infty$.  We start with the next proposition, where we study the behavior of $Q_{ij}$ as $\lambda/\tau \to \infty$.  The proof is in Appendix \ref{app:proofs}.  To state the proposition, we define:
\[ g_{\gamma,i}(\theta ,j) = \frac{\alpha_{\gamma,i}(\theta,j) v_{\gamma,i}(\theta, j)}{\epsilon_{\gamma,i}}. \]

\begin{proposition}
\label{prop:Qij_supp_constrained}
Fix all system parameters except $\lambda$ and $\tau$, and consider a sequence of systems in which $\lambda/\tau \to \infty$.  Along this sequence, the following limit holds:
\begin{equation}
\label{eq:Qij_supp_constrained}
\frac{1}{\tau}Q_{ij}(\infty | a_C, a_L) \to \sum_\theta \left(\frac{\sum_\gamma \phi_{\gamma,i} g_{\gamma,i}(\theta,j)}{\sum_{i' = 0,1} \sum_{\gamma}\phi_{\gamma,i'} g_{\gamma,i'}(\theta,j)}\right) \rho(\theta,j) \nu(\theta).
\end{equation}
\end{proposition}
As before, the expression on the right hand side depends on both $a_C$ and $a_L$ through $\phi_{\gamma,i}$ and $\v{\rho}$ respectively.  In particular, we recall that $\phi_{\gamma,1} = a_C\phi_\gamma$, $\phi_{\gamma,0} = (1-a_C) \phi_\gamma$, and $\rho(\theta,1) = a_L \rho(\theta)$, $\rho(\theta,0) = (1-a_L)\rho(\theta)$.

A key intermediate result we employ is to demonstrate that in the steady-state in this limit, $s^*(\theta, j|a_C, a_L) \to 0$ for all $\theta,j$.  We know that in the steady state of the mean field limit, the rate at which occupied listings become available must match the rate at which available listings become occupied (flow conservation).  We use this fact to show that to first order in $\lambda/\tau$, in the limit where $\lambda/\tau \to \infty$ we have:
\[ s^*(\theta, j | a_C, a_L) \approx \frac{1}{\lambda/\tau} \cdot \frac{\rho(\theta, j) \nu(\theta)}{\sum_{\gamma} \sum_{i = 0,1} \phi_{\gamma,i} g_{\gamma,i}(\theta, j)}. \]
The proposition follows by using this limit to characterize the choice probabilities.

The proof of the preceding proposition reveals that in the limit where $\lambda/\tau \to \infty$, we have:
\[ p_{\gamma,i}(\theta,j | \v{s}^*(a_C, a_L)) \approx g_{\gamma,i}(\theta,j) s^*(\theta,j|a_C,a_L) = \frac{ \alpha_{\gamma,i}(\theta,j) v_{\gamma,i}(\theta,j) s^*(\theta,j|a_C, a_L)}{ \epsilon_{\gamma,i}}. \]
This preceding expression is the formalization of our intuition that, in the limit where the market is heavily supply-constrained, it is as if each arriving customer seeing an available listing compares only that listing to the outside option; there is no longer any competition {\em between} listings.

We can use the preceding proposition to understand the behavior of the $\GTE$, the naive $\LR$ estimator, and the naive $\CR$ estimator in steady-state, as $\lambda/\tau \to \infty$.  For simplicity, we hold $\tau$ constant and take the limit $\lambda \to \infty$.  In this case, the preceding proposition shows that:
\[ Q_{11}(\infty | 1, 1) \to  \tau \sum_\theta  \rho(\theta) \nu(\theta) ; \ \  Q_{00}(\infty | 0, 0) \to  \tau \sum_\theta  \rho(\theta) \nu(\theta). \]
The global treatment effect $\GTE \to 0$ in this limit. 
Bookings occur essentially instantaneously after a listing becomes available, which happens at rate $\tau \sum_\theta  \rho(\theta) \nu(\theta)$. 

We also note that:
\[ Q_{11}(\infty | 1,a_L) \to a_L \tau \sum_{\theta} \rho(\theta) \nu(\theta); \ \ Q_{10}(\infty | 1,a_L) \to (1-a_L) \tau \sum_{\theta} \rho(\theta) \nu(\theta).\]
The preceding two expressions reveal that the steady-state naive $\LR$ estimator $\widehat{\GTE}^{\LR}(\infty|a_L)$ in this setting approaches zero, matching the $\GTE$; thus it is asymptotically unbiased.  

It is also now straightforward to see why the $\CR$ design will be biased.  Note that:
\[ Q_{11}(\infty | a_C,1) \to a_C \tau \sum_\theta \left(\frac{\sum_\gamma \phi_{\gamma} g_{\gamma,1}(\theta,1)}{\sum_{\gamma}\sum_{i' = 0,1} \phi_{\gamma,i'} g_{\gamma,i'}(\theta,1)}\right) \rho(\theta) \nu(\theta). \] 
An analogous expression holds for $Q_{01}(\infty | a_C, 1)$.  We see that the right hand side reflects the dynamic interference created between treatment and control customers: just as in our simple example, whether or not an available listing is seen by, e.g., a control customer depends on whether it has previously been booked by a treatment customer. That is, customers compete for listings.
As in the example, the naive $\CR$ estimator will remain nonzero in general in the limit, even though the $\GTE$ approaches zero.

We summarize our discussion in the following theorem.
\begin{theorem}
\label{thm:supp_constrained}
Consider a sequence of systems where $\lambda/\tau \to \infty$.  Then $\GTE/\tau \to 0$, and for all $a_L$ such that $0 < a_L < 1$, there also holds $\widehat{\GTE}^{\LR}(\infty | a_L)/\tau \to 0$.  However, for $0 < a_C < 1$, generically over parameter values we have
$\lim\ \widehat{\GTE}^{\CR}(\infty|a_C)/\tau - \GTE/\tau \neq 0$.
\end{theorem}

Although the preceding theorem shows that the {\em absolute} bias of the naive $\LR$ estimator approaches zero, in fact in general the {\em relative} bias $(\widehat{GTE}^{\LR}(\infty|a_L) - \GTE)/\GTE$ will not generally approach zero; this is because the $\GTE$ is also approaching zero, and so the second-order behavior of the naive $\LR$ estimator matters.  This is in contrast to the behavior of the naive $\CR$ estimator in the demand-constrained limit: in that limit, the $\GTE$ remains nonzero in general, and so the naive $\CR$ estimator is both absolutely and relatively unbiased.  Nevertheless, note that relative bias of the naive $\LR$ estimator will be significantly smaller than the relative bias of the naive $\CR$ estimator in the supply-constrained limit, since the naive $\CR$ estimator has a nonzero absolute bias in this limit while the $\GTE$ approaches zero.

We finish this subsection by noting Theorems \ref{thm:demand_constrained} and \ref{thm:supp_constrained} are driven by fundamental competition dynamics in the respective limiting regimes, and therefore, we believe they hold under a much more general class of models than the ones considered here. We leave this for future investigation.

\subsubsection{An example: Homogeneous customers and listings.}
\label{ssec:Qij_homogeneous}

To more clearly understand the behavior of the bias, in this section we apply Propositions \ref{prop:Qij_demand_constrained} and \ref{prop:Qij_supp_constrained} to a simpler setting where 
both listings and customers are homogeneous, i.e., there is only one type of customer and one type of listing.  This example illustrates the symmetry between the two sides of the market and the resulting implications for bias in marketplace experiments. 

Let $v$ denote the control utility 
and $\tilde{v}$ the treatment utility of a customer for a listing. Let $\epsilon$ denote the outside option value of both control and treatment customers,  $\alpha_0(0) = \alpha_1(1) = 1$, and  $\nu(0) = \nu(1) = 1$. %
In this example, we consider two limits: one where $\lambda$ is fixed and $\tau \to \infty$ (the demand-constrained regime), and one where $\tau$ is fixed and $\lambda \to \infty$ (the supply-constrained regime).  

In the first case, when $\tau \to \infty$ with $\lambda$ fixed, if we apply Proposition \ref{prop:Qij_demand_constrained}, we obtain:
\begin{equation}
\label{eq:Qij_homogeneous_demand_const}
\begin{aligned}[c]
Q_{00}(\infty | a_C, a_L) \to & \lambda \cdot \frac{(1-a_C)(1-a_L) \rho v}{\epsilon + \rho v}; %
\\
Q_{10}(\infty | a_C, a_L) \to &\lambda \cdot \frac{a_C(1-a_L)\rho v}{\epsilon + (1-a_L)\rho v + a_L \rho \tilde{v}};%
\\
\end{aligned}
\qquad
\begin{aligned}[c]
Q_{01}(\infty | a_C, a_L) \to &\lambda \cdot \frac{(1-a_C)a_L \rho v}{\epsilon + \rho v};%
\\
Q_{11}(\infty | a_C, a_L) \to & \lambda \cdot \frac{a_C a_L \rho \tilde{v}}{\epsilon + (1-a_L)\rho v + a_L \rho \tilde{v}}.%
\end{aligned}
\end{equation}
In this limit,
\[ \GTE \to \lambda \cdot \left(\frac{\rho \tilde{v}}{\epsilon + \rho \tilde{v}} - \frac{\rho v}{\epsilon + \rho v}\right). \]
From these expressions it is clear that the naive $\CR$ estimator is unbiased, while the naive $\LR$ estimator is biased.  Further, the expressions reveal that listing-side randomization creates interference across listings.

In the second case, when $\lambda \to \infty$ with $\tau$ fixed, if we apply Proposition \ref{prop:Qij_supp_constrained}, we obtain:
\begin{equation}
\label{eq:Qij_homogeneous_supp_const}
\begin{aligned}[c]
    Q_{00}(\infty | a_C, a_L) \to &\tau (1-a_C)(1-a_L) \rho; %
    \\
    Q_{10}(\infty | a_C, a_L) \to &\tau a_C(1-a_L)\rho;%
\end{aligned}
\qquad
\begin{aligned}[c]
    Q_{01}(\infty | a_C, a_L) \to &\tau \cdot \frac{(1-a_C)v}{(1-a_C)v + a_C \tilde{v}} a_L \rho; %
    \\
    Q_{11}(\infty | a_C, a_L) \to &\tau \cdot \frac{a_C\tilde{v}}{(1-a_C)v + a_C \tilde{v}} a_L \rho.%
\end{aligned}
\end{equation}
In this limit,  $\GTE \to 0$.  From these expressions it is clear that the naive $\CR$ estimator is biased, while the naive $\LR$ estimator is unbiased.  Further, these expressions also reveal that customer-side randomization creates interference across customers.

Interestingly, these expressions highlight a remarkable symmetry.  As expected, in the limit of a highly demand-constrained market, customers choose among listings; thus there is competition for customers among listings, and this is the source of potential interference in $\LR$ designs.  The expressions reveal that in the limit of a highly supply-constrained market, it is {\em as if} listings choose among customers; thus there is competition among customers, and this is the source of potential interference in $\CR$ designs.  Indeed, the expressions for $Q_{01}$ and $Q_{11}$ in  \eqref{eq:Qij_homogeneous_supp_const} take the form of a multinomial logit choice model of listings for customers.  We believe this type of symmetry provides important insight into the nature of experimental design in two-sided markets, and in particular the roots of the interference typically observed in such settings.

\subsubsection{Sign of the bias in $\CR$ and $\LR$ estimates.}
Theorems \ref{thm:demand_constrained} and \ref{thm:supp_constrained} state that the $\LR$ estimate is biased in the demand constrained limit and the $\CR$ estimate is biased in the supply constrained limit, but make no claim as to whether the estimators overestimate or underestimate the $\GTE$. In general, we cannot provide guarantees for the sign of the bias, as it depends on the distribution of listings, the rates at which listing replenish, and the lift on the individual $\alpha_\gamma(\theta)$ and $v_\gamma(\theta)$ induced by the interventions. However, for a broad class of interventions, we can show that the $\LR$ estimate overestimates in the demand constrained limit and $\CR$ overestimates in the supply constrained limit. In such cases where we know the bias to be positive, $\CR$ and $\LR$ experiments can be used to bound the size of the $\GTE$. 

We call an intervention \textit{positive} if $\tilde{\alpha}_\gamma(\theta)\tilde{v}_\gamma(\theta) > \alpha_\gamma(\theta)v_\gamma(\theta)$ for all $\gamma$ and $\theta$. Such an intervention can be viewed as an improvement on the platform for all customer and listing type pairs, since for each pair at least one of the customer's consideration probability or utility for the listing type must increase. Note that this class of interventions is broad enough to allow for heterogeneous treatment effects across different $(\gamma, \theta)$ pairs.\footnote{A symmetric analysis can be applied for ``negative" interventions, where $\tilde{\alpha}_\gamma(\theta)\tilde{v}_\gamma(\theta) < \alpha_\gamma(\theta)v_\gamma(\theta)$ for all $\gamma$ and $\theta$; though, of course, interventions known to be negative in advance are less likely to be desirable from the platform's perspective.}

For positive interventions, straightforward applications of Propositions \ref{prop:Qij_demand_constrained} and \ref{prop:Qij_supp_constrained} show that $\widehat{\GTE}^{\LR}$ overestimates the $\GTE$ in the demand constraint limit and $\widehat{\GTE}^\CR$ overestimates the $\GTE$ in the supply constrained limit. The result follows from the fact that in a customer-side (resp., listing-side) experiment in a supply constrained (resp., demand-constrained) setting, the individuals in the treatment group face less competition than they would in the global treatment setting, whereas the individuals in the control group face more competition than in the global control setting. 

\begin{proposition}
\label{prop:cr_lr_positive_bias}
Suppose that the treatment is positive, i.e., $\tilde{\alpha}_\gamma(\theta)\tilde{v}_\gamma(\theta) > \alpha_\gamma(\theta)v_\gamma(\theta)$ for all $\gamma, \theta$.
Then we have the following. 
\begin{enumerate}
    \item $\LR$ bias when demand constrained: Consider a sequence of systems where $\lambda/\tau \to 0$. For any $0 < a_L < 1$, we have
    $\lim\ \widehat{\GTE}^{\LR}(\infty|a_C)/\lambda - \GTE/\lambda > 0$. 
    \item $\CR$ bias when supply constrained: Consider a sequence of systems where $\lambda/\tau \to \infty$. For any $0 < a_C < 1$,  we have $\lim\ \widehat{\GTE}^{\CR}(\infty|a_C)/\tau - \GTE/\tau > 0$.
\end{enumerate}
\end{proposition}
Further, we find through simulations that $\CR$ and $\LR$ overestimate the $\GTE$ with positive treatments in intermediate ranges of market balance, for all parameter regimes that we study in the examples in this section (see Figure \ref{fig:simulations_hom} as well as  Appendix \ref{app:simulations}). We do find in some cases that the $\TSRI$ estimators underestimate the $\GTE$, and so, since we plot bias on a log scale, we report the absolute value of the bias.  %

\subsection{Discussion: Violation of SUTVA}

Our results on the bias of the naive $\CR$ and $\LR$ experiments
can be interpreted through the lens of the classical potential outcomes model. An important result from this literature is that when the {\em stable unit treatment value assumption} (SUTVA) holds, then naive estimators of the sort we consider will be unbiased for the true treatment effect.  SUTVA requires that the treatment condition of units other than a given customer or listing should not influence the potential outcomes of that given customer or listing.  The discussion above illustrates that in the limit where $\lambda \to 0$, there is no interference across customers in the $\CR$ design; this is why the naive $\CR$ estimator is unbiased.  Similarly, in the limit where $\lambda \to \infty$, there is no interference across listings in the $\LR$ design; this is why the naive $\LR$ estimator is unbiased.  On the other hand, the cases where each estimator is biased involve interference across experimental units.

\subsection{Estimation with the $\TSR$ design}
\label{ssec:TSR_est}

The preceding sections reveal that each of the naive $\LR$ and $\CR$ estimators has its virtues, depending on market balance conditions.  In this section, we explore whether we can develop $\TSR$ designs and estimators in which $a_C$ and $a_L$ are chosen {\em as a function of} $\lambda/\tau$, to obtain the beneficial asymptotic performance of the naive $\CR$ estimator in the highly-demand constrained regime, as well as the $\LR$ estimator in the highly supply-constrained regime. We also expect that an appropriate interpolation should yield a bias for $\TSR$
that is comparable to, if not lower than, $\CR$ and $\LR$ in intermediate regimes of market balance.

Recall the naive $\TSRN$ estimator defined in \eqref{eq:naive_TSR_est}, and in particular the steady-state version of this estimator.  Suppose the platform observes $\lambda/\tau$; note that this is reasonable from a practical standpoint as this is a measure of market imbalance involving only the overall arrival rate of customers and the average rate at which listings become available.  For example, consider the following heuristic choices of $a_C$ and $a_L$ for the $\TSR$ design, for some fixed values of $\b{a}_C$ and $\b{a}_L$:
\begin{equation}
a_C(\lambda/\tau) = \left(1 - e^{-\lambda/\tau}\right) + \b{a}_C e^{-\lambda/\tau}; \quad a_L(\lambda/\tau) = \b{a}_L\left(1 - e^{-\lambda/\tau}\right) + e^{-\lambda/\tau}. \label{eq:aC_aL_naive_TSR}
\end{equation}
Then as $\lambda/\tau \to 0$, we have $a_C(\lambda/\tau) \to \b{a}_C$ and $a_L(\lambda/\tau) \to 1$, while as $\lambda/\tau \to \infty$ we have $a_C(\lambda/\tau) \to 1$ and $a_L(\lambda/\tau) \to \b{a}_L$.\footnote{Our choice of exponent here is somewhat arbitrary; the same analysis follows even if we replace $e^{-\lambda/\tau}$ with $e^{-c\lambda/\tau}$ for any value of $c > 0$.}. With these choices, it follows that in the highly demand-constrained limit ($\lambda/\tau \to 0$), the $\TSRN$ estimator becomes equivalent to the naive $\CR$ estimator, while in the highly supply-constrained limit ($\lambda/\tau \to \infty$), the $\TSRN$ estimator becomes equivalent to the naive $\LR$ estimator.  In particular, using Propositions \ref{prop:Qij_demand_constrained} and \ref{prop:Qij_supp_constrained}, it is straightforward to show that the steady-state naive $\TSRN$ estimator is unbiased in {\em both} limits; we state this as the following theorem, and omit the proof.%
\begin{corollary}
\label{cor:naive_TSR}
For each $\lambda/\tau$, consider the $\TSR$ design with $a_C$ and $a_L$ defined as in \eqref{eq:aC_aL_naive_TSR}.  Consider a sequence of systems where either $\lambda/\tau \to 0$, or $\lambda/\tau \to \infty$.  Then in either limit:
\[ \widehat{\GTE}^{\TSRN}(\infty | a_C(\lambda/\tau), a_L(\lambda/\tau)) - \GTE\to 0. \]
\end{corollary}

\begin{figure} 
    \centering
    \includegraphics[scale=0.35]{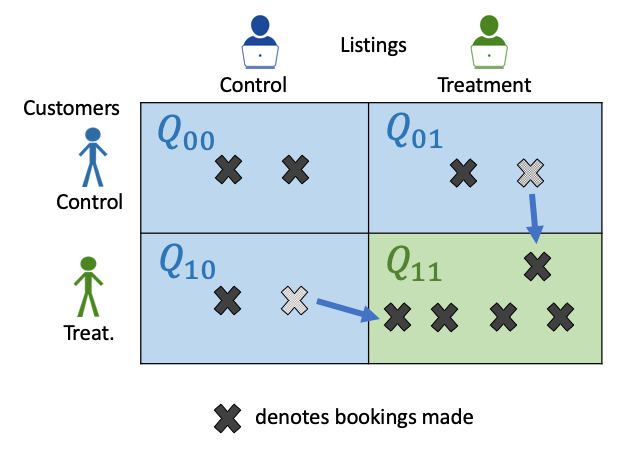}
    \caption{Illustration of $\TSR$ design and competition effects. Intervention only applies when treatment customers view treatment listings (green cell). Suppose that the intervention makes the listing more attractive. %
    }
    \label{fig:tsr_comp_terms}
\end{figure}

\begin{figure}
    \centering
    \includegraphics[width=0.5\textwidth]{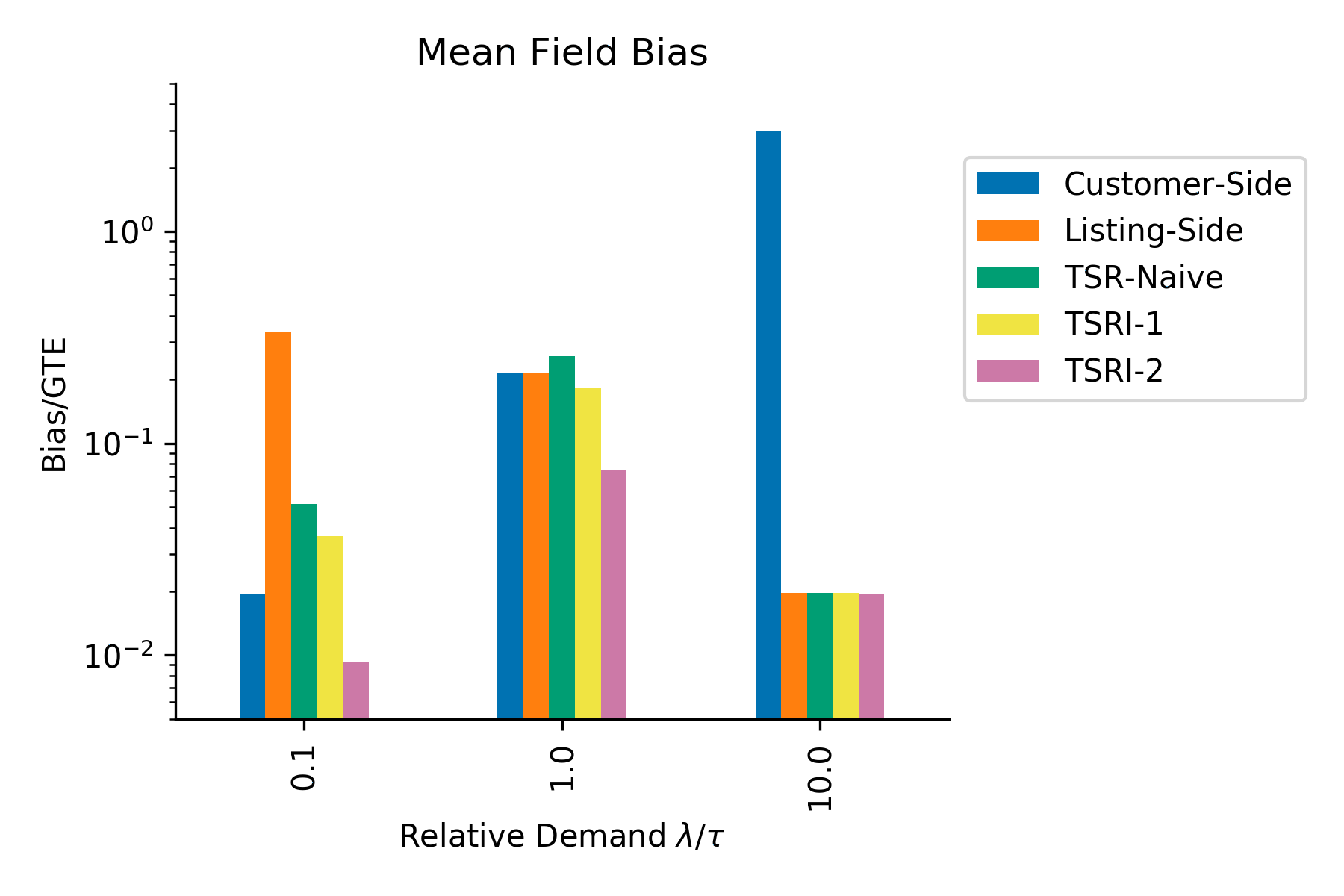}
    \caption{Difference between estimator and $\GTE$ in steady state.  We consider variation in $\lambda/\tau$ by fixing $\tau = 1$ and varying $\lambda$; analogous results are obtained if $\lambda$ is fixed and $\tau$ is varied. We consider a market with homogeneous customers and homogeneous listings, pre-treatment. We set $\epsilon=1$ and $\alpha=0.5$. In the $\CR$ design, $a_C=0.5$. In the $\LR$ design, $a_L=0.5$. 
    Customers have utility $v=0.315$ for control listings and $\tilde{v}=0.394$ for treatment listings, which corresponds to a steady state booking probability of 20 percent in global control and 23 percent global treatment when $\lambda=\tau$. }
    \label{fig:numerics_homogeneous}
\end{figure}

We are also led to ask whether we can improve upon the naive $\TSRN$ estimator when the market is moderately balanced.  %
Note that the $\TSRN$ estimator does not explicitly correct for {\em either} the fact that there is interference across listings, or the fact that there is interference across customers. We now suggest a heuristic for correcting these effects, which we use to define two improved interpolating $\TSR$ estimators; these estimators are the fourth and fifth estimators appearing in Figure \ref{fig:numerics_homogeneous}, which we call ``$\TSR$-Improved (1)'' and ``$\TSR$-Improved (2)". These effects are visualized in Figure \ref{fig:tsr_comp_terms}.

First, abusing notation, let $\widehat{\GTE}^{\CR}(T|a_C, a_L)$ denote the estimator in \eqref{eq:naive_CR} using the same terms from a $\TSR$ design, and  dividing through by $a_L$ on both terms as normalization.  Similarly abusing notation,  let $\widehat{\GTE}^{\LR}(T|a_C, a_L)$ denote the estimator in \eqref{eq:naive_LR} using the same terms from a $\TSR$ design, and dividing through by $a_C$ on both terms as normalization.  Motivated by these naive estimators, we explicitly consider an interpolation between the $\LR$ and $\CR$ estimators of the form:
\begin{align}
\label{eq:interpolatingTSR}
 \beta \widehat{\GTE}^{\CR}(T|a_C,a_L)+(1-\beta)\widehat{\GTE}^{\LR}(T|a_C,a_L)& \\ = \beta \left(\frac{Q_{11}(T | a_C,a_L)}{a_Ca_L} - \frac{Q_{01}(T | a_C,a_L)}{(1-a_C)a_L}\right) &+ (1-\beta) \left(\frac{Q_{11}(T | a_C,a_L)}{a_Ca_L} - \frac{Q_{10}(T | a_C,a_L)}{a_C(1-a_L)}\right). \notag
 \end{align}

Now, consider the quantity $Q_{00}(T|a_C,a_L)/((1-a_C)(1-a_L)) - Q_{10}(T|a_C,a_L)/((1-a_C)a_L)$ in a $\TSR$ design. This is the (appropriately normalized) difference between the rate at which control customers book control listings, and the rate at which treatment customers book control listings.   Note that both treatment and control customers have the same utility for control listings, due to the $\TSR$ design, but potentially different utilities for treatment listings. Hence, the difference in steady-state rates of booking among control and treatment customers on {\em control listings} must be driven by the fact that treatment customers substitute bookings from control listings to treatment listings (or vice versa). This difference captures the ``cannibalization" effect (i.e., interference) that was found in $\LR$ designs in the  demand-constrained regime.

Thus motivated, we can think of this difference as a ``correction term'' for the $\LR$ design from our interpolating $\TSR$ estimator in \eqref{eq:interpolatingTSR}. %
Using a symmetric argument we can also consider an appropriately weighted correction term associated to interference across customers in a $\CR$ design:  $Q_{00}(T|a_C,a_L)/((1-a_L)(1-a_C))-Q_{01}(T|a_C,a_L)/((1-a_C)a_L)$.  (Similar estimates were also studied in \cite{bajari2019double}; see the related work for further details on this work.)  See Figure \ref{fig:tsr_comp_terms} for an illustration of these competition effect estimates. %

We can weight these correction terms with different factors $k > 0$ to control their impact.  In addition, we can choose these weights in a market-balance-dependent fashion, based on the direction of market balance in which we have seen that the respective interference grows.  Combining these insights, for $\beta \in (0,1)$ and $k > 0$ we define a class of improved $\TSR$ estimators given by:
\begin{multline}
\label{eq:TSRI-A}
\widehat{\GTE}^{\TSRIk}(T|a_C, a_L) = \\
\beta \left[\frac{Q_{11}(T | a_C,a_L)}{a_Ca_L} - \frac{Q_{01}(T | a_C,a_L)}{(1-a_C)a_L}-k(1-\beta)\left(\frac{Q_{00}(T|a_C, a_L)}{(1-a_C)(1-a_L)}-\frac{Q_{01}(T|a_C, a_L)}{(1-a_C)a_L}\right) \right] \\
+(1- \beta) \left[ \frac{Q_{11}(T | a_C,a_L)}{a_Ca_L} - \frac{Q_{10}(T | a_C,a_L)}{a_C(1-a_L)}-k\beta\left(\frac{Q_{00}(T|a_C,a_L)}{(1-a_C)(1-a_L)}-\frac{Q_{10}(T|a_C,a_L)}{a_C(1-a_L)}\right) \right] \ .
\end{multline}
Given market balance $\lambda/\tau$, we set $\beta=e^{-\lambda/\tau}$, and we choose $a_C$ and $a_L$ as in \eqref{eq:aC_aL_naive_TSR}.

In the limit where $\lambda/\tau \to 0$, note that $\widehat{\GTE}^{\TSRI-k} (T|a_C(\lambda/\tau),a_L(\lambda/\tau))$ approaches $\widehat{\GTE}^{\CR}(T|\b{a}_C)$ as expected.  Similarly, in the limit where $\lambda/\tau \to \infty$, $\widehat{\GTE}^{\TSRIk} (T|a_C(\lambda/\tau),a_L(\lambda/\tau))$ approaches $\widehat{\GTE}^{\LR}(T|\b{a}_L)$.  It is straightforward to show that  $\widehat{\GTE}^{\TSRIk}$ for any $k$ is  unbiased  in {\em both} the highly demand-constrained and highly supply-constrained regimes, since the correction terms play no role in the limits.

For moderate values of market balance,  both the cannibalization correction terms kick in, which lead to improvements over naive $\TSRN$ as seen in Figure \ref{fig:numerics_homogeneous}. 
To simplify the exposition, we only consider two factors $k=1,2$; we see that $\TSRIo$ has lower bias than the naive $\TSR$, but $\TSRIt$, which has a higher weight in front of the correction terms, has a lower bias than both naive $\TSRN$ and $\TSRIo$, as well as the naive $\CR$ and $\LR$ estimators.  In Appendix \ref{app:simulations}, we explore the robustness of our results to other model primitives, specifically scenarios with smaller or larger utilities and the introduction of heterogeneity on one or both sides of the market. We find that the bias of $\CR$ and $\LR$ estimators can increase with the introduction of these factors, but remarkably the bias of the $\TSRI$ estimators remains low across the ranges that we study.  We emphasize the fact that the three $\TSR$ estimators presented here are examples to illustrate the potential for bias reduction using this new design. There is of course a much broader range of both $\TSR$ designs and estimators; some of these may offer even better performance.%

We conclude this section with two additional observations.  First, note that all our analysis in this section has been carried out in the mean field steady state; in particular, Figure \ref{fig:numerics_homogeneous} shows the bias of the estimators in steady state. 
For practical implementation, it is also important to consider the relative bias in the candidate estimators in the transient system, since experiments are typically run for relatively short time horizons. For discussion of the transient behavior for a finite time horizon, see Appendix \ref{app:transient}.  Second, in the next section, we discuss variance of the estimators we have studied.  There we find that the $\TSR$ estimators with the lowest bias also have the highest variance; in other words, there is a {\em bias-variance tradeoff}.

\section{Bias-variance tradeoff of estimators}
\label{sec:variance}

Our mean field model is deterministic, so it does not allow us to study the {\em variance} of the different estimators. In practice, however, markets consist of finitely many listings and experiments are run for finite time horizon $T$, and so the variance of any estimator will be nonzero.\footnote{We note that even if the system only consists of a finite number of listings $N$, as $T \to \infty$ the standard error of the various estimators proposed in this paper will converge to zero.  However, for finite $T$, this is not the case; since A/B tests are always run to a finite horizon $T$, this nonzero variance will impact the accuracy of any estimates obtained.}
In particular, the variance of estimators becomes an important consideration alongside bias, particularly in choosing between multiple estimators with similar bias.  The variance of the $\TSR$ estimators is especially important, given the earlier discussion that many heuristics that platforms use to minimize bias do so at the cost of increased variance, leading to under-powered experiments  (see Section \ref{sec:related}).

With this background as motivation, in this section we provide a preliminary yet suggestive simulation study of variance.  %
The simulations highlight two important considerations that a platform must take into account when designing and analyzing an experiment. First, similar to the results from Section \ref{sec:bias} on bias, we find that
the estimator with the lowest variance depends on market balance. Second, we see a bias-variance tradeoff between the $\CR$, $\LR$ and $\TSR$ estimators, with the $\TSR$ estimators offering bias improvements at the cost of an increase in variance. 
We emphasize the point that whether a platform should care more about bias or variance depends on the size of the platform (number of listings $N$) and the time horizon on which the experiment is run. The bias of the experiment is relatively unaffected by changes in these two factors, but of course variance decreases in the size of the market and the length of the time horizon. Thus these two factors dictate whether bias or variance contribute more to the overall $\RMSE$.

Full details of the simulation environment and parameters are in Appendix \ref{app:simulations}, which we briefly summarize here.  We simulate marketplace experiments with varying market parameters for a finite system with a number of listings $N=5000$ and fixed time horizon $T$. For each run of the simulation, we fix an experiment design (e.g., $\CR$, $\LR$, $\TSR$) and simulate customer arrivals and booking decisions until time $T$.  System evolution is simulated according to the continuous time Markov chain specified in \eqref{eq:R1}-\eqref{eq:R2}. We calculate the estimator corresponding to the experiment design, defined in \eqref{eq:naive_CR}-\eqref{eq:naive_TSR_est} and \eqref{eq:TSRI-A} (for $k=1,2$) for the time interval $[T_0, T]$, where $T_0$ is chosen to eliminate the transient burn-in period.  We then simulate multiple runs and compare the bias and standard error of the estimators across runs. 
Note that we report the true standard errors, calculated across simulation runs. For discussion on the estimation of standard errors, see Section \ref{sec:conclusion}.

\begin{figure}
    \centering
        \includegraphics[height=0.3\textwidth ]{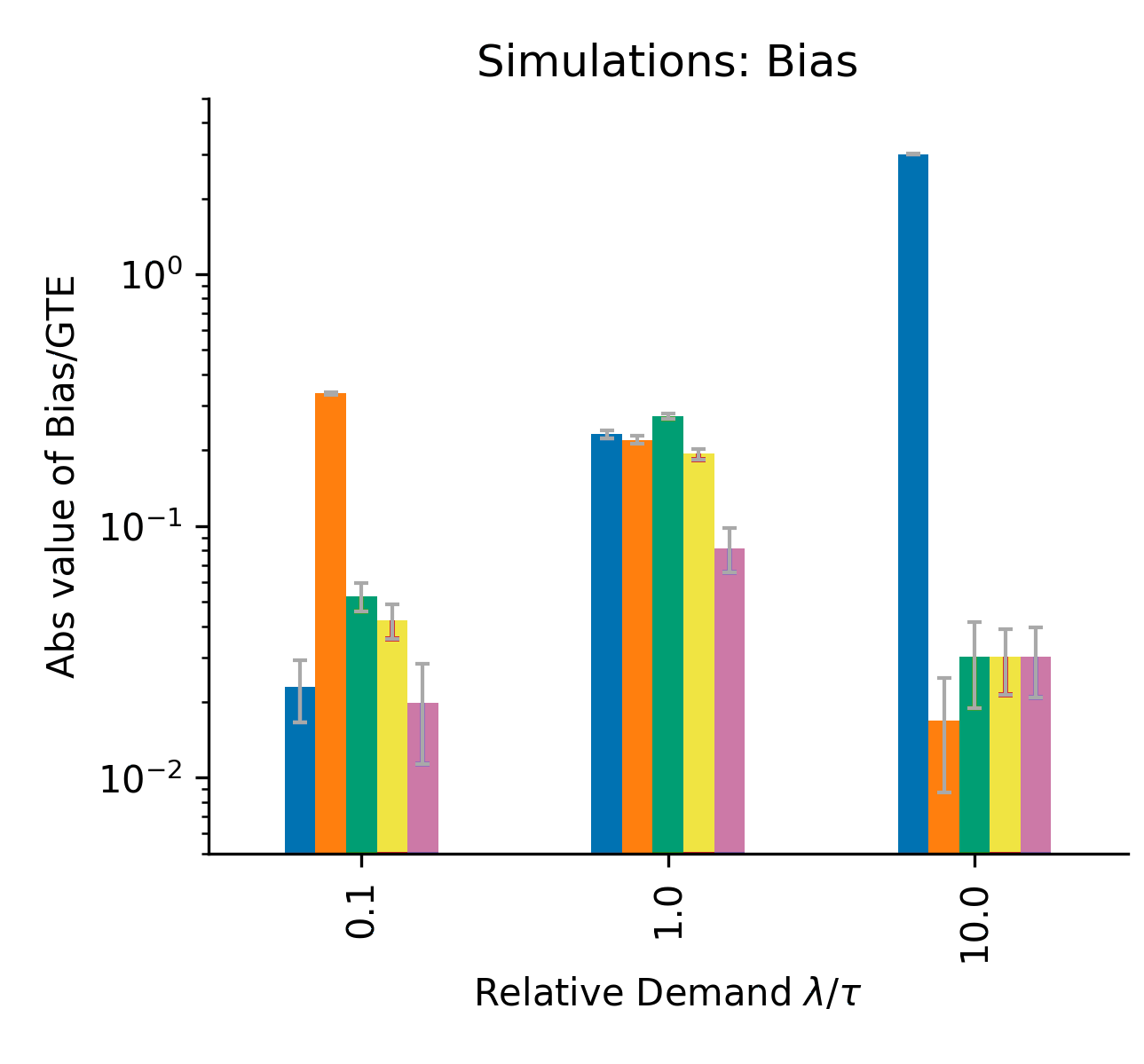}
        \includegraphics[height=0.3\textwidth]{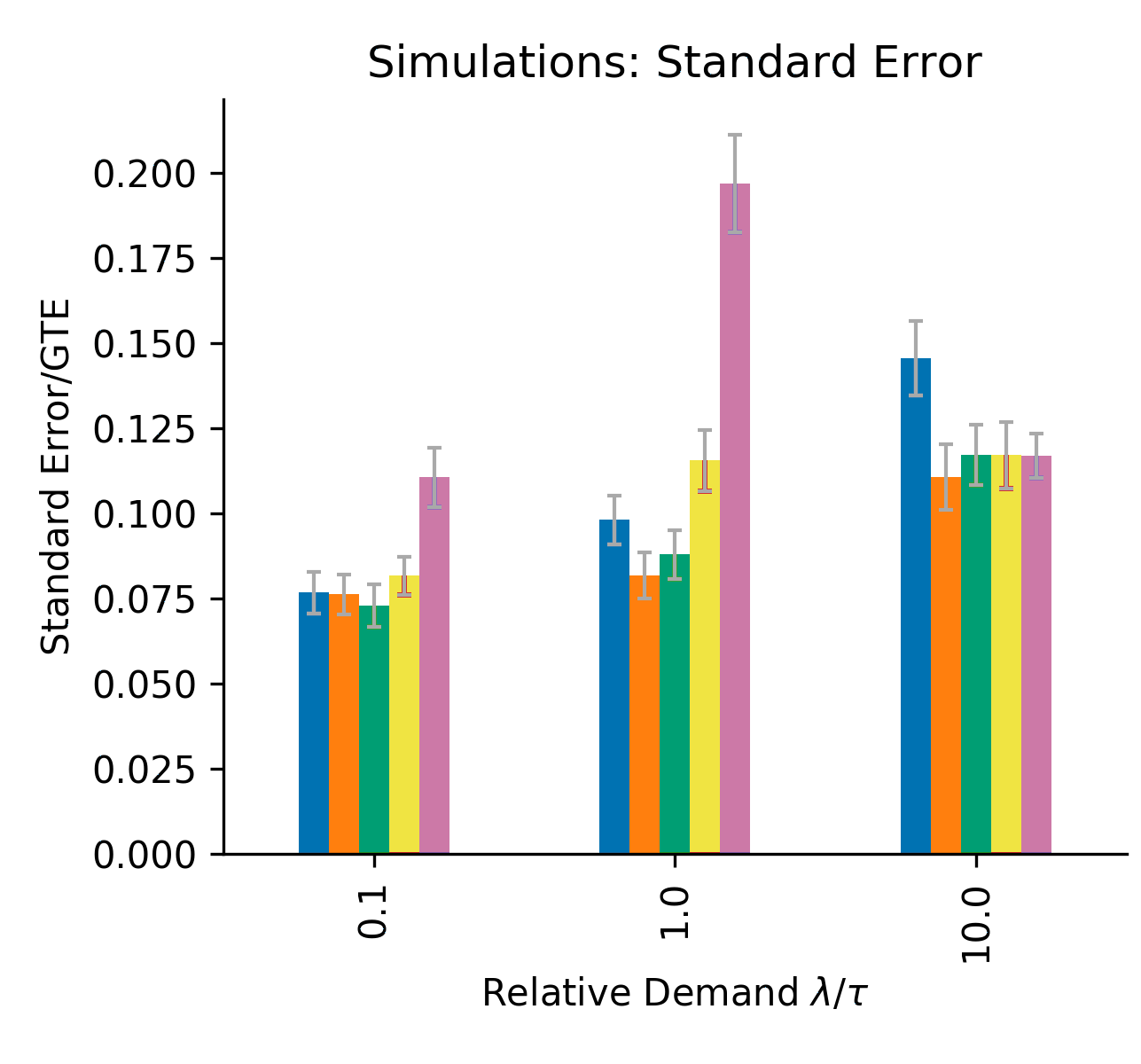}
        \\
        \hspace{2cm}\includegraphics[height=0.3\textwidth]{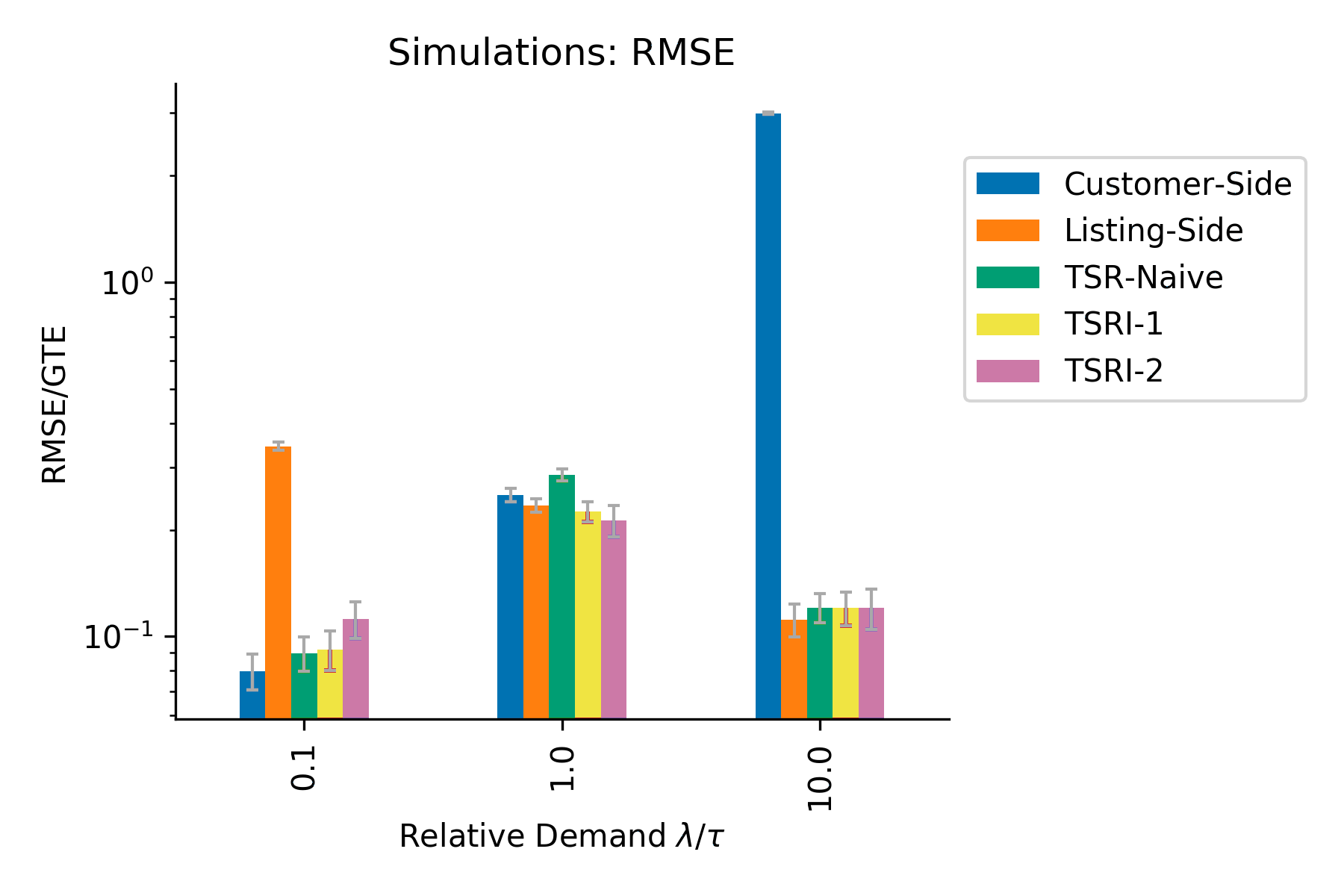}

    \caption{(Homogeneous listings and customers.) Top Left: Bias of each estimator. 
    Top right: Standard error of estimates. 
    Bottom: $\RMSE$ of the estimates. Statistics are normalized by $\GTE$. All statistics are calculated across 500 runs, with bootstrapped 95 percentile confidence intervals provided for each statistic.
    We consider a setting with homogeneous listings and customers, with the same utilities as defined in Figure \ref{fig:numerics_homogeneous}. In the $\CR$ design, $a_C = 0.5$. In the $\LR$ design, $a_L = 0.5$.  Simulation parameters are defined in Appendix \ref{app:simulations}.
    }
    \label{fig:simulations_hom}
\end{figure}

Figure \ref{fig:simulations_hom} shows simulations for a homogeneous system with only one customer and one listing type, with the same parameters as the mean field numerics presented in Figure \ref{fig:numerics_homogeneous}. Note that the bias of the estimators in these large market simulations echo the qualitative insights about bias obtained from the mean field model. Similar findings are obtained in more general scenarios, cf.~Appendix \ref{app:simulations}, where we investigate the effect of heterogeneity in the marketplace.

These simulations point to a bias variance-tradeoff between $\TSR$ estimators and the naive $\CR$ and $\LR$ estimators, as well as between the three $\TSR$ estimators themselves. The $\TSR$ estimators, as discussed earlier, offer benefits over the naive $\CR$ and $\LR$ estimators with respect to bias, but they do so at the cost of an increase in variance. Moreover, among the three $\TSR$ estimators that we explore, those with lower bias also have higher variance.  The naive $\TSRN$ estimator has similar variance with the lowest of $\CR$ and $\LR$, but the bias of this estimator is also similar to the lowest bias of $\CR$ and $\LR$. On the other hand, $\TSRIt$ shows a substantial improvement in bias over both $\CR$ and $\LR$ for several market conditions, but this estimator also has the largest variance among all five estimators, especially in the regime of intermediate market balance (cf. Appendix \ref{app:simulations}).

Further, the minimum variance estimator depends on market conditions. For example, in a demand-constrained market with $\lambda/\tau = 0.1$, $\widehat{\GTE}^{\CR}$ has the lowest standard error, whereas in a supply-constrained market with $\lambda/\tau=10$, $\widehat{\GTE}^{\CR}$ has the highest standard error.

We conclude this section by highlighting the potential for the class of $\TSR$ experiments, which open up a large class of both designs and estimators. Among the three estimators we explore, we see that there is a $\TSR$ estimator with low bias and another $\TSR$ estimator with low variance. It is possible that, with further optimization of these designs and estimators, one can devise a new estimator that optimizes this bias-variance tradeoff. %

\section{Comparison with cluster-randomized experiments}
\label{sec:cluster}

In this section we compare the $\TSR$ approach to existing approaches to reduce bias in marketplace experiments.  One such approach is to run a cluster-randomized experiment, which changes the unit of randomization in order to reduce interference effects across units. The typical approach is to divide the marketplace into clusters, such as geographical regions, such that there is less interaction of market participants across different clusters. All participants within a cluster receive the same treatment condition. The platform then estimates the $\GTE$ by comparing the outcomes within the treatment clusters versus the outcomes in the control clusters. It is important to note that many markets and social networks are highly connected and it is not possible to avoid all interference across clusters; see, e.g., \cite{holtz2020reducing} for an example in the context of Airbnb. Thus cluster-randomized experiments will reduce but not fully remove the bias.

To compare the performance of the cluster-randomized and $\TSR$ approaches, we use our existing model to define a regime that gives the best-case performance for cluster-randomized experiments, where there are tightly clustered preferences in the marketplace and the platform knows ex-ante the true clusters (without having to learn them).

The simulations suggest that cluster-randomized estimators offer substantial bias reductions over when the market is tightly clustered, but these improvements diminish if the market becomes more interconnected. The $\TSR$ estimators, however, offer bias reductions in both clustered and interconnected markets and, in interconnected markets, are less biased than the cluster-randomized estimator. 
In our simulations, the variance of the cluster-randomized estimator is lower than that of $\TSRIt$ in our example, although the variance of the cluster-randomized estimator will likely change if we deviate from this best-case scenario with perfect knowledge of the clusters, identical listings within clusters, and uniform treatment effects across different clusters.  Hence, while our model for market clusters is stylized, we believe our results suggest that $\TSR$ designs can be a useful alternative to cluster-randomized experiments in interconnected markets.

\subsection{Market setup for cluster randomization}

We consider the case where clusters are defined on the listing side, so that the cluster-randomized experiment is expected to improve upon a listing-randomized experiment.\footnote{Because we analyze a model where customers are short-lived while listings remain on the platform, it is likely that the platform has more information on the listings, and is better able to learn clusters on the listing side.} To induce a clustered structure, we model a setting with two customer types and two listing types, where each type of customer prefers a different type of listing. Formally, there are customer types $\{\gamma_1, \gamma_2\}$ and listing types $\{\theta_1, \theta_2\}$. All customers consider all listings ($\alpha_\gamma(\theta)=1$ for all $\gamma, \theta$) but customers have different utilities for different listings.\footnote{Alternatively, we can induce a clustered structure by modifying the consideration probabilities $\alpha_\gamma(\theta)$. Both approaches of modifying the $\alpha_\gamma(\theta)$ and modifying the $v_\gamma(\theta)$ are equivalent in the mean field model.} The global control utilities $v_\gamma(\theta)$ have the following form:
\medskip

\begin{center}
\begin{tabular}{ c | c c } 
 & $\theta_1$ & $\theta_2$ \\
\hline
$\gamma_1$ & $x$ & $y$ \\
$\gamma_2$ & $y$ & $x$ \\
\end{tabular}
\end{center}
\smallskip
where $x\geq y \geq 0$. 
Note that if $y=0$, then the market can be perfectly decomposed into two sub-markets, where customers of type $\gamma_1$ (resp., $\gamma_2$) \textit{only} book listings of type $\theta_1$ (resp., $\theta_2$). If $y=x$, then each customer prefers both listings equally. Thus we can interpret the ratio $y/x$ as a measure of how equally a customer prefers both products, where intuitively the market is tightly clustered when $y/x$ is small. We call $y/x$ the \textit{preference ratio}.

The platform then runs a cluster randomized experiment where it first assigns listings to clusters and then randomizes entire clusters to either treatment or control. 
In practice, the platform must \textit{learn} how to create the clusters, likely through observational data in the global control setting,\footnote{See \cite{holtz2020reducing} in the context of marketplaces and \cite{Ugander13} in the context of social networks.} but in these simulations, we assume that the platform observes the cluster structure perfectly. The platform assigns all $\theta_1$ listings to one cluster and $\theta_2$ listings to another, and runs a completely randomized design on the clusters, assigning one of the clusters to treatment and one to control. For simplicity, assume that the intervention has a multiplicative lift $\delta>1$ on all customer-listing pairs, so that the treatment utilities satisfy $\tilde{v}_\gamma(\theta) = \delta v_\gamma(\theta)$. %

The cluster-randomized estimator $\widehat{\GTE}^{\cluster}$, with clusters defined on the listing side, compares the (scaled) rate of bookings of listings in treatment clusters to the rate of bookings of listings in control clusters.
Formally, in the mean field setting, once the clusters are randomized, let $Z$ denote the mass of listings assigned to a treatment cluster. %
Then
\begin{equation}
\label{eq:naive_cluster}
 \widehat{\GTE}^{\cluster}(T | Z) = \frac{Q_{11}(T | 1,Z)}{Z} - \frac{Q_{10}(T | 1,Z)}{1-Z}. 
\end{equation}
We can similarly define an analogous estimator in the finite model.

The full set of parameters is as follows. The market has an equal number of listings of both types, so that $m^{(N)}(\theta_1) = m^{(N)}(\theta_2) = 0.5$, and the same arrival rate for both customers types ($\lambda_{\gamma_1}^{(N)} = \lambda_{\gamma_2}^{(N)}$). We set $x=0.5$ and vary $y \in [0,1]$. We set $\delta=1.3$. We fix $\tau=1$ and  consider $\lambda \in \{0.1, 1, 10\}$. 

\subsection{Results}

\begin{figure}
    \centering
    \begin{tabular}{l l }
        \includegraphics[height=.28\textwidth]{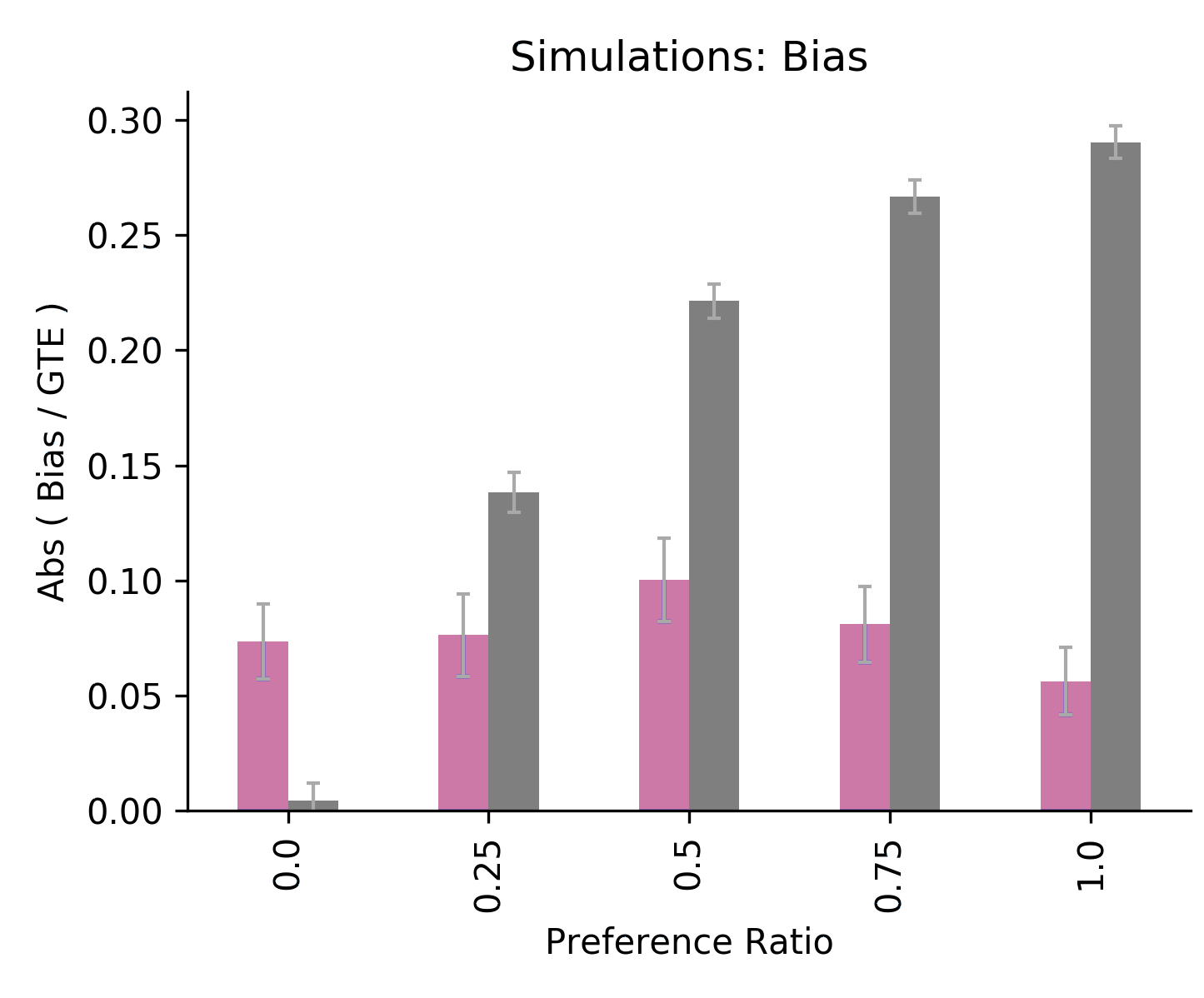}
        &
        \includegraphics[height=.28\textwidth]{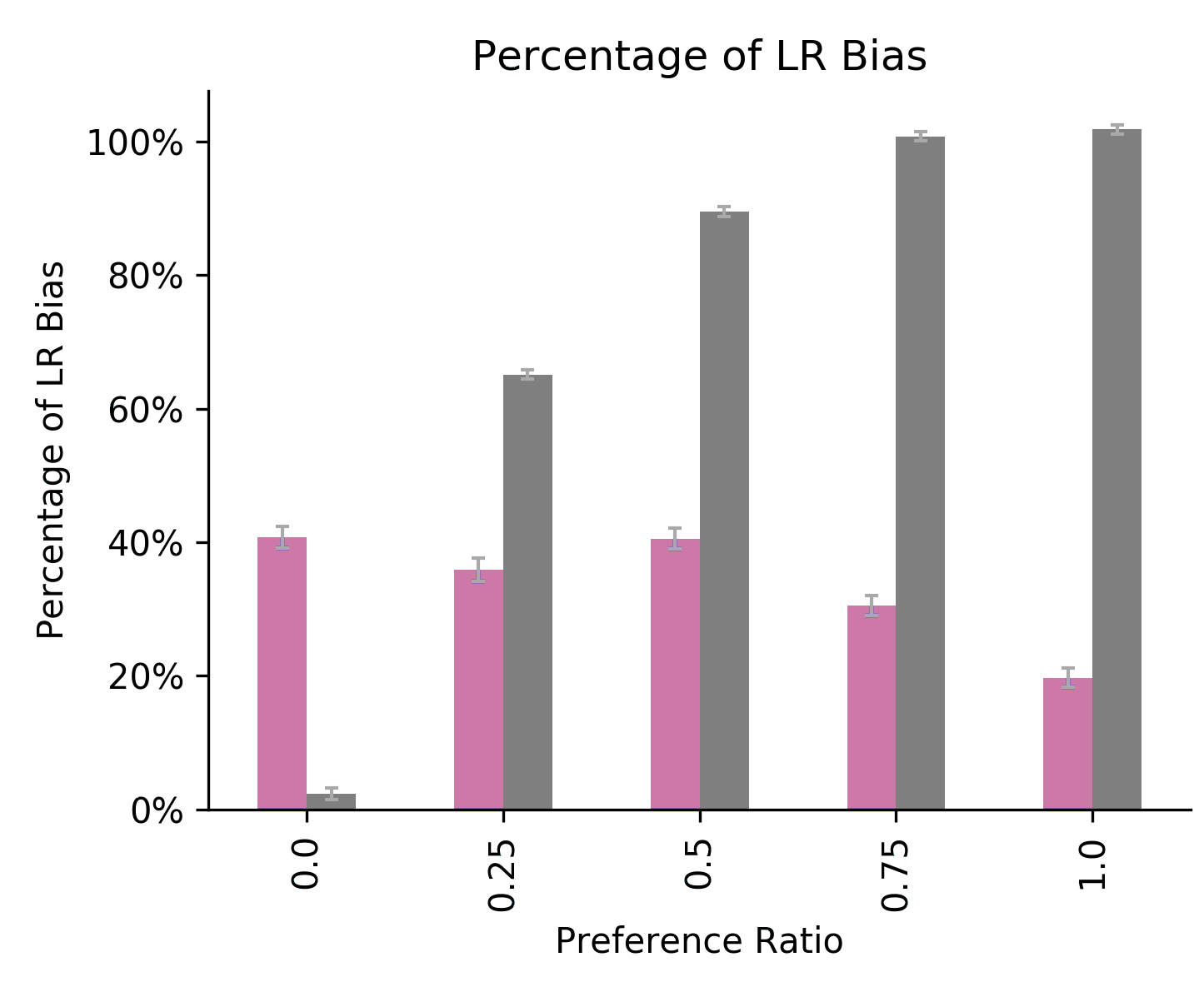}
        \\
        \includegraphics[height=.28\textwidth]{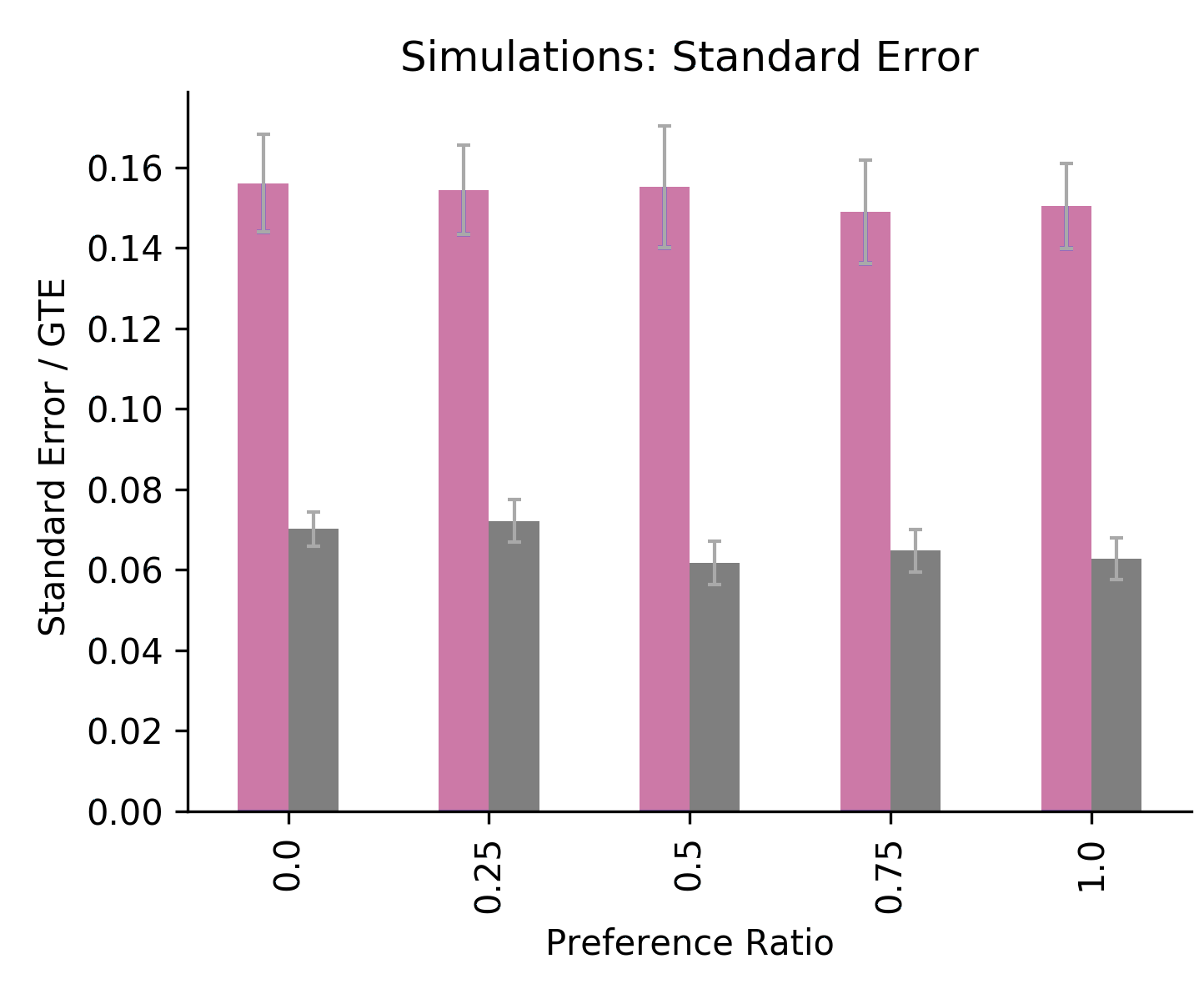}
        &
        \includegraphics[height=.28\textwidth]{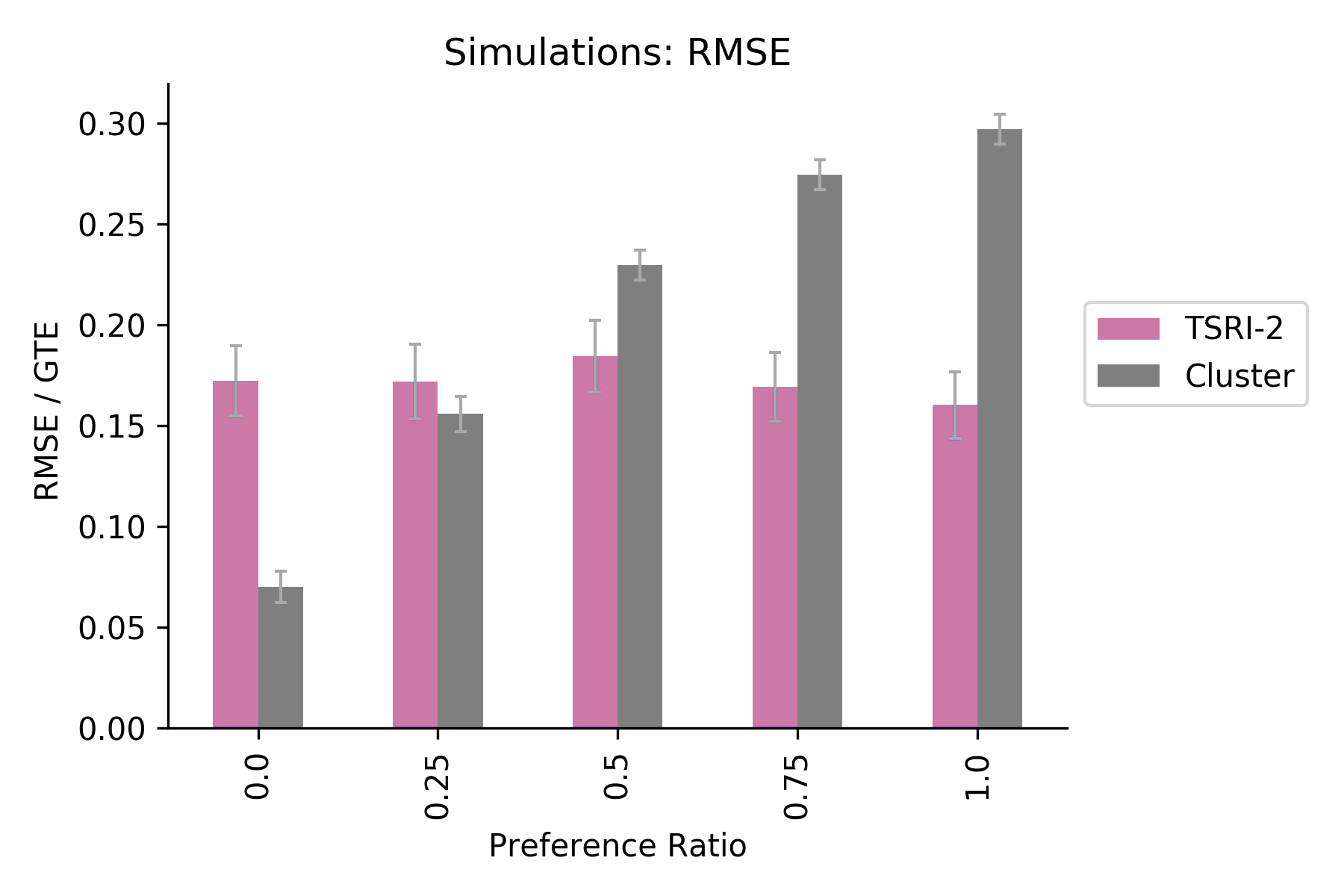}
    \end{tabular}
    \caption{Simulations of cluster-randomized estimator and TSR estimators as preference ratio $y/x$ varies. Relative demand is fixed at $\lambda/\tau=1$. We fix $x=0.5$ and vary $y$. Bootstrapped 95 percentile confidence intervals are provided for each statistic.
    }
    \label{fig:cluster_vary_pref_ratio}
\end{figure}

Figure \ref{fig:cluster_vary_pref_ratio} shows how the performance of the estimators change when we vary the preference ratio (at a fixed market balance). We choose $\TSRIt$ as a representative estimator for the $\TSR$ approach; see Figure \ref{fig:cluster_vary_pref_ratio_all_estimators} in Appendix \ref{app:cluster} for a comparison with all estimators. 

We see a clear takeaway that, perhaps unsurprisingly, the cluster randomized estimator offer substantial bias improvements when the market is tightly clustered (i.e., a customer strongly prefers one type of item over another) but offer little reduction in bias when the market is more interconnected.  In particular, the cluster-randomized estimator outperforms the $\TSR$ estimators when the market is tightly clustered, while the $\TSR$ estimators outperform the cluster-randomized estimator when the market is more connected.

The standard error changes little across the preference ratios, with the standard error of the cluster-randomized estimate lower than that of $\TSRIt$, although the variance of the cluster-randomized estimator will likely change if we deviate from this best-case scenario with perfect knowledge of the clusters, identical listings within clusters, and uniform treatment effects across clusters.  In our scenario, the cluster-randomized estimator has lower $\RMSE$ for tightly clustered markets and higher otherwise.

\section{Conclusion}
\label{sec:conclusion}

This paper proposes a general mean field framework to study the dynamics of inventory bookings in two-sided platforms, and we leverage this framework to study the design and analysis of a number of different experimental designs and estimators.  We study both commonly used designs and estimators ($\CR$, $\LR$), and also introduce a family of more general two-sided randomization designs and estimators ($\TSR$).  Our work sheds light on the market conditions in which each approach to estimation performs best, based on the relative supply and demand balance in the marketplace.

For bias minimization, we suggest two directions for future work. The first is further optimization of the $\TSR$ design as a standalone experiment design. We have proposed three natural $\TSR$ estimators, but the space of both designs and estimators is much richer and it is worth asking which are optimal with respect to bias and variance, as well as how this answer may change with differing market conditions. The second direction is to develop the $\TSR$ design as a method to debias one-sided experiments. The design allows us to measure competition effects between customers and between listings; this observation suggests that these measurements can be used to approximately debias existing $\CR$ and $\LR$ experiments, providing another route for platforms to utilize two-sided randomization designs.

To make technical progress in this paper, we employed several simplifying assumptions on the choice model and booking behavior. %
We believe that our core insight of market balance mediating competition effects, and thus affecting the resulting bias in an experiment, extends to other settings as well. 
We hypothesize, however, that some of our results are more robust to modeling choices than others. For example, the result that the $\CR$ estimator is unbiased in the demand constrained limit depends only on the fact that booked listings are replenished in between customer arrivals, and likely extends to any reasonable choice model. On the other hand, the result that the absolute bias of the $\LR$ estimator approaches 0 in the supply constrained limit may be more sensitive to different choice models, and in particular how the customer weighs options on the platform compared to the outside option. In this supply constrained limit, we conjecture that the relative performance of the two experiment types still holds even in modified settings; that is, the $\LR$ estimator has lower bias than the $\CR$ bias. Further, we believe that the approach of using $\TSR$ to observe competition effects and heuristically debias estimators also extends beyond our model.

Another practical consideration for platforms is that not all experiment designs are suitable for all types of interventions. In this paper, we have largely focused on interventions such as user interface changes that change how an \textit{individual} customer perceives an \textit{individual} listing. There are also other interventions that operate between \textit{subsets} of customers or \textit{subsets} of listings. For example, a modification in the ranking algorithm over the listings changes operates not on an individual customer-listing pair, but rather changes how a customer interacts with a subset of listings. This intervention is more conducive to a $\CR$ experiment than an $\LR$ or $\TSR$ experiment. Beyond these feasibility constraints, it remains an open question whether certain types of experiments lead to lower bias in different classes of interventions.

Finally, we emphasize the importance of inference in these settings, which we do not study in this paper. In practice, standard errors are also estimated ``naively": they are typically computed assuming independence of observations. However, because of interference, observations are clearly not independent. %
In these settings, how biased might the standard error estimates be? How can experimenters derive valid confidence intervals in these settings? Such questions are critical for any platforms controlling the false positive and false negative results arising from their experiments.

\ACKNOWLEDGMENT{We would like to thank Navin Sivanandam from Airbnb for numerous fruitful conversations. We would also like to thank Guillaume Basse, Dominic Coey, Nikhil Garg, David Holtz, Michael Luca, Johan Ugander, and Nicolas Stier for their feedback, as well as seminar participants at ACM EC'20, Facebook, Marketplace Algorithms and Design, CODE 2020, INFORMS 2020, Lyft, Vinted, Marketplace Innovations Workshop 2021, MSOM 2021, and Stanford. Finally, we thank the ACM EC'20 review team for their thoughtful comments and engagement with our work. \gw{Complete}

This work was supported by the National Science Foundation under grants 1931696 and 1839229 and the Dantzig-Lieberman Operations Research Fellowship.}


\bibliographystyle{informs2014} 
\bibliography{bibliography} 

\begin{thebibliography}{23}
\providecommand{\natexlab}[1]{#1}
\providecommand{\url}[1]{\texttt{#1}}
\providecommand{\urlprefix}{URL }

\bibitem[{Athey et~al.(2018)Athey, Eckles, \protect\BIBand{} Imbens}]{Athey18}
Athey S, Eckles D, Imbens GW (2018) Exact p-values for network interference.
  \emph{Journal of the American Statistical Association} 113(521):230--240,
  \urlprefix\url{http://dx.doi.org/10.1080/01621459.2016.1241178}.

\bibitem[{Bajari et~al.(2019)Bajari, Burdick, Imbens, McQueen, Richardson,
  \protect\BIBand{} Rosen}]{bajari2019double}
Bajari P, Burdick B, Imbens G, McQueen J, Richardson T, Rosen I (2019) Multiple
  randomization designs for interference
  \urlprefix\url{https://assets.amazon.science/c1/94/0d6431bf46f7978295d245dd6e06/double-randomized-online-experiments.pdf}.

\bibitem[{Basse et~al.(2019)Basse, Feller, \protect\BIBand{} Toulis}]{Basse19}
Basse GW, Feller A, Toulis P (2019) {Randomization tests of causal effects
  under interference}. \emph{Biometrika} 106(2):487--494, ISSN 0006-3444,
  \urlprefix\url{http://dx.doi.org/10.1093/biomet/asy072}.

\bibitem[{Basse et~al.(2016)Basse, Soufiani, \protect\BIBand{}
  Lambert}]{Basse16}
Basse GW, Soufiani HA, Lambert D (2016) Randomization and the pernicious
  effects of limited budgets on auction experiments. Gretton A, Robert CC,
  eds., \emph{Proceedings of the 19th International Conference on Artificial
  Intelligence and Statistics, {AISTATS} 2016, Cadiz, Spain, May 9-11, 2016},
  volume~51 of \emph{{JMLR} Workshop and Conference Proceedings}, 1412--1420
  (JMLR.org), \urlprefix\url{http://proceedings.mlr.press/v51/basse16b.html}.

\bibitem[{Blake \protect\BIBand{} Coey(2014)}]{Blake14}
Blake T, Coey D (2014) Why marketplace experimentation is harder than it seems:
  The role of test-control interference. \emph{Proceedings of the Fifteenth ACM
  Conference on Economics and Computation}, 567–582, EC ’14 (New York, NY,
  USA: Association for Computing Machinery), ISBN 9781450325653,
  \urlprefix\url{http://dx.doi.org/10.1145/2600057.2602837}.

\bibitem[{Boyd(2008)}]{boyd}
Boyd S (2008) Lecture 12 basic lyapunov theory.
  \urlprefix\url{https://web.stanford.edu/class/ee363/lectures/lyap.pdf}.

\bibitem[{Chamandy(2016)}]{chamandy16}
Chamandy N (2016) Experimentation in a ridesharing marketplace.
  \urlprefix\url{https://eng.lyft.com/experimentation-in-a-ridesharing-marketplace-f75a9c4fcf01}.

\bibitem[{Darling(2002)}]{darling2002fluid}
Darling RWR (2002) Fluid limits of pure jump markov processes: a practical
  guide.

\bibitem[{Edwards et~al.(2014)Edwards, Penney, \protect\BIBand{}
  Calvis}]{Edwards14}
Edwards CH, Penney DE, Calvis D (2014) \emph{Differential Equations and
  Boundary Value Problems: Computing and Modeling} (Pearson), 5th edition, ISBN
  0321796985.

\bibitem[{Fradkin(2015)}]{Fradkin2015SearchFA}
Fradkin A (2015) Search frictions and the design of online marketplaces.
  \emph{AMMA 2015}.

\bibitem[{Fradkin(2019)}]{Fradkin_2019}
Fradkin A (2019) A simulation approach to designing digital matching platforms.
  \emph{SSRN Electronic Journal} ISSN 1556-5068,
  \urlprefix\url{http://dx.doi.org/10.2139/ssrn.3320080}.

\bibitem[{Ha-Thuc et~al.(2020)Ha-Thuc, Dutta, Mao, Wood, \protect\BIBand{}
  Liu}]{hathuc2020counterfactual}
Ha-Thuc V, Dutta A, Mao R, Wood M, Liu Y (2020) A counterfactual framework for
  seller-side a/b testing on marketplaces. \emph{Proceedings of the 43rd
  International ACM SIGIR Conference on Research and Development in Information
  Retrieval}, 2288–2296, SIGIR ’20 (New York, NY, USA: Association for
  Computing Machinery), ISBN 9781450380164,
  \urlprefix\url{http://dx.doi.org/10.1145/3397271.3401434}.

\bibitem[{Holtz(2018)}]{Holtz18}
Holtz D (2018) \emph{Limiting bias from test-control interference in online
  marketplace experiments}. Ph.D. thesis,
  \urlprefix\url{https://dspace.mit.edu/handle/1721.1/117999}.

\bibitem[{Holtz et~al.(2020)Holtz, Lobel, Liskovich, \protect\BIBand{}
  Aral}]{holtz2020reducing}
Holtz D, Lobel R, Liskovich I, Aral S (2020) Reducing interference bias in
  online marketplace pricing experiments.

\bibitem[{Imbens \protect\BIBand{} Rubin(2015)}]{ImbensRubin15}
Imbens GW, Rubin DB (2015) \emph{Causal Inference for Statistics, Social, and
  Biomedical Sciences: An Introduction} (USA: Cambridge University Press), ISBN
  0521885884.

\bibitem[{Kurtz(1970)}]{kurtz1970}
Kurtz TG (1970) Solutions of ordinary differential equations as limits of pure
  jump markov processes. \emph{Journal of Applied Probability} 7(1):49--58,
  ISSN 00219002, \urlprefix\url{http://www.jstor.org/stable/3212147}.

\bibitem[{Manski(2013)}]{Manski13}
Manski CF (2013) Identification of treatment response with social interactions.
  \emph{The Econometrics Journal} 16(1):S1--S23,
  \urlprefix\url{http://dx.doi.org/10.1111/j.1368-423X.2012.00368.x}.

\bibitem[{Pouget-Abadie et~al.(2019)Pouget-Abadie, Aydin, Schudy, Brodersen,
  \protect\BIBand{} Mirrokni}]{pouget2019variance}
Pouget-Abadie J, Aydin K, Schudy W, Brodersen K, Mirrokni V (2019) Variance
  reduction in bipartite experiments through correlation clustering.
  \emph{Advances in Neural Information Processing Systems}, 13288--13298.

\bibitem[{Saveski et~al.(2017)Saveski, Pouget-Abadie, Saint-Jacques, Duan,
  Ghosh, Xu, \protect\BIBand{} Airoldi}]{Saveski17}
Saveski M, Pouget-Abadie J, Saint-Jacques G, Duan W, Ghosh S, Xu Y, Airoldi EM
  (2017) Detecting network effects: Randomizing over randomized experiments.
  \emph{Proceedings of the 23rd ACM SIGKDD International Conference on
  Knowledge Discovery and Data Mining}, 1027–1035, KDD ’17 (New York, NY,
  USA: Association for Computing Machinery), ISBN 9781450348874,
  \urlprefix\url{http://dx.doi.org/10.1145/3097983.3098192}.

\bibitem[{Sneider et~al.(2019)Sneider, Tang, \protect\BIBand{}
  Tang}]{sneider19}
Sneider C, Tang Y, Tang Y (2019) Experiment rigor for switchback experiment
  analysis.
  \urlprefix\url{https://doordash.engineering/2019/02/20/experiment-rigor-for-switchback-experiment-analysis/}.

\bibitem[{Ugander et~al.(2013)Ugander, Karrer, Backstrom, \protect\BIBand{}
  Kleinberg}]{Ugander13}
Ugander J, Karrer B, Backstrom L, Kleinberg J (2013) Graph cluster
  randomization: Network exposure to multiple universes. \emph{Proceedings of
  the 19th ACM SIGKDD International Conference on Knowledge Discovery and Data
  Mining}, 329–337, KDD ’13 (New York, NY, USA: Association for Computing
  Machinery), ISBN 9781450321747,
  \urlprefix\url{http://dx.doi.org/10.1145/2487575.2487695}.

\bibitem[{Wager \protect\BIBand{} Xu(2019)}]{Wager19}
Wager S, Xu K (2019) Experimenting in equilibrium.
  \urlprefix\url{https://arxiv.org/abs/1903.02124}.

\bibitem[{Zigler \protect\BIBand{} Papadogeorgou(2018)}]{zigler2018bipartite}
Zigler CM, Papadogeorgou G (2018) Bipartite causal inference with interference.
  \emph{arXiv preprint arXiv:1807.08660} .

\end{thebibliography}

\newpage

\clearpage
\pagenumbering{arabic}
\renewcommand*{\thepage}{A\arabic{page}}
\begin{APPENDICES}

\section{Proofs}

\label{app:proofs}

\proof{Proof of Proposition \ref{prop:ODE_trajectory}.}
For each $\theta \in \Theta$, define $f_\theta(\v{s})$ to be the right hand side of \eqref{eq:ODE}:
\[ f_\theta(\v{s}) = (\rho(\theta) - s(\theta))\tau(\theta) - \lambda \sum_\gamma \phi_\gamma  p_\gamma(\theta | \v{s}). \]
Let $\v{f}$ denote the $|\Theta|$-dimensional vector-valued function where each component is defined by $f_\theta$.
For a fixed $c > 0$ define the set $I = (- c, \infty)$ of times $t$ for which we wish to show the solution is unique.   We require that $\v{s}_t \in \mcal{S}$ for all $t \in I$ and $\v{s}_0 = \hat{\v{s}}$.

By the Picard-Lindel\"{o}f theorem \cite{Edwards14}, if $\v{f}(\v{s})$ is Lipschitz continuous for all $\v{s} \in \mcal{S}$, then there exists a unique solution $\{\v{s}_t : t \in I\}$ on the entire time interval $I$ with the desired initial condition. We will show that each component $f_\theta(\v{s})$ satisfies the Lipschitz condition, which then implies that the vector-valued function $\v{f}(\v{s})$ satisfies the condition.

Consider the partial derivatives of $f_\theta(\v{s})$ with respect to each $s(\theta')$:
\[ 
\frac{\partial f_\theta(\v{s})}{\partial s(\hat{\theta})} = \left \{ \begin{array}{ll}

- \lambda \sum_{\gamma} \phi_\gamma \cdot 
        \frac{(\epsilon_\gamma +  \sum_{\theta'}\alpha_\gamma(\theta') s(\theta')v_\gamma(\theta') ) \cdot \alpha_\gamma(\theta) v_\gamma(\theta) - \alpha_\gamma(\theta)^2 s(\theta) v_\gamma^2(\theta)}{(\epsilon_\gamma +  \sum_{\theta'} \alpha_\gamma(\theta') s(\theta') v_\gamma(\theta'))^2} - \tau(\theta),& \ \ \hat{\theta} = \theta;\\
\lambda \sum_\gamma \phi_\gamma \cdot \frac{\alpha_\gamma(\theta)^2 s(\theta)v_\gamma(\theta) s(\hat{\theta}) v_\gamma(\hat{\theta}) }{(\epsilon_\gamma +  \sum_{\theta'}\alpha_\gamma(\theta') s(\theta') v_\gamma(\theta'))^2},& \ \ \hat{\theta} \neq \theta.
\end{array}    \right .
\]
The partial derivatives of $f_\theta(\v{s})$ are continuous and (since $\epsilon_\gamma > 0$ for all $\gamma$) are bounded on $\mcal{S}$, and so $f_\theta(\v{s})$ is Lipschitz continuous on $\mcal{S}$.  It follows then that $\v{f}(\v{s})$ is Lipschitz on $\mcal{S}$ and so there exists a unique solution $\{\v{s}_t: t \geq 0\}$ to \eqref{eq:ODE} in $\mcal{S}$ such that $\v{s}_0 = \hat{\v{s}}$.\hfill$\Box$
\endproof

\proof{Proof of Theorem \ref{thm:ODE_ss}.}
Let $\mcal{Y} = \{ \v{y} : y(\theta) \leq \log \rho(\theta) \}$.  For $\v{y} \in \mcal{Y}$, define $\omega_{\v{y}}(\theta) = e^{y(\theta)}$; note that $\v{\omega}_{\v{y}} \in \mcal{S}$.  For $\v{y} \in \mcal{Y}$, we define $V(\v{y}) = W(\v{\omega}_{\v{y}})$, i.e.:
\begin{equation}
\label{eq:V}
 V(\v{y}) =  \lambda \sum_\gamma \left[ \phi_\gamma \log(\epsilon_\gamma +  \sum_\theta \alpha_\gamma(\theta) e^{y(\theta)} v_\gamma(\theta)) \right] - \sum_\theta \tau(\theta) \rho(\theta) y(\theta) + \sum_\theta \tau(\theta) e^{y(\theta)}.
\end{equation}

We prove the theorem in a sequence of steps.  

{\em Step 1: $V(\v{y})$ is strictly convex for $\v{y} \in \mcal{Y}$.}  The first term is the log-sum-exp function, which is strictly convex (recall $\epsilon_\gamma > 0$ for all $\gamma$); the second term is linear; and the last term is strictly convex.  

{\em Step 2: $V(\v{y})$ possesses a unique minimum $\v{y}^*$ on $\mcal{Y}$.}  Note that as $y(\theta) \to -\infty$, we have $V(\v{y}) \to \infty$ (recall that $\epsilon_\gamma > 0$ for all $\gamma$).  Therefore $V$ must possess a minimizer on $\mcal{Y}$; since $V$ is strictly convex, this minimizer is unique.

{\em Step 3: Define $\v{s}^* = \omega_{\v{y}^*}$, i.e., $s^*(\theta) = e^{y^*(\theta)}$.  Then $\v{s}^*$ is the unique solution to \eqref{eq:ss_objective}-\eqref{eq:ss_constraint}.}  Since $V(\v{y}) = W(\v{\omega}_{\v{y}})$, and the mapping $\v{y} \mapsto \v{\omega}_{\v{y}}$ maps $\mcal{Y}$ to $\{ \v{s} : 0 < s(\theta) \leq \rho(\theta) \} \subset \mcal{S}$, it suffices to show that \eqref{eq:ss_objective}-\eqref{eq:ss_constraint} cannot be minimized at any $\v{s}$ such that $s(\theta) = 0$ for some $s(\theta)$.  To see this, note that since $V(\v{y}) \to \infty$ as $y(\theta) \to -\infty$, it follows that $W(\v{s}) \to \infty$ as $s(\theta) \to 0$.  It follows that $\v{s}^*$ is the unique solution to \eqref{eq:ss_objective}-\eqref{eq:ss_constraint}.

{\em Step 4: $\v{y}^*$ lies in the interior of $\mcal{Y}$, and thus $\v{s}^*$ lies in the interior of $\mcal{S}$.} We have already shown that $s^*(\theta) > 0$ for all $\theta$.  It is straightforward to check that if $y(\theta) = \log \rho(\theta)$, the derivative of $V(\v{y})$ becomes positive, because the derivative of the first term of $V(\v{y})$ with respect to $y(\theta)$ is always positive, and the derivatives of the last two terms cancel when $y(\theta) = \log \rho(\theta)$.  Therefore we must have $y^*(\theta) < \rho(\theta)$ for all $\theta$, which suffices to establish the claim.\\

For the next step, fix an initial condition $\v{s}_0 \in \mcal{S}$ with $s_0(\theta) > 0$ for all $\theta$, and let $\v{s}_t$ be the resulting trajectory of \eqref{eq:ODE}.  We first observe that the right hand side of \eqref{eq:ODE} is equal to $\tau(\theta) \rho(\theta)$ when $s(\theta) = 0$, and this is positive; therefore, we must have  $s_t(\theta) > 0$ for all $t \geq 0$.  Define $y_t(\theta) = \log s_t(\theta)$, and let $\v{y}_t = (y_t(\theta), \theta \in \Theta)$.\\

{\em Step 5: $V$ is a Lyapunov function for $\{ \v{y}_t : t \geq 0\}$.  Further, $\v{y}^*$ is the unique limit point of $\{ \v{y}_t : t \geq 0 \}$, and it is globally asymptotically stable over all $\v{y}_0 \in \mcal{Y}$.}  We consider the function $V(\v{y}_t)$ as a function of $t$.  By the chain rule, we have:
\begin{align*}
    \frac{d}{dt} V(\v{y}_t) 
    &= \sum_\theta \frac{\partial V(\v{y_t})}{\partial y(\theta)} 
    \cdot \frac{d y_t(\theta)}{ds_t(\theta)} 
    \cdot \frac{ds_t(\theta)}{dt} \\
    &= \sum_\theta \left( \sum_\gamma \left( \frac{\lambda \phi_\gamma \alpha _\gamma(\theta) e^{y_t(\theta)} v_\gamma(\theta)}{ \epsilon_\gamma +  \sum_{\theta'} \alpha_\gamma(\theta') e^{y_t(\theta')} v_\gamma(\theta')} \right)- \tau(\theta) \rho(\theta) + \tau(\theta) e^{y_t(\theta)} \right)
    \cdot \frac{1}{s_t(\theta)} \\
    & \quad \quad \cdot \left( \sum_\gamma \left( \frac{-\lambda \phi_\gamma \alpha_\gamma(\theta) e^{y_t(\theta)} v_\gamma(\theta)}{ \epsilon_\gamma +  \sum_{\theta'} \alpha_\gamma(\theta') e^{y_t(\theta')} v_\gamma(\theta')} \right) + \tau(\theta) \rho(\theta) - \tau(\theta) e^{y_t(\theta)} \right) \\
    &= -\sum_\theta \frac{1}{s_t(\theta)} \left( \sum_\gamma \left( \frac{\lambda \phi_\gamma \alpha_\gamma(\theta) e^{y_t(\theta)} v_\gamma(\theta)}{ \epsilon_\gamma + \sum_{\theta'} \alpha_\gamma(\theta') e^{y_t(\theta')} v_\gamma(\theta')} \right) - \tau(\theta) \rho(\theta) + \tau(\theta) e^{y_t(\theta)} \right) ^2 \\
&=  -\sum_\theta \frac{1}{s_t(\theta)} \left( \frac{\partial V(\v{y}_t)}{\partial y(\theta)}\right)^2.
\end{align*}
It follows that $dV(\v{y}_t)/dt < 0$ whenever $\v{y}_t \neq \v{y}^*$, and $dV(\v{y}_t)/dt = 0$ if and only if $\v{y}_t = \v{y}^*$.  $V$ is clearly positive definite, since it is strictly convex; and as shown, it is minimized at $\v{y}^*$.  Thus it is a Lyapunov function for $\v{y}_t$, as required \cite{boyd}.

{\em Step 6: $\v{s}^*$ is the unique limit point of $\{ \v{s}_t : t \geq 0 \}$,  and it is globally asymptotically stable over all $\v{s}_0 \in \mcal{S}$.}   This follows from the preceding observation, as long as $\v{s}_0$ satisfies $s_0(\theta) > 0$ for all $\theta$.  If $s_0(\theta) = 0$ for some $\theta$, then again because the right hand side of \eqref{eq:ODE} is positive when $s_t(\theta) = 0$, we must have $s_t(\theta) \geq 0$ for all $t > 0$.  In this case we need only define $y_t(\theta) = \log s_t(\theta)$ for $t > 0$, the desired result follows from the preceding step.  This completes the proof of the theorem.\hfill$\Box$
\endproof

\proof{Proof of Theorem \ref{thm:mf_convergence}.}

Kurtz's theorem (see \cite{kurtz1970}
and \cite{darling2002fluid}) provides conditions under which a sequence of Markov jump processes converges to the mean field (or fluid) limit.  In our proof, we follow the development of \cite{darling2002fluid} to establish the claimed convergence in the theorem.

First we define some necessary notation. Let $T^{(N)}_i$ denote the time of the $i$'th jump; in the proof we suppress the index $(N)$ and instead write $T_i$. 
Let %
$$\mu^{(N)}(\v{y})=\E[\v{Y}^{(N)}_{T_{i+1}} - \v{Y}^{(N)}_{T_i} | \v{Y}^{(N)}_{T_{i+1}} = \v{y}]$$ 
and 
$$\Sigma^{(N)}(\v{y}) = \Var( \v{Y}^{(N)}_{T_{i+1}} - \v{Y}^{(N)}_{T_i} | \v{Y}^{(N)}_{T_{i+1}} = \v{y} )$$ 
denote the mean and covariance matrix of the increments.  Let  $c^{(N)}(\v{y})$ be the rate function, i.e.,  $T_{i+1} - T_i \sim \exp(c^{(N)}(\v{y}))$ if $\v{Y}_{T_i}^{(N)} = \v{y}$. Define 
$$b^{(N)}(\v{y}) = c^{(N)}(\v{y}) \mu^{(N)}(\v{y}).$$ 

Note that for each $N$ we have $I_N \subseteq \mcal{S}$, where $I_N$ is the state space of $\v{Y}_t^{(N)}$ (cf.~\eqref{eq:I_N}) and $\mcal{S}$ is the state space of the mean field ODE (cf.~\eqref{eq:S}).

Following the presentation of Kurtz' theorem in \cite{darling2002fluid}, the desired result follows if we establish the following conditions.

\begin{enumerate}
    \item {\em Convergence of initial conditions.} For all $\delta>0$, there exists $\kappa_1(\delta)$ such that for all $N$,
\begin{equation}
\label{eq:cond_initial_condition}
    \P[ \| \v{Y}_0^{(N)}  - \hat{\v{s}} \| > \delta ] \leq \kappa_1(\delta)/N,
\end{equation}
where $\v{Y}_0^{(N)}(\theta)$ denotes the initial condition of the chain.  (Recall from the theorem statement that $\hat{\v{s}}$ is the initial condition of the mean field ODE.) 
    \item {\em Convergence of mean dynamics.}
There exists a Lipschitz vector field $b:\mcal{S} \rightarrow \mcal{S}$ such that
\begin{equation}
\label{eq:cond_mean_dynamics}
    \sup_{y \in I_N} \| b^{(N)}(\v{y}) - b(\v{y}) \| \rightarrow 0.
\end{equation}
    \item {\em Convergence of noise to zero.} There exists $\kappa_2, \kappa_3$ such that the following two conditions hold:
\begin{align}
    \label{eq:cond_rate_function_linear_n}
    \sup_{y \in I_N} & \ c^{(N)}(y) \leq \kappa_2 N \\
    \label{eq:cond_variance}
    \sup_{y \in I_N} & \  \text{Trace}[\Sigma^{(N)}(\v{y})] + \| \mu^{(N)}(\v{y}) \|^2 \leq \kappa_3 N^{-2}.
\end{align}
\end{enumerate}

Once we show that \eqref{eq:cond_initial_condition}-\eqref{eq:cond_variance} are satisfied, Theorem \ref{thm:mf_convergence} follows from an application of Kurtz's theorem to this setting.

We assume without loss of generality for the remainder of the proof that there exists a constant $C$ such that for all $\gamma$ the customer arrival rates satisfy
$$ \frac{\lambda_\gamma^{(N)}}{N} \leq C \lambda_\gamma.$$

{\em Step 1: Find $\kappa_1(\delta)$ such that \eqref{eq:cond_initial_condition} holds.}
Note that there is no randomness in our initial conditions and so the probability on the left hand side of \eqref{eq:cond_initial_condition} is either 0 or 1. 

For $\delta > 0$, choose $\kappa_1(\delta)$ such that if $N > \kappa_1(\delta)$, then $\| \v{Y}_0^{(N)} - \hat{\v{s}} \| < \delta$.  Then for $N \leq \kappa_1(\delta)$, the right hand side of \eqref{eq:cond_initial_condition} is greater than or equal to $1$, and so the condition trivially holds.  On the other hand, for $N > \kappa_1(\delta)$, the left hand side of \eqref{eq:cond_initial_condition} is zero, and so again the condition trivially holds.

{\em Step 2: Find $b$ such that \eqref{eq:cond_mean_dynamics} holds.}
Define $b:\mcal{S} \rightarrow \mcal{S}$ to be the right hand side of the ODE defined in \eqref{eq:ODE}, so each component of $b$ is given by
\begin{equation}
b_\theta(\v{y}) = (\rho(\theta) - y(\theta))\tau(\theta) - \sum_\gamma \lambda_\gamma  p_\gamma(\theta | \v{y}),\ \ \theta \in \Theta
\end{equation}
and 
\begin{equation}
p_\gamma( \theta | \v{y}) \triangleq \frac{\alpha_\gamma(\theta) v_\gamma(\theta) y(\theta) }{\epsilon_\gamma + \sum_{\theta'} \alpha_\gamma(\theta') v_\gamma(\theta') y(\theta')}.
\end{equation}
We have already shown in the proof of Proposition \ref{prop:ODE_trajectory} that $b$ is Lipschitz on $\mcal{S}$ and so we need only to prove convergence. 

In the $N$'th system, since increments are proportional to $1/N$, we have
\begin{align*}
    b_\theta^{(N)}(\v{y}) &= c^{(N)}(\v{s}) \mu^{(N)}_\theta(\v{s}) \\
    &= \frac{1}{N} \left( m^{(N)}(\theta) - N y(\theta) \right) \tau(\theta) - \frac{1}{N} \sum_\gamma \lambda_\gamma^{(N)}  r_\gamma^{(N)}(\theta | \v{y})
\end{align*}
with
\begin{equation*}
    r_\gamma^{(N)}(\theta | \v{y})
    = \E \left[
    \frac{ D_\gamma^{(N)}(\theta | \v{y}) 
    v_{\gamma}(\theta)}{\epsilon_{\gamma}^{(N)} + \sum_{\theta'} D_\gamma^{(N)}(\theta' | \v{y}) 
    v_{\gamma}(\theta')}\right],
\end{equation*}
where we have overloaded notation to define $D_\gamma^{(N)}(\theta | \v{y}) \sim \text{Binomial}(N y(\theta), \alpha_\gamma(\theta))$.  

Condition \eqref{eq:cond_mean_dynamics} follows from these definitions of $b_\theta^{(N)}$ and $b_\theta$ once we show that
\begin{equation}
\label{eq:mnl_convergence_sup_norm}
    \sup_{\v{y} \in I_N} | r_\gamma^{(N)}(\theta|\v{y}) - p_\gamma(\theta|\v{y}) | \rightarrow 0
\end{equation}
for all $\theta$ and $\gamma$. 

We start with the following lemma.

\begin{lemma}
\label{lem:logitbound_outsideoption}
Fix nonnegative constants $A, B$, and  $\epsilon, \epsilon' > 0$.  Then:
\[ \left| \frac{A}{\epsilon + A + B} - \frac{A}{\epsilon' + A + B}\right| \leq \frac{A}{\epsilon \epsilon'} \cdot | \epsilon - \epsilon'|. \]
\end{lemma}

\proof{Proof of lemma.}
We have: 
\[ \frac{A}{\epsilon + A + B} - \frac{A}{\epsilon' + A + B} = \frac{A(\epsilon - \epsilon')}{(\epsilon + A + B)(\epsilon' + A +B)}. \]
The proof follows since $A, B \geq 0$  and $\epsilon, \epsilon' > 0$.\hfill$\Box$
\endproof

With this lemma in hand, define:
\[ p_\gamma^{(N)}( \theta | \v{y}) \triangleq \frac{\alpha_\gamma(\theta) v_\gamma(\theta) y(\theta) }{\epsilon^{(N)}_\gamma/N + \sum_{\theta'} \alpha_\gamma(\theta') v_\gamma(\theta') y(\theta')}. \]
Since $\epsilon_\gamma^{(N)}/N \to \epsilon$ as $N \to \infty$, it follows from Lemma \ref{lem:logitbound_outsideoption} that:
\[ \sup_{\v{y} \in \mcal{S}} | p_\gamma^{(N)}(\theta | \v{y}) - p_\gamma(\theta | \v{y}) | \to 0 \]
as $N \to \infty$.  Thus it suffices to show that:
\begin{equation}
\label{eq:mnl_convergence_sup_norm_mod}
    \sup_{\v{y} \in I_N} | r_\gamma^{(N)}(\theta|\v{y}) - p_\gamma^{(N)}(\theta|\v{y}) | \rightarrow 0.
\end{equation}

We now introduce two further lemmas.

\begin{lemma}
\label{lem:logitbound}
Let $X$ and $Z$ be nonnegative random variables with means $\mu_X$ and $\mu_Z$ respectively.  Fix $\epsilon > 0$.  Then:
\[ \left| \E\left[ \frac{ X}{\epsilon + X + Z} \right] - \frac{\mu_X}{\epsilon + \mu_X + \mu_Z} \right|  \leq \frac{(\epsilon + \mu_Z) \E[|X-\mu_X|] + \mu_X \E[|Z - \mu_Z|]}{\epsilon^2}. \]
\end{lemma}

\proof{Proof of lemma.}
Observe that since $\epsilon > 0$ and both $X$ and $Z$ are nonnegative, we have:
\[  \frac{ X}{\epsilon + X + Z}  - \frac{\mu_X}{\epsilon + \mu_X + \mu_Z} \leq \frac{(\epsilon + \mu_Z)(X - \mu_X) + \mu_X (\mu_Z - Z)}{\epsilon^2}. \]
By taking the expectation of the absolute value of both sides and using the triangle inequality, we have:
\[ \E \left[ \left| \frac{ X}{\epsilon + X + Z}  - \frac{\mu_X}{\epsilon + \mu_X + \mu_Z} \right| \right] \leq \frac{(\epsilon + \mu_Z)\E[|X - \mu_X|] + \mu_X \E[|\mu_Z - Z)|]}{\epsilon^2}. \]
Finally, the lemma follows since by Jensen's inequality, the absolute value of the expectation is less than or equal to the expectation of the absolute value.\hfill$\Box$
\endproof

\begin{lemma}
\label{lem:binomial_concentration}
Fix $B > 0$ and $p \in [0,1]$.  For each $y$ such that $0 \leq y \leq B$ and $yN \in \mathbb{Z}$, let $X^{(N)}(y)$ be a binomial random variable with success probability $p$ and number of trials $Ny$.  Then:
\[ \sup_{0 \leq y \leq B : Ny \in \mathbb{Z}} \E \left[ \left|  \frac{X^{(N)}(y)}{N} - py \right| \right] \to 0 \]
as $N \to \infty$.
\end{lemma}

\proof{Proof of lemma.}
If $p = 0$ or $p = 1$, the result is immediate, so we assume without loss of generality that $0 < p < 1$.  

Fix $a > 0$.  Note that for all $N$ and for all $y \leq a$ we have:
\begin{equation}
\label{eq:y_less_than_a}
\E \left[ \left|  \frac{X^{(N)}(y)}{N} - py \right| \right] \leq (1 + p)a. 
\end{equation}
On the other hand, for $y$ such that $a<y\leq B$ we have:
\begin{equation}
\label{eq:concentration_abound}
 \E \left[ \left|  \frac{X^{(N)}(y)}{N} - py \right| \right] \leq B \cdot  \E\left[ \left|  \frac{X^{(N)}(y)}{N y} - p \right| \right].
 \end{equation}

Fix $\delta$ such that $0 < \delta < 1$. Using a standard Chernoff bound, 
\begin{equation*}
    \P\left( \left| \frac{X^{(N)}(y)}{N y} - p \right|> \delta p \right) \leq 2 e^{-\delta^2 p N y / 3}.
\end{equation*}

This concentration inequality implies that
\begin{equation*}
    \E \left[ \left|  \frac{X^{(N)}(y)}{N y} - p \right| \right]
    \leq \left( 1 - 2 e^{-\delta^2 p N y / 3} \right) \delta p + 2 e^{-\delta^2 p N y / 3} p.
\end{equation*}
Thus combining with \eqref{eq:concentration_abound}, we have:
\begin{equation*}
    \sup_{a<y \leq B} 
    \E \left[ \left|  \frac{X^{(N)}(y)}{N} - py \right| \right]
    \leq \left( 1 - 2 e^{-\delta^2 p N a / 3} \right) \delta B p + 2 e^{-\delta^2 p N a / 3} B p
\end{equation*}
and so 
\begin{equation}
\label{eq:y_greater_than_a}
    \lim_{N \rightarrow \infty} \sup_{y>a} \E \left[ \left|  \frac{X^{(N)}(y)}{N } - py \right| \right] \leq \delta B p.
\end{equation}

Since equations \eqref{eq:y_less_than_a} and \eqref{eq:y_greater_than_a} hold for all $a>0$ and $0<\delta<1$, by taking $a \rightarrow 0$ and $\delta \rightarrow 0$, we conclude that 
\[ \sup_{0 \leq y \leq B : Ny \in \mathbb{Z}} \E \left[ \left|  \frac{X^{(N)}(y)}{N} - py \right| \right] \to 0, \]
as required.
 \hfill$\Box$
\endproof

We can combine the preceding two lemmas to complete the proof of this step, as follows.  If we divide through the numerator and denominator of $r_\gamma^{(N)}(\theta | \v{y})$  by $N$, and apply Lemma \ref{lem:logitbound} and the triangle inequality, we obtain:
\begin{multline}
| r_\gamma^{(N)}(\theta|\v{y}) - p_\gamma^{(N)}(\theta|\v{y}) | \leq \\ \frac{(\epsilon_\gamma^{(N)}/N + \sum_{\theta' \neq \theta} \alpha_\gamma(\theta') v_\gamma(\theta') y(\theta') )}{(\epsilon_\gamma^{(N)}/N)^2} 
\cdot v_\gamma(\theta) \E\left[\left|\frac{D_\gamma^{(N)}(\theta | \v{y})}{N} - \alpha_\gamma(\theta) y(\theta) \right|\right] \\ 
+ \frac{v_\gamma(\theta) \alpha_\gamma(\theta) y(\theta)}{(\epsilon_\gamma^{(N)}/N)^2} \sum_{\theta'\neq \theta} v_\gamma(\theta') \E\left[ \left|\frac{D_\gamma^{(N)}(\theta' | \v{y})}{N} - \alpha_\gamma(\theta') y(\theta') \right|\right]  .
\end{multline}

By applying Lemma \ref{lem:binomial_concentration}, we have
$$\sup_{\v{y} \in I_N} \E\left[\left|\frac{D_\gamma^{(N)}(\theta | \v{y})}{N} - \alpha_\gamma(\theta) y(\theta) \right|\right] \rightarrow 0$$
for all $\theta$. Since $\epsilon_\gamma^{(N)}/N \rightarrow \epsilon_\gamma>0$ for all $\gamma$ and $\alpha_\gamma(\theta), v_\gamma(\theta)$, and $y(\theta)$ are bounded for all $\gamma$ and $\theta$, we conclude that 
\begin{equation*}
    \sup_{\v{y} \in I_N } |r_\gamma^{(N)}(\theta|\v{y}) - p_\gamma^{(N)}(\theta|\v{y}) | \rightarrow 0
\end{equation*}
as $N \rightarrow \infty$, as required.

{\em Step 3: Find $\kappa_2$ such that \eqref{eq:cond_rate_function_linear_n} holds.}

For each $\v{y} \in I_N$ the rate function is bounded by
\begin{align*}
    c^{(N)}(\v{y}) 
    &= \sum_\theta \left[ \left( m^{(N)}(\theta) - N y(\theta)\right) \tau(\theta) + \sum_\gamma \lambda_\gamma^{(N)}  r_\gamma(\theta | \v{y})
    \right]  \\
    &\leq N \cdot |\Theta| \cdot \tau(\theta) + N \cdot C  \sum_\gamma \lambda_\gamma
\end{align*}
and so \eqref{eq:cond_rate_function_linear_n} is satisfied with $\kappa_2 = \Theta \tau(\theta) + C \sum_\gamma \lambda_\gamma$.

{\em Step 4: Find $\kappa_3$ such that \eqref{eq:cond_variance} holds.}

Here we rely on the fact that increments at any jump of $Y_t^{(N)}(\theta)$ are of size $1/N$.  Observe that:
\[ \text{Trace}\left[ \Sigma^{(N)}(\v{y}) \right] + \| \mu_\theta^{(N)}(\v{y})\|^2 = \sum_{\theta } \E\left[ (Y_{T_{i+1}}^{(N)}(\theta) - Y_{T_i}^{(N)}(\theta))^2 | \v{Y}_{T_i} = \v{y} \right], \]
since the diagonal entries of $\Sigma^{(N)}(\v{y})$ are just the variances of the increments at each jump.  Now at any jump, $Y_{T_{i+1}}^{(N)}(\theta) - Y_{T_i}^{(N)}(\theta)$ is at most $1/N$.  Thus the right hand side of the preceding expression is at most $\Theta / N^2$, and so \eqref{eq:cond_variance} holds with $\kappa_3 = \Theta$.

\medskip 

Thus conditions \eqref{eq:cond_initial_condition}-\eqref{eq:cond_variance} hold, and Theorem \ref{thm:mf_convergence} follows from Kurtz's Theorem (cf.~Theorem 2.8 in \cite{darling2002fluid}).\hfill$\Box$
\endproof

\proof{Proof of Proposition \ref{prop:Qij_demand_constrained}. }
Throughout the proof, to simplify notation we fix $a_C, a_L$, and then suppress them throughout the proof (e.g., instead of $s^*(\theta, j | a_C, a_L)$, we simply write $s^*(\theta, j)$).  We also assume for simplicity that $0 < a_C < 1$ and $0 < a_L < 1$.  This assumption can be made without loss of generality: if one or more of these inequalities fails, we can reduce the type space by eliminating one or more of the treatment conditions, and then replicate the argument below.  

We recall from \eqref{eq:Qijinfty} that $Q_{ij}(\infty)$ is:
\begin{equation}
\label{eq:Qijinfty_proof}
 Q_{ij}(\infty)  = \lambda \sum_\theta \sum_\gamma \phi_{\gamma,i} p_{\gamma,i}(\theta,j|\v{s}^*),
\end{equation}
where the choice probability is:
\begin{equation}
\label{eq:choice_proof}
p_{\gamma,i}(\theta,j | \v{s}^*) = \frac{ \alpha_{\gamma,i}(\theta,j) v_{\gamma,i}(\theta,j) s^*(\theta,j)}{\epsilon_{\gamma,i} + \sum_{\theta'} \sum_{j' = 0,1}\alpha_{\gamma,i} (\theta',j')v_{\gamma,i}(\theta',j') s^*(\theta',j')}.
\end{equation}

We also make use of the following flow conservation condition cf.~\eqref{eq:FOC}, which we rewrite here for the experimental setting:
\begin{equation}
\label{eq:FOC_proof}
 (\rho(\theta,j) - s^*(\theta,j))\tau \nu(\theta) = \lambda \sum_{\gamma} \sum_{i = 0,1} \phi_{\gamma,i}
 p_{\gamma,i}(\theta,j | \v{s}^*).
\end{equation}

{\em Step 1: We have $s^*(\theta,j) \rightarrow \rho(\theta, j)$ for all $\theta, j$.} Divide both sides of \eqref{eq:FOC_proof} by $\tau \nu(\theta) s^*(\theta, j)$. The left hand side of the equation becomes $\rho(\theta,j)/s^*(\theta,j) - 1$. The right hand side of the equation becomes
\begin{equation*}
\frac{\lambda}{\tau \nu(\theta)} \sum_\gamma \sum_{i=0,1} \frac{ \phi_{\gamma,i}\alpha_{\gamma,i}(\theta,j) v_{\gamma,i}(\theta,j)}{\epsilon_{\gamma,i} + \sum_{\theta'} \sum_{j' = 0,1}\alpha_{\gamma,i} (\theta',j')v_{\gamma,i}(\theta',j') s^*(\theta',j')}  \ ,
\end{equation*}
where we used the definition of the choice probability. Note that each term in the sum is bounded by one, and there are finitely  many terms, so the entire expression approaches zero as $\lambda/\tau \rightarrow 0$. Thus we have $\rho(\theta, j)/s^*(\theta,j) \rightarrow 1$.

{\em Step 2: For all $\gamma, \theta, i, j$, we have $p_{\gamma, i} (\theta,j|\v{s}^*) \rightarrow p_{\gamma, i} (\theta,j|\v{\rho})$.}
This follows because the choice probabilities $p_{\gamma, i} (\theta,j|\v{s}^*)$ are continuous in $\v{s}^*$. 

{\em Step 3: Completing the proof.}
The limit in \eqref{eq:Qij_demand_constrained} follows immediately from Step 2 and the definition of $Q_{ij}(\infty)$. 
\endproof

\bigskip

\proof{Proof of Theorem \ref{thm:demand_constrained}}
Consider a sequence of systems where $\lambda/\tau \to 0$.  Using the Proposition \ref{prop:Qij_demand_constrained}, we observe that:
\begin{equation}
\label{eq:GTE_demand_constrained} 
 \frac{1}{\lambda}{\GTE} = \frac{1}{\lambda}Q_{11}(\infty | 1,1) - \frac{1}{\lambda}Q_{00}(\infty| 0,0) \to \sum_{\theta} \sum_{\gamma} \phi_{\gamma} p_{\gamma,1}( \theta,1 | \v{\rho}(1) ) - \sum_{\theta} \sum_{\gamma} \phi_{\gamma} p_{\gamma,0}( \theta,0 | \v{\rho}(0) ).
\end{equation}

We now use Proposition \ref{prop:Qij_demand_constrained} to show that the steady-state naive $\CR$ estimator is unbiased in the limit as $\lambda/\tau \rightarrow 0$, while the steady-state naive $\LR$ estimator remains biased. First we consider a $\CR$ experiment paired with the naive $\CR$ estimator.

Using Proposition \ref{prop:Qij_demand_constrained}, it follows that:
\[
\frac{1}{\lambda} \widehat{\GTE}^{\CR}(\infty | a_C) \to
\frac{1}{a_C}\sum_{\theta} \sum_{\gamma} a_C \phi_\gamma p_{\gamma,1}( \theta,1 | \v{\rho}(1) ) 
- \frac{1}{1-a_C}\sum_{\theta} \sum_{\gamma} (1-a_C) \phi_\gamma p_{\gamma,0}( \theta,1 | \v{\rho}(1) ).
\]
Now note that $\rho(\theta,1|1) = \rho(\theta)$ and $\rho(\theta,0|1) = 0$ when $a_L = 1$; similarly, $\rho(\theta,0|0) = \rho(\theta)$, and $\rho(\theta,1|0) = 0$ when $a_L = 0$.  Thus, from the definition of the $\TSR$ design in \eqref{eq:v_expt}-\eqref{eq:epsilon_expt} and the definition of the choice probability in \eqref{eq:meanfieldchoice}, the choice probability of a control customer for a treatment listing at $\v{\rho}(1)$ is the same as the choice probability of a control customer for a control listing at $\v{\rho}(0)$:
\[
p_{\gamma,0}(\theta,1|\v{\rho}(1)) = p_{\gamma,0}(\theta,0|\v{\rho}(0)).
\]

These choice probabilities are the same because (1) {\em all} listings are in treatment in the $\CR$ design, with the mass of each type $\theta$ equal to $\rho(\theta)$;  and (2) control customers have the same choice model parameters for these listings regardless of whether they are in treatment or control. %
Thus it follows that $\widehat{\GTE}^{\CR}(\infty | a_C)/\lambda - \GTE/\lambda \to 0$ as $\lambda/\tau \to 0$, i.e., the steady-state naive $\CR$ estimator is asymptotically unbiased. 

On the other hand, consider the steady-state naive $\LR$ estimator.  Observe that: 
\[ \frac{1}{\lambda} \widehat{\GTE}^{\LR}(\infty | a_L) \to \frac{1}{a_L}\sum_{\theta} \sum_{\gamma} \phi_{\gamma} p_{\gamma,1}( \theta,1 | \v{\rho}(a_L) ) 
- \frac{1}{1-a_L}\sum_{\theta} \sum_{\gamma} \phi_\gamma p_{\gamma,1}( \theta,0 | \v{\rho}(a_L) ). \]
Since $\v{\rho}(a_L)$ is different from both $\v{\rho}(1)$ (all listings in treatment) and $\v{\rho}(0)$, in general, this limit will {\em not} be equivalent to the $\GTE$; i.e., the naive $\LR$ estimator is asymptotically biased.  Thus the difference between $\widehat{\GTE}^{\LR}(\infty|a_L)$ and the $\GTE$ does not converge to zero in general as $\lambda/\tau \to 0$; i.e., the naive $\LR$ estimator is biased.

\endproof

\proof{Proof of Proposition \ref{prop:Qij_supp_constrained}.} 
We prove the proposition in a sequence of steps.  We adopt the same conventions as in the proof of Proposition \ref{prop:Qij_demand_constrained}: to simplify notation we fix $a_C, a_L$, and then suppress them throughout the proof (e.g., instead of $s^*(\theta, j | a_C, a_L)$, we simply write $s^*(\theta, j)$).  We also again assume that $0 < a_C < 1$ and $0 < a_L < 1$; as before, this assumption is without loss of generality.

{\em Step 1: We have $s^*(\theta,j) \to 0$ for all $\theta,j$.}  Suppose instead that for some $\theta, j$ pair, the limit inferior of $s^*(\theta,j)$ is positive along the sequence of systems considered.  Divide both sides of \eqref{eq:FOC_proof} by $\lambda$, and take the limit inferior of each side.  The left hand side approaches zero.  On the other hand, the right hand side remains positive (because $\phi$, $\epsilon$, $\alpha$, and $v$ are all positive).  Thus we have a contradiction, establishing the claim.

{\em Step 2: The following limit holds:}
\[ \frac{s^*(\theta, j)}{ 1/(\lambda/\tau) } \to \frac{\rho(\theta, j) \nu(\theta)}{\sum_{\gamma} \sum_{i = 0,1} \phi_{\gamma,i} g_{\gamma,i}(\theta,j)}. \]
To prove this, divide both sides of \eqref{eq:FOC_proof} by $\lambda s^*(\theta, j)$.  The left hand side becomes 
\[\frac{1}{\lambda/\tau} \cdot \frac{(\rho(\theta, j) - s^*(\theta,j))\cdot \nu(\theta) }{  s^*(\theta,j)}.\]
The left hand side will then have the same limit as:
\[ \frac{\rho(\theta,j) \nu(\theta)}{\lambda/\tau} \cdot \frac{1}{s^*(\theta,j)}. \]
The limit of the right hand side becomes $\sum_{\gamma} \sum_{i = 0,1} \phi_{\gamma,i} g_{\gamma,i}(\theta,j)$, establishing the desired result.

{\em Step 3: For all $\gamma,\theta,i,j$, the following limit holds:}
\[ \frac{p_{\gamma,i}(\theta,j|\v{s}^*)}{1/(\lambda/\tau)} \to \frac{\rho(\theta,j) \nu(\theta) g_{\gamma,i}(\theta,j)}{\sum_{\gamma'} \sum_{i'= 0,1} \phi_{\gamma',i'} g_{\gamma',i'}(\theta,j)}. \]
This follows by the definition of $p_{\gamma,i}$, and the previous step.

{\em Step 4: Completing the proof.} Given the definition of $Q_{ij}$ in \eqref{eq:Qijinfty_proof}, the preceding step completes the proof.
\endproof

\section{Cluster-randomized experiments}
\label{app:cluster}
Figure \ref{fig:cluster_vary_pref_ratio_all_estimators} compares the performance of all estimators in a clustered market. We have the same simulation set-up as Figure \ref{fig:cluster_vary_pref_ratio} and we add the $\CR$ and $\LR$ estimators, as well as $\TSRN$ and $\TSRIo$. We find that the bias of $\TSRIo$ is more sensitive to changes in the strength of the clustering than $\TSRIt$, though not as sensitive as the cluster randomized estimator, $\CR$, $\LR$, or $\TSRN$. In particular, we see that $\TSRIo$ has larger bias than the cluster randomized estimator for preference ratios $y/x < 0.5$, similar bias for $y/x = 0.5$, and lower bias for $y/x > 0.5$. Further, $\TSRIo$ and the cluster randomized estimator have similar standard errors. The simulations suggest that $\TSRI$ estimators can reduce bias in highly interconnected markets where cluster randomized experiments cannot. 

\begin{figure}[H]
    \centering
    \begin{tabular}{l l }
        \includegraphics[height=.3\textwidth]{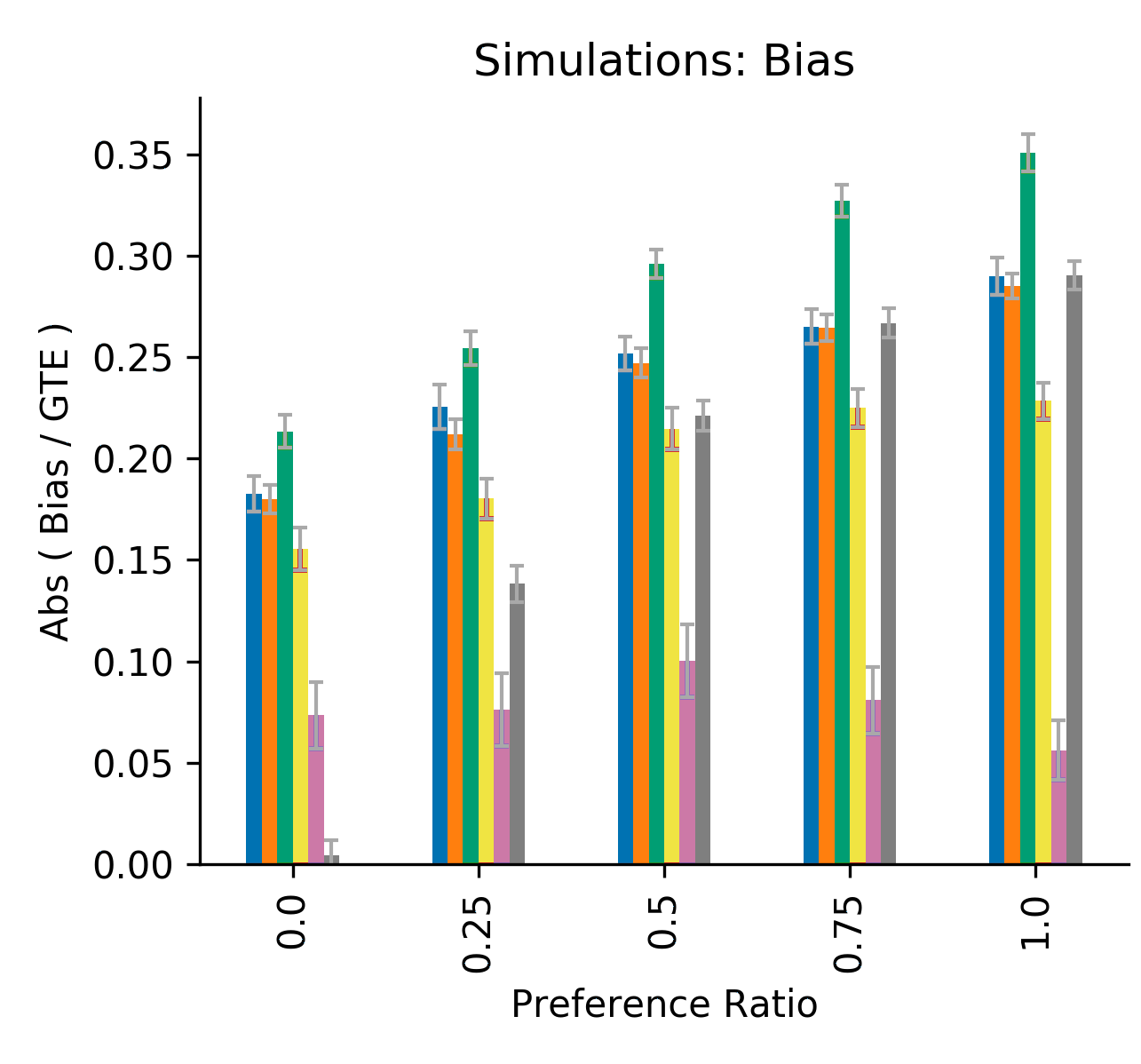}
        &
        \includegraphics[height=.3\textwidth]{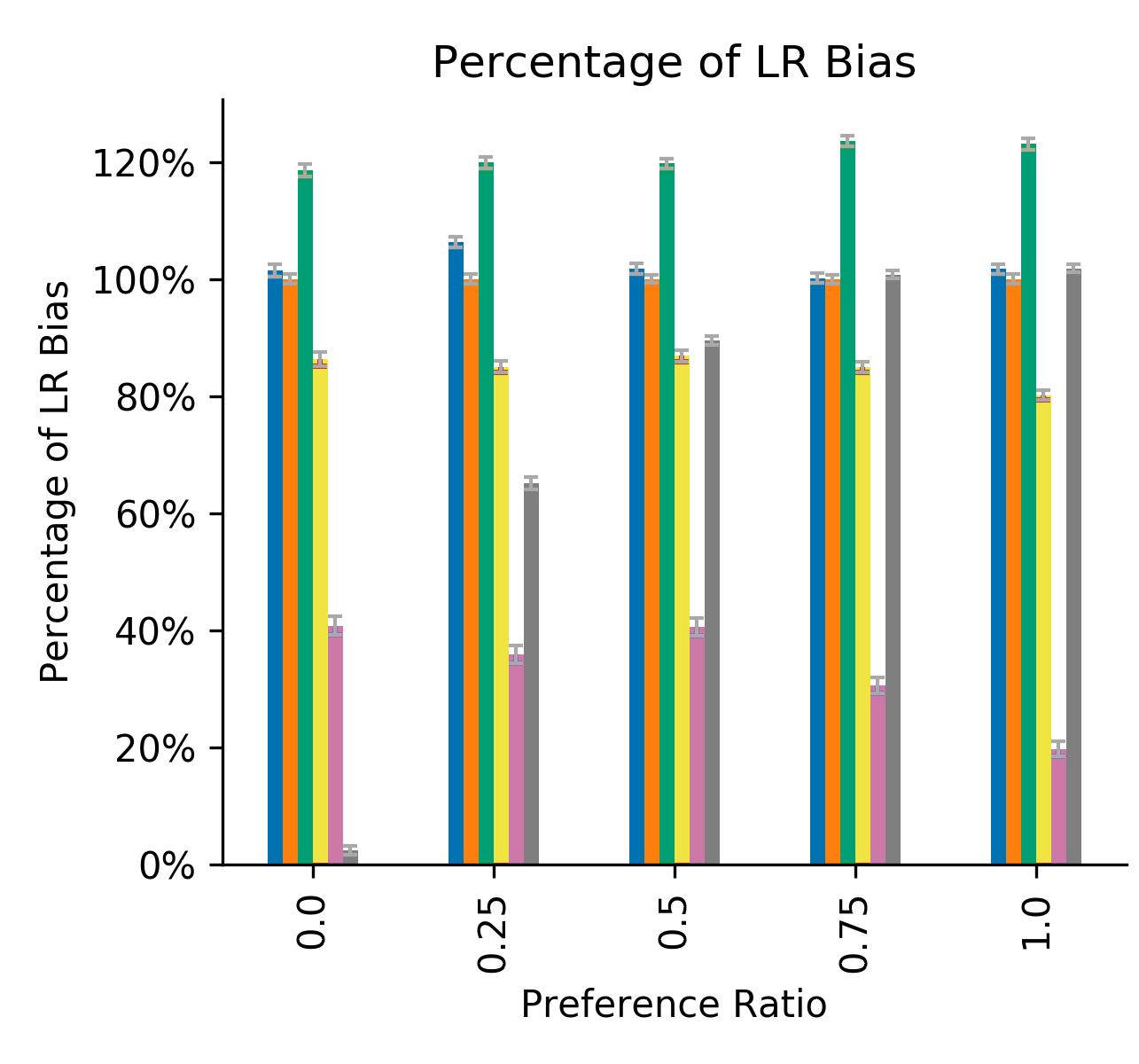}
        \\
        \includegraphics[height=.3\textwidth]{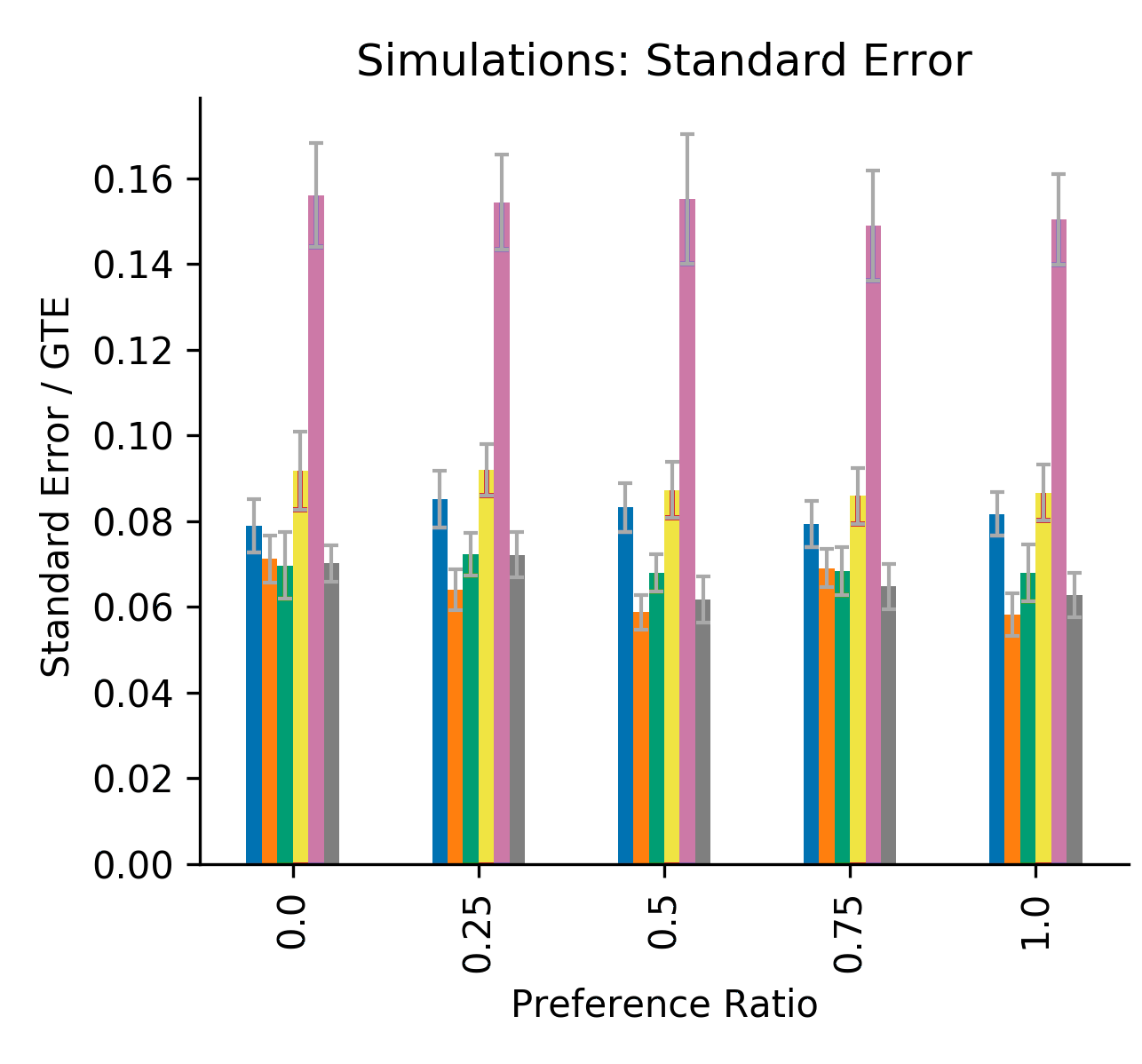}
        &
        \includegraphics[height=.3\textwidth]{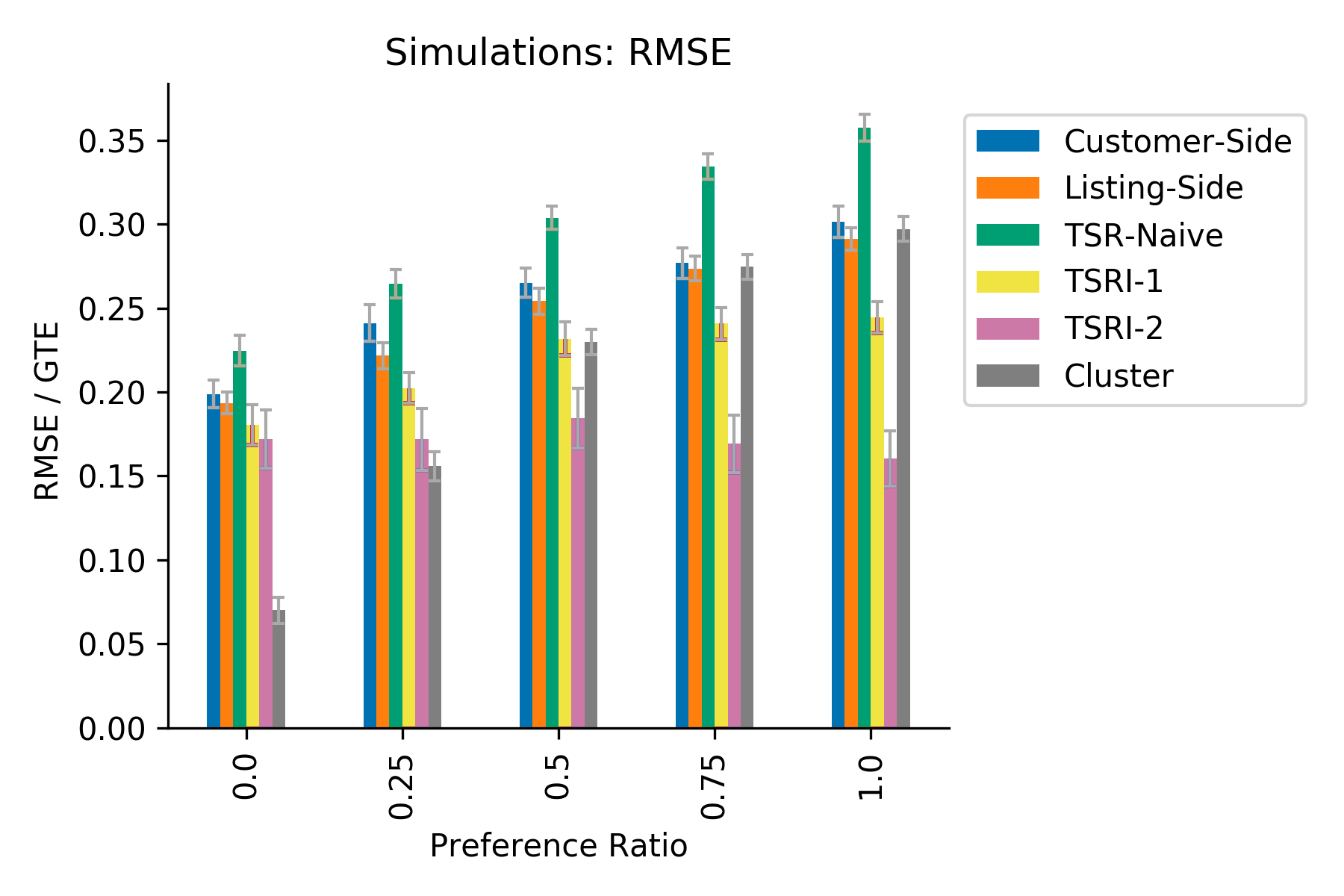}
    \end{tabular}
    \caption{Comparison of all estimators in a clustered market as the preference ratio $y/x$ varies. Relative demand is fixed at $\lambda/\tau=1$. We fix $x=0.5$ and vary $y$. Bootstrapped 95 percentile confidence intervals are provided for each statistic. }
    \label{fig:cluster_vary_pref_ratio_all_estimators}
\end{figure}

Figure \ref{fig:cluster_vary_lambda} shows the bias of $\TSRIt$ and the cluster-randomized estimator when the preference ratio is fixed at $y/x=0$ and market balance $\lambda/\tau$ varies. In this setting with a highly clustered market, we find qualitatively similar behavior that in the bias and standard error of the two estimators, though the magnitude of the difference may differ. We omit the plot depicting the improvement over the $\LR$ in this figure since the bias of the $\LR$ estimate approaches 0 as $\lambda/\tau \to \infty$. 

\begin{figure}
    \centering
    \begin{tabular}{c c }
        \includegraphics[height=.3\textwidth]{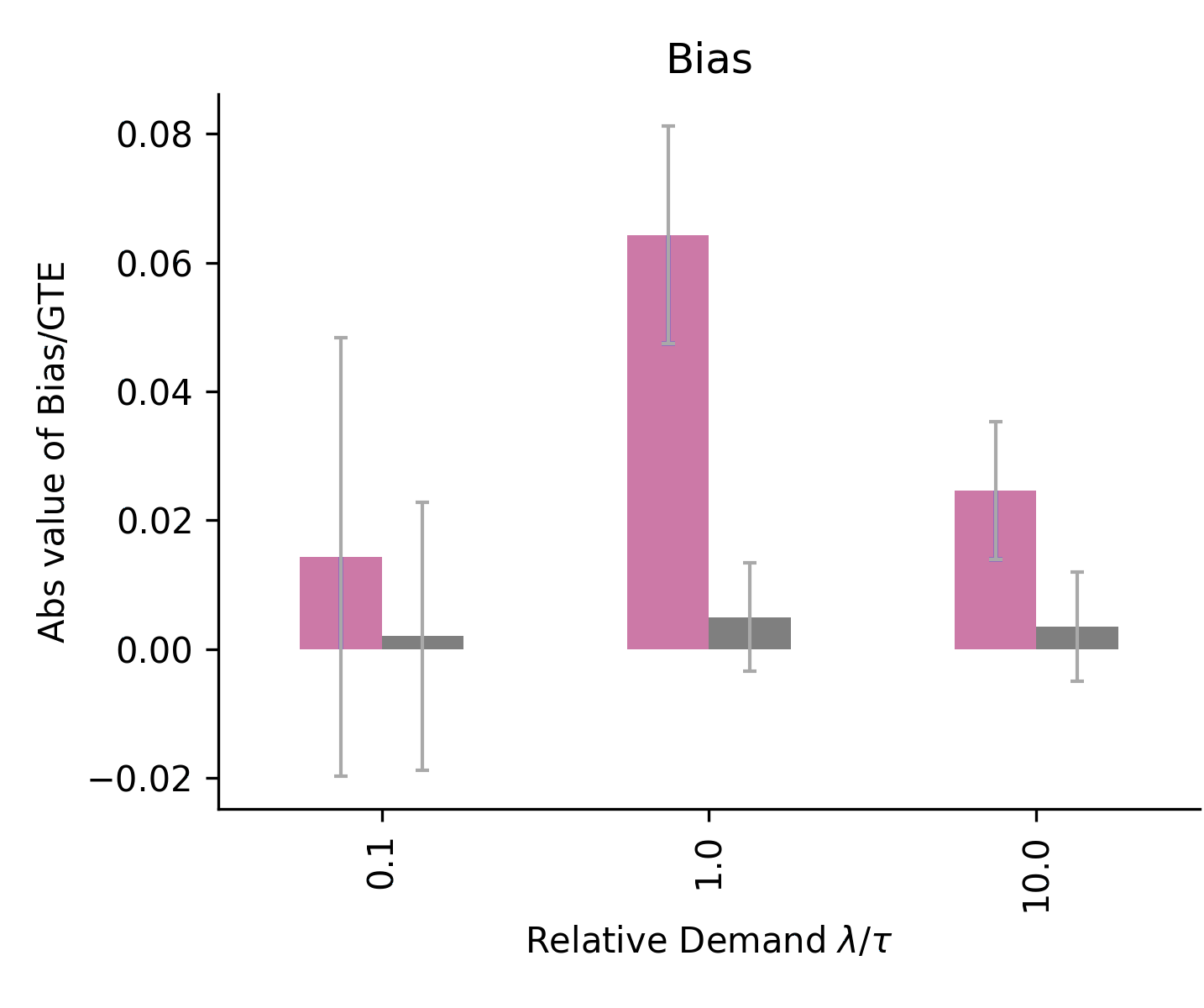}
        &
        \includegraphics[height=.3\textwidth]{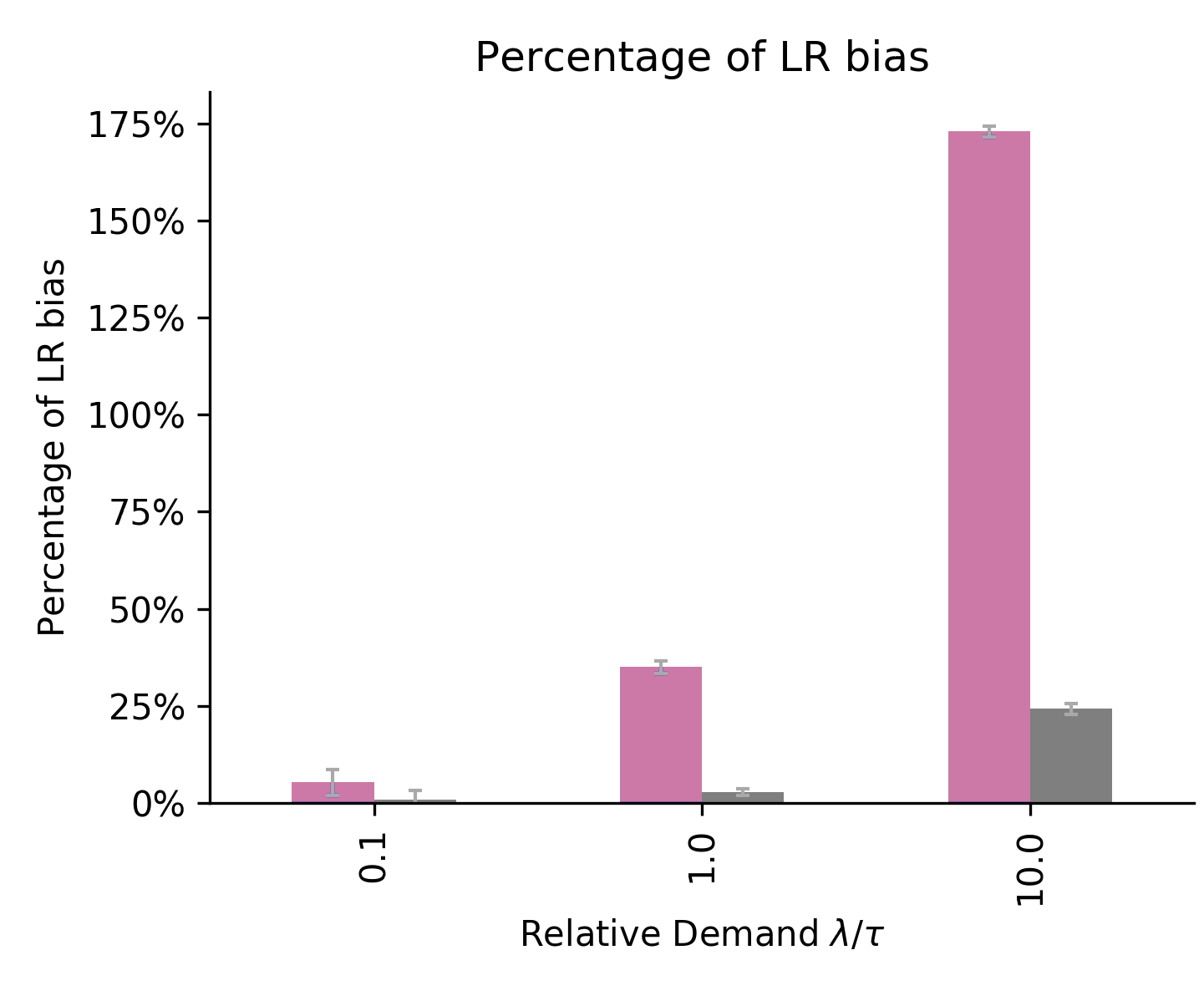}
        \\
        \includegraphics[height=.3\textwidth]{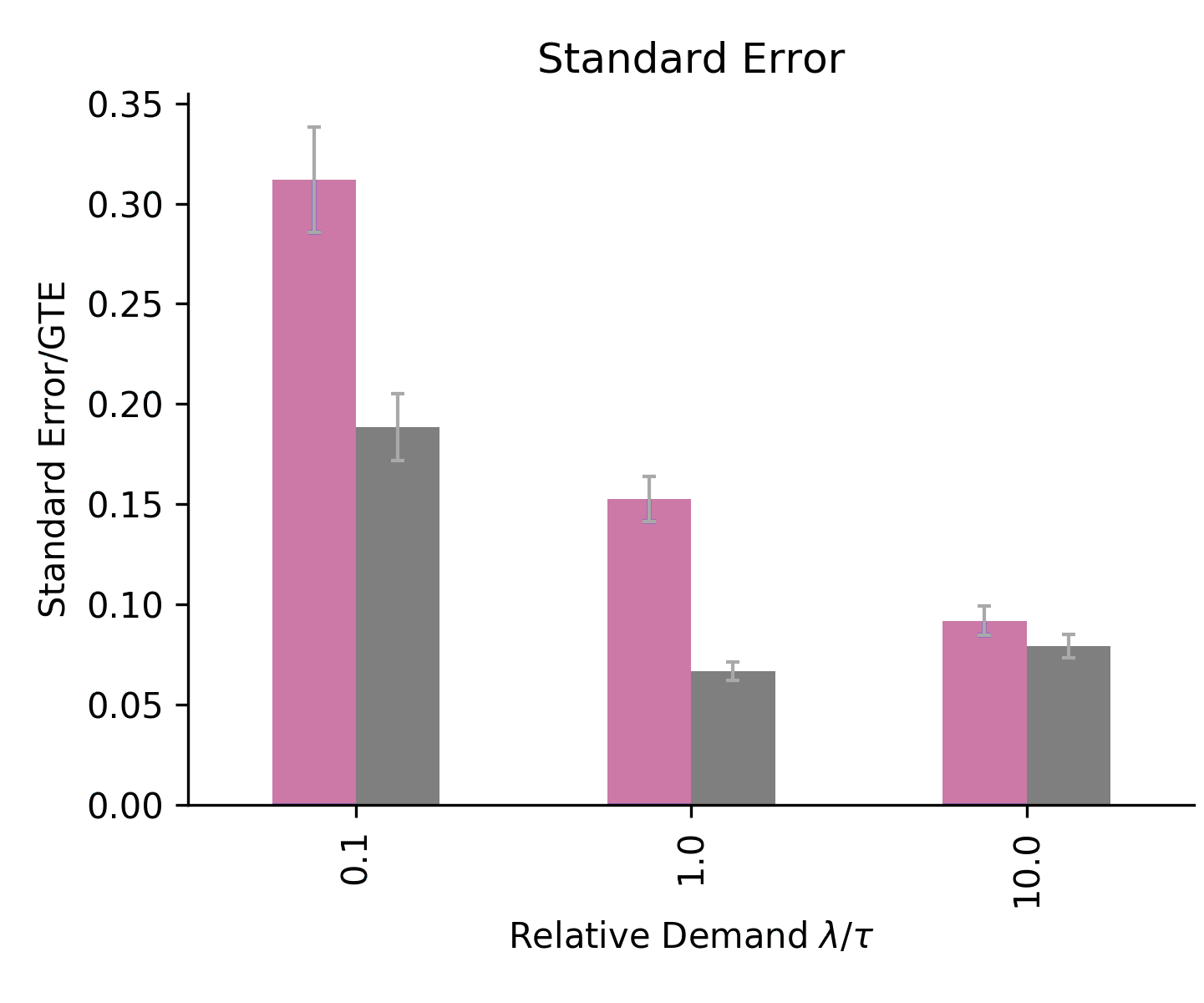}
        &
        \includegraphics[height=.3\textwidth]{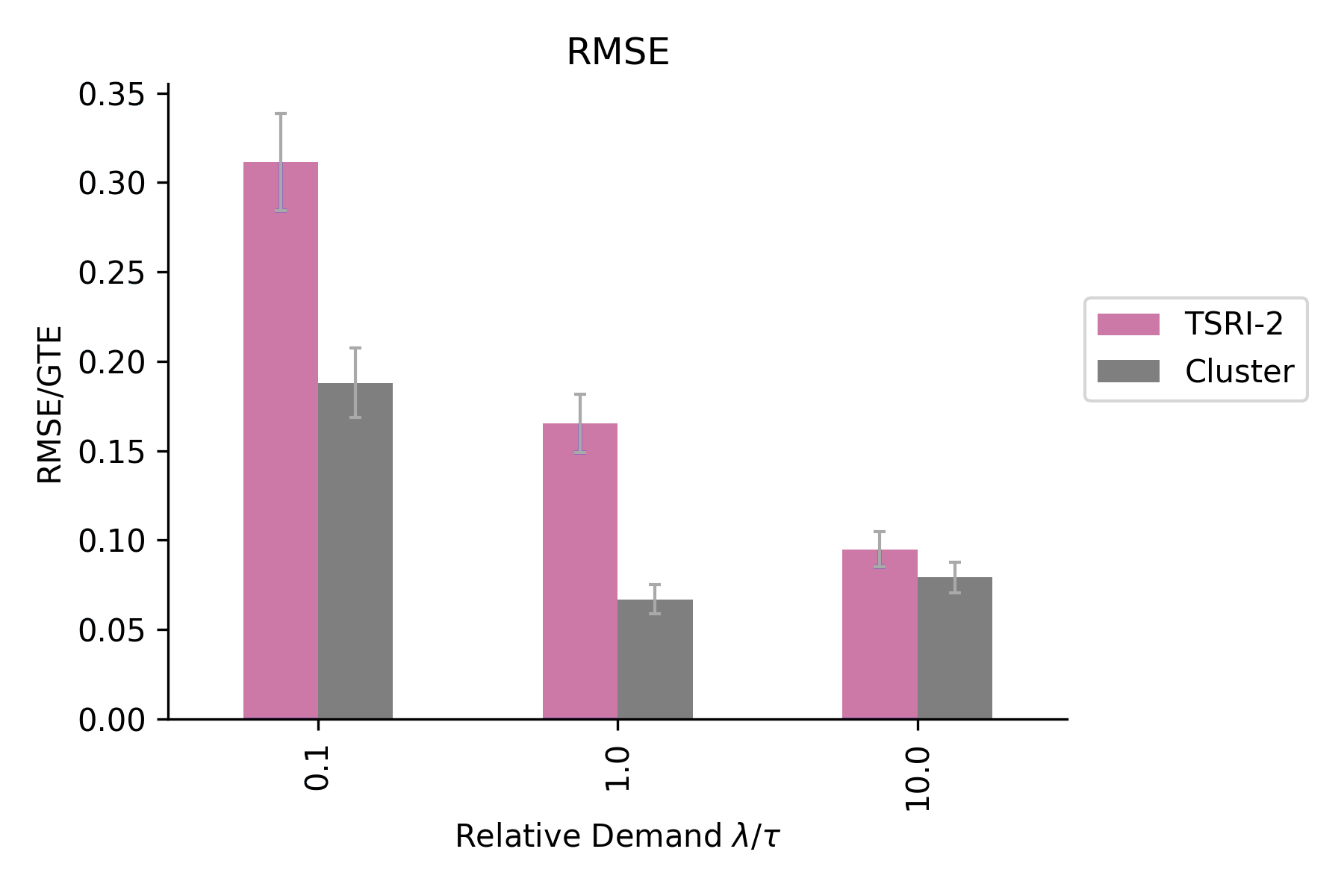}
    \end{tabular}
    \caption{Simulations of cluster-randomized estimator and TSR estimators as relative demand $\lambda/\tau$ varies. Preference ratio is fixed at $y/x=0$.  We fix $\tau=1$ and vary $\lambda$. Note that at $\lambda=1$, the $\LR$ bias is extremely small (1.7 percent of the $\GTE$) and so the bias of the $\TSR$ estimators, while higher than that of the $\LR$ estimator, is less than 3 percent of the $\GTE$. Bootstrapped 95 percentile confidence intervals are provided for each statistic. }
    \label{fig:cluster_vary_lambda}
\end{figure}

\section{Numerics and simulations across different market parameters}
\label{app:simulations}
In this section, we investigate bias, standard error, and $\RMSE$ of the estimators as we change market parameters and introduce heterogeneity into the system. The purpose of this investigation is two-fold, to verify that the results presented in the main text hold in heterogeneous systems and to analyze the relative performance of the estimators in varying market conditions.
We replicate the mean field and simulation results shown in Figure \ref{fig:numerics_homogeneous} and Figure \ref{fig:simulations_hom} in settings with different utilities, customer heterogeneity, listing heterogeneity, and heterogeneous treatment effects. We also consider a modification of the consideration set formation process where customers sample a fixed $K$ number of listings into their consideration set.

Across our range of simulations, we see qualitatively similar behavior to our findings in the main text: among the two one-sided experiments, the naive $\CR$ estimator has lower bias for small $\lambda/\tau$ and the naive $\LR$ estimator has lower bias for large $\lambda/\tau$, while the $\TSRIo$ and $\TSRIt$ estimators offer bias reductions at both extremes and moderate levels of market balance.  The $\TSR$ estimators offer these bias reductions at the cost of higher variance.   %

When we fix $\lambda/\tau$ and vary parameters regarding customer choice, we find that increasing booking probability and introducing heterogeneity can increase the bias of $\CR$ and $\LR$ estimators, but remarkably the bias of the $\TSRI$ estimators remains low across the ranges that we study.

In each of these cases, $\TSRIt$ consistently has the lowest bias, though it also has the highest standard error. $\TSRIt$ minimizes $\RMSE$ in these simulations, though we note that this depends on the size of the market and the time horizon of the experiment. The size of the standard error decreases as the size of the market increases and as the time horizon increases, and the bias remains constant. Thus in a large enough market with a long enough time horizon (here with $N=5000$ and $T=25$), $\TSRIt$ minimizes $\RMSE$.

\subsection{Simulation details}

For every setting, we run 500 simulations with $N=5000$. We fix $\alpha=1$, $\epsilon=1$, and $\tau=1$ across all settings and for each we consider three levels of demand: $\lambda= 0.1, 1, 10$. The simulation runs for $T_1$ time units. We drop the burn in time and calculate the value of the estimator on the time interval $[T_0, T_1]$ for some $0 < T_0 < T_1$. We set $T_0=5$ and $T_1=25$. 

We further scale time by $\min\{\lambda, \tau\}$ so that the number of ``events'' (i.e., bookings) that occur in the time interval is consistent across different values of $\lambda$ and $\tau$. More specifically, for a given $(\lambda, \tau)$ pair, we calculate the value of the estimator on the time interval $[T_0 / \min\{\lambda, \tau\}, \  T_1/\min\{\lambda, \tau\}]$. 
To see why this is necessary, note that in the case where there is no outside option and all customers book a listing if one is available, the rate of bookings is $\min(\lambda, \tau)$. We heuristically rescale time so that if we removed the effect of choice set dynamics and utilities and all customers book any available listing, the number of bookings would be consistent across market balance levels. 
This rescaling allows us to achieve similar precision in our estimates of bias across different levels of market balance; that is, under this rescaling the size of the standard error of the estimates is similar across different levels of market balance. Because of this rescaling, it is difficult to compare how standard errors of the estimators change when we fix the time horizon and change market balance. We suggest that the reader use the standard error figures to compare the standard error of different estimators \textit{within} the same level of market balance.

For the base setting shown in Section \ref{sec:variance} with homogeneous listings and customers, a customer has utility $v = 0.315$ for a control listing and $\tilde{v}=0.3937$ for a treatment listing, corresponding to a mean field steady state booking probability of 20 percent in the global control model and 23 percent in the global treatment model. This change corresponds to a 25 percent increase in the utility, due to treatment.
Unless otherwise noted, all sets of market parameters are chosen to maintain the 20 and 23 percent booking probabilities in global control and treatment, respectively. 

We fix $a_C=0.5$ in the $\CR$ experiments and $a_L=0.5$ in the $\LR$ experiments. For the $\TSR$ experiments, we vary $a_C$ and $a_L$ as defined in Section \ref{ssec:TSR_est}. For a general $\TSR$ experiment with randomization parameters $(a_C, a_L)$, we simulate a completely randomized design on the listing side, fixing $\lfloor N \cdot a_L \rfloor$ listings in treatment and $\lceil N \cdot (1-a_L) \rceil$ listings in control. Since customers arrive over time, we cannot run a completely randomized design and so we simulate Bernoulli randomization on the customers, randomizing each customer to treatment independently with probability $a_C$. 

For each statistic (bias, standard error, and RMSE), bootstrapped 95th percentile confidence intervals are presented. In each setting, we re-sample 500 simulation runs, with replacement, from the original set of 500 simulations. We calculate the value of the bias, standard error, and RMSE for each estimator and present the 95th percentile intervals of each. 

\subsection{Market scenarios}
\label{ssec:robustness_market_scenarios}

In this section, we analyze the effect of changes in customers' choice set parameters on the behavior of the estimators, while holding the market balanced fixed with $\lambda = \tau = 1$. In all of these scenarios, $\TSRIt$ has the lowest bias and the highest variance. $\TSRIt$ is the estimator that minimizes $\RMSE$ in these simulations, though this depends on the relative magnitude of the bias and standard error, as a result of the size of the market. For a larger market, the standard error is smaller the magnitude of the bias, and vice versa for a smaller market. 

\textbf{Varying average utility.}
In addition to the homogeneous system with one customer type and one listing type presented in Figure \ref{fig:simulations_hom}, we consider two additional settings, again in a homogeneous system, but now scaling the utility $v$. We fix the effect of the intervention such that the lift in utility $\tilde{v}/v = 1.25$ is constant in all three settings, and consider settings with smaller $v$ and larger $v$. Rescaling $v$ will change the steady state booking probabilities in global treatment and control so that we no longer have 23 percent of booking in global treatment and 20 percent in global control. 

Note that changes to $\alpha$ and/or $\epsilon$ can be equivalently viewed as a rescaling of the utility, so changing the utility also allows us to explore the effect of changes in $\alpha$ or $\epsilon$ as well.

\begin{itemize}
    \item Low utility: $v = 0.155, \tilde{v} = 0.1938$. 
    \item Medium utility: $v = 0.315, \tilde{v} = 0.3937$.
    \item High utility: $v = 0.62, \tilde{v} = 0.775$.
\end{itemize}

Results are presented in Figure \ref{fig:robustness_vary_avg_utility}. As the average utility increases (and so does the steady state booking probability of arriving customers), the bias for $\CR, \LR, \TSR-naive$ and $\TSRIo$ all increase, whereas the bias $\TSRIt$ is largely consistent across different levels of utility. The standard errors of the estimators do not change notably as we vary the average utility. A consequence of the change in bias is that the $\RMSE$ minimizing estimator depends on the utility level, with $\TSRIt$ having the highest $\RMSE$ for $v=0.15$ but the smallest for $v=0.62$. 

\begin{figure}
    \centering
    \begin{tabular}{l l }
        \includegraphics[height=.26\textwidth]{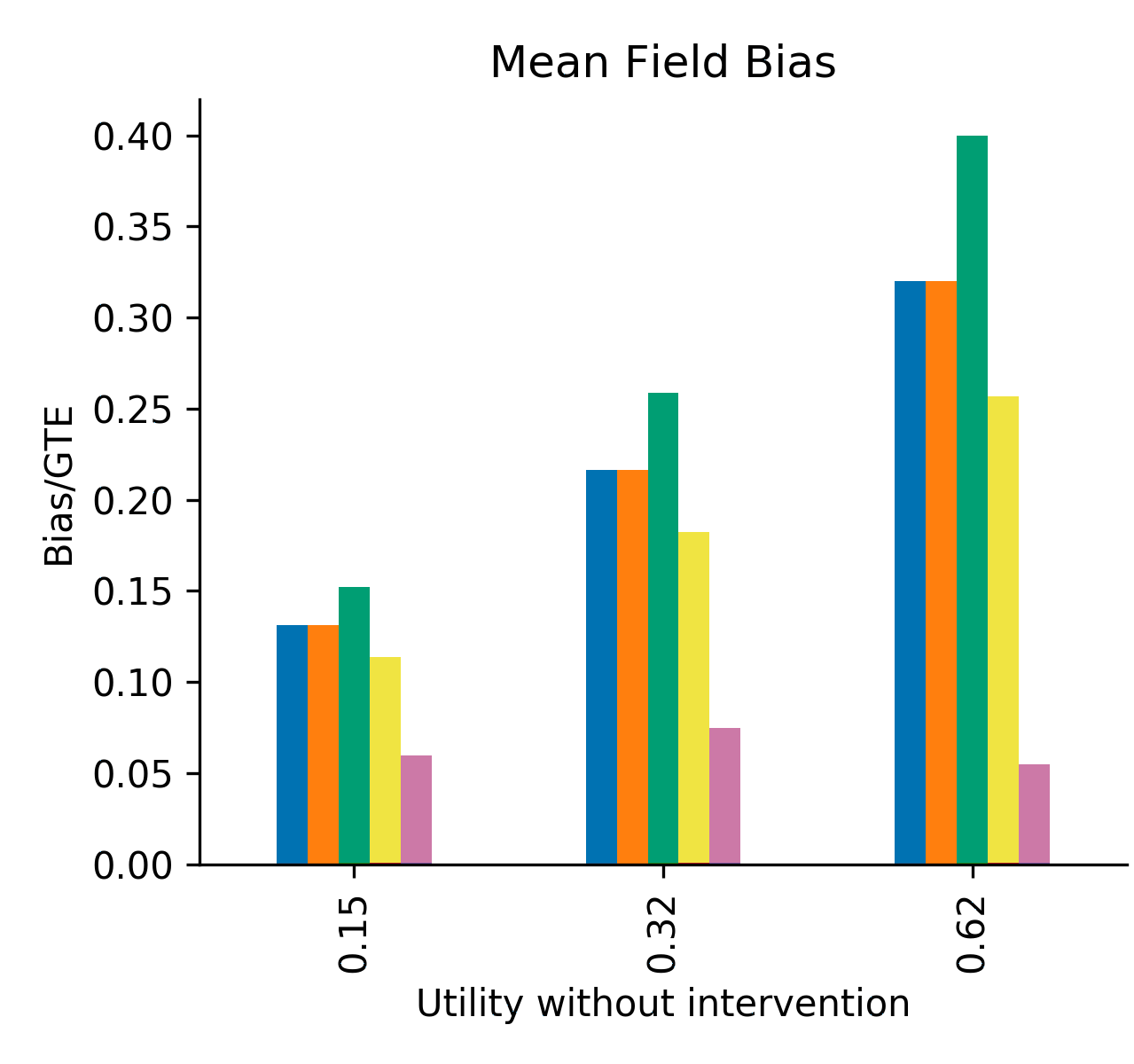}
        &
        \includegraphics[height=.26\textwidth]{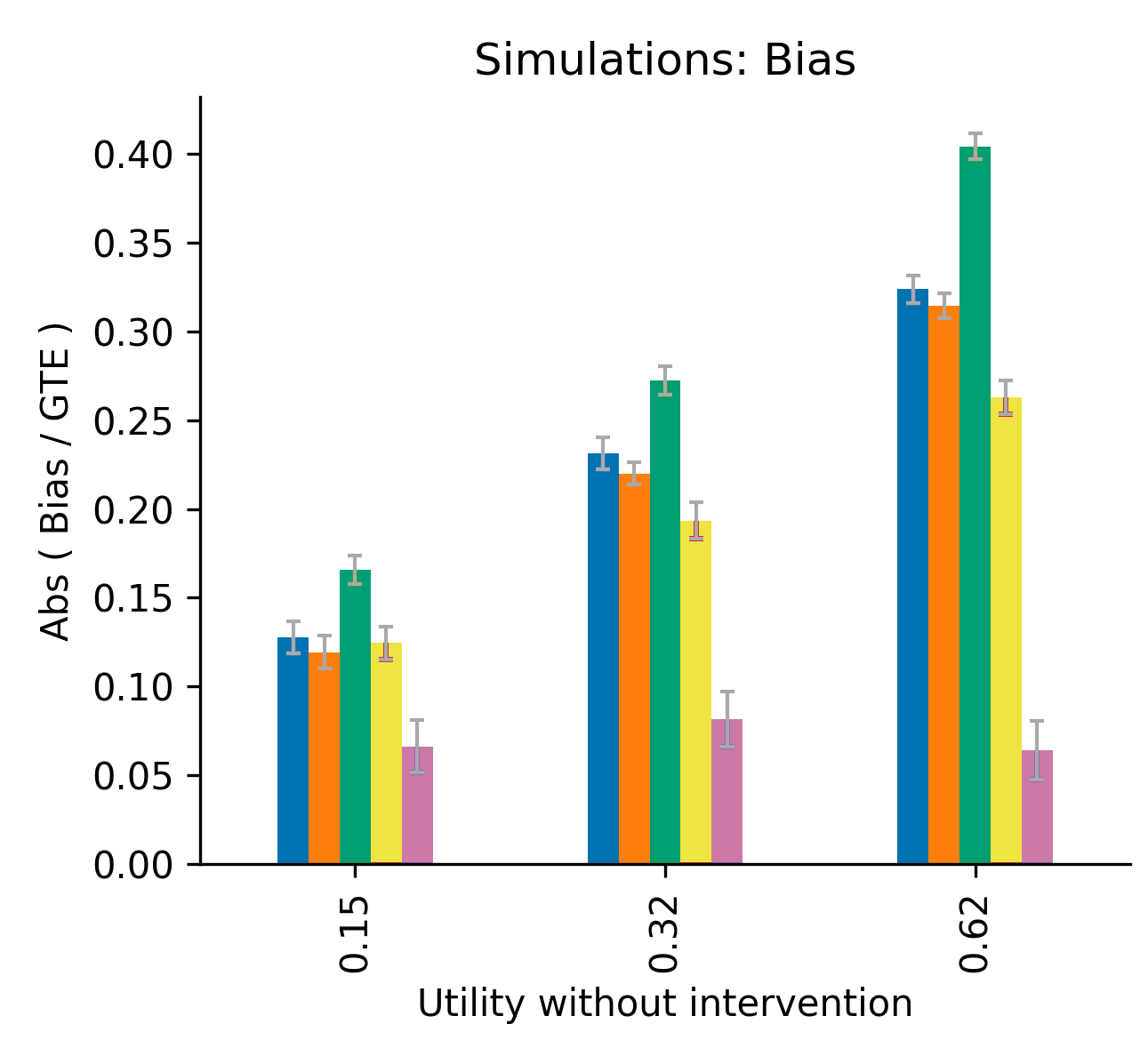}
        \\
        \includegraphics[height=.26\textwidth]{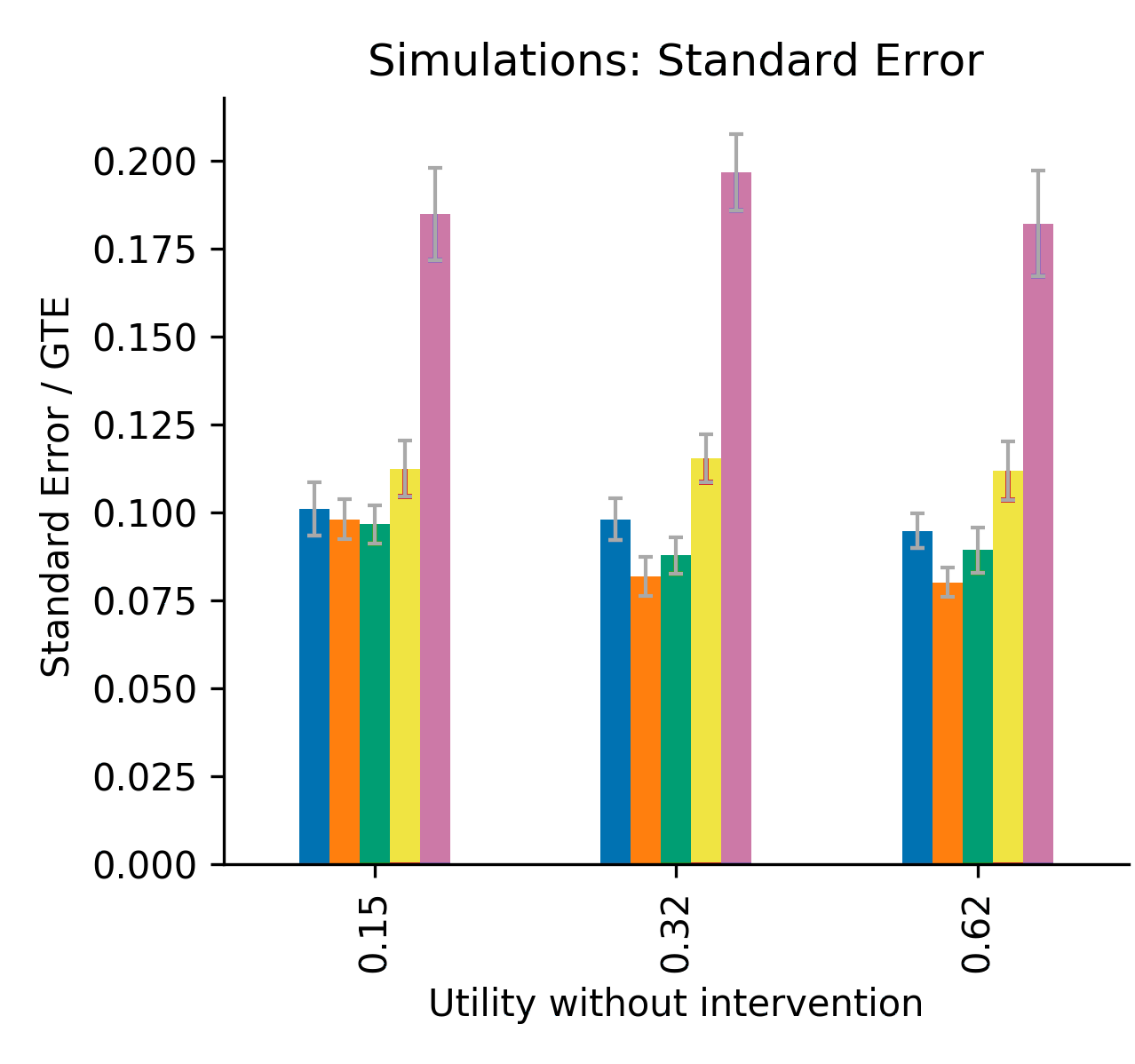}
        &
        \includegraphics[height=.26\textwidth]{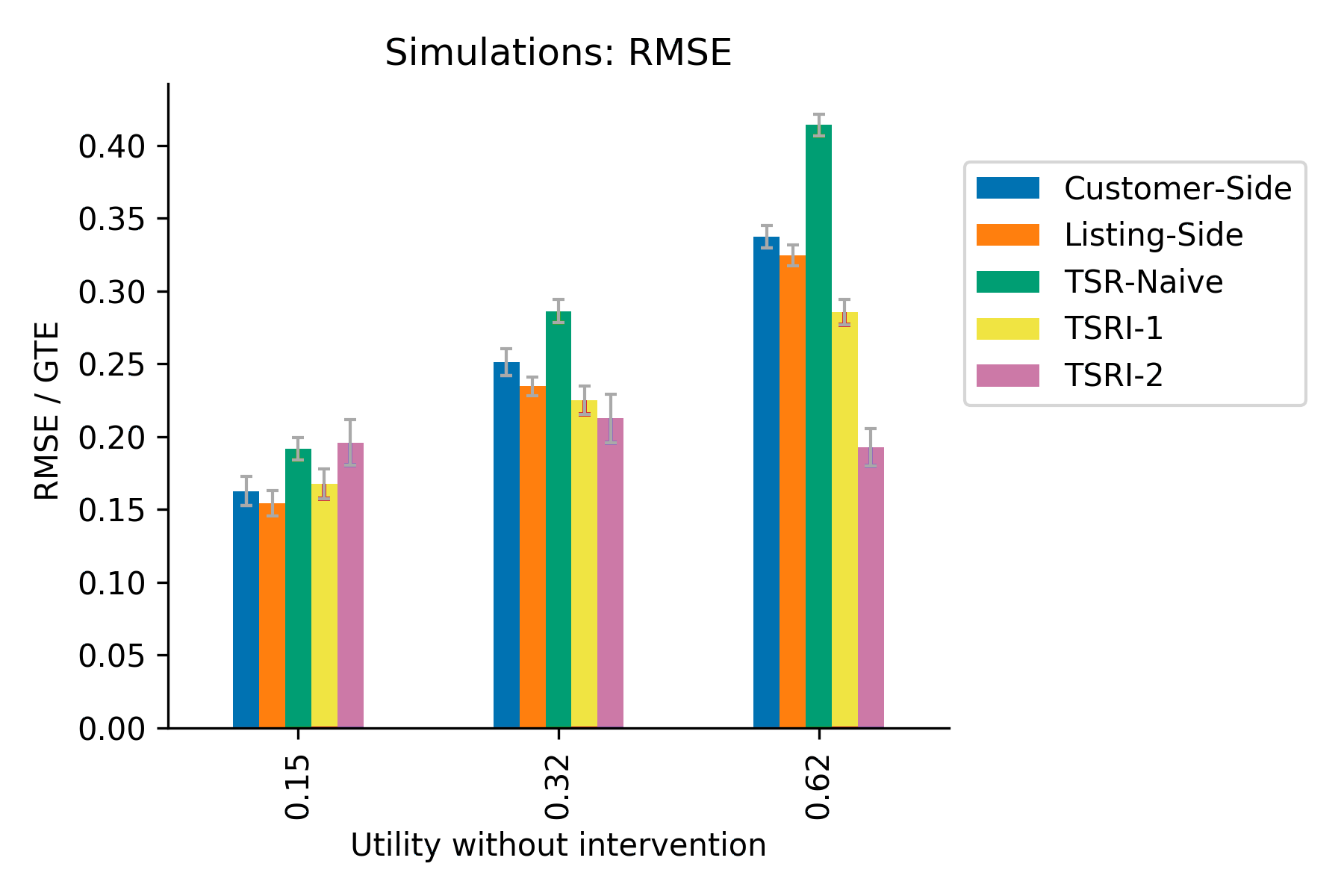}
    \end{tabular}
    \caption{(Varying average utility.) Results in balanced market with $\lambda=\tau=1$. 
    Top left: Bias of each estimator in the mean field model (normalized by GTE). 
    Top right: Bias of each estimator in simulations, averaged across 500 runs (normalized by GTE). 
    Bottom left: Standard error of estimates, calculated across 500 runs (normalized by GTE). 
    Bottom right: $\RMSE$ of the estimates, calculated across $500$ runs (normalized by GTE). 
    }
    \label{fig:robustness_vary_avg_utility}
\end{figure}

\textbf{Heterogeneity of customers.} There is one listing type $\theta$ and two customer types $\gamma_1$, $\gamma_2$. We fix the size of the treatment utility increase such that $\tilde{v}_{\gamma_1}(\theta) = 1.25 \cdot  v_{\gamma_1}(\theta)$ and $\tilde{v}_{\gamma_2}(\theta) = 1.25 \cdot v_{\gamma_2}(\theta)$. We additionally fix the steady state booking probabilities in global control and global treatment to be 20 percent and 23 percent, respectively.

Customers of type $\gamma_1$ have higher utility for the listing than customers of type $\gamma_2$. We vary the heterogeneity of the customers by varying the ratio $v_{\gamma_2}(\theta)/ v_{\gamma_1}(\theta)$.
\begin{itemize}
    \item Homogeneous: $v_{\gamma_2}(\theta)/ v_{\gamma_1}(\theta)=1$, with $v_{\gamma_1}(\theta) = 0.315,  v_{\gamma_2}(\theta) = 0.315$. 
    \item Low level of heterogeneity (Het-L): $v_{\gamma_2}(\theta)/ v_{\gamma_1}(\theta)=3$, with $v_{\gamma_1}(\theta) = 0.17,  v_{\gamma_2}(\theta) = 0.51$. 
    \item High level of heterogeneity (Het-H): $v_{\gamma_2}(\theta)/ v_{\gamma_1}(\theta)=3.83$, with $v_{\gamma_1}(\theta) = 0.12,  v_{\gamma_2}(\theta) = 0.46$. 
\end{itemize}

Results are presented in Figure \ref{fig:robustness_vary_customer_het}. As heterogeneity of the customers increases, the $\CR$ estimator has similar levels of bias, while $\LR, \TSRN$, and $\TSRIo$ slightly increase, though not as appreciably as the increase seen when varying the average utility level. In all cases, $\TSRIt$ has the lowest bias and highest standard error. In these simulations, $\TSRIt$ is the estimator that minimizes $\RMSE$, although this can change depending on the size of the market and the relative sizes of the bias and the standard error.

\begin{figure}
    \centering
    \begin{tabular}{l l }
        \includegraphics[height=.26\textwidth]{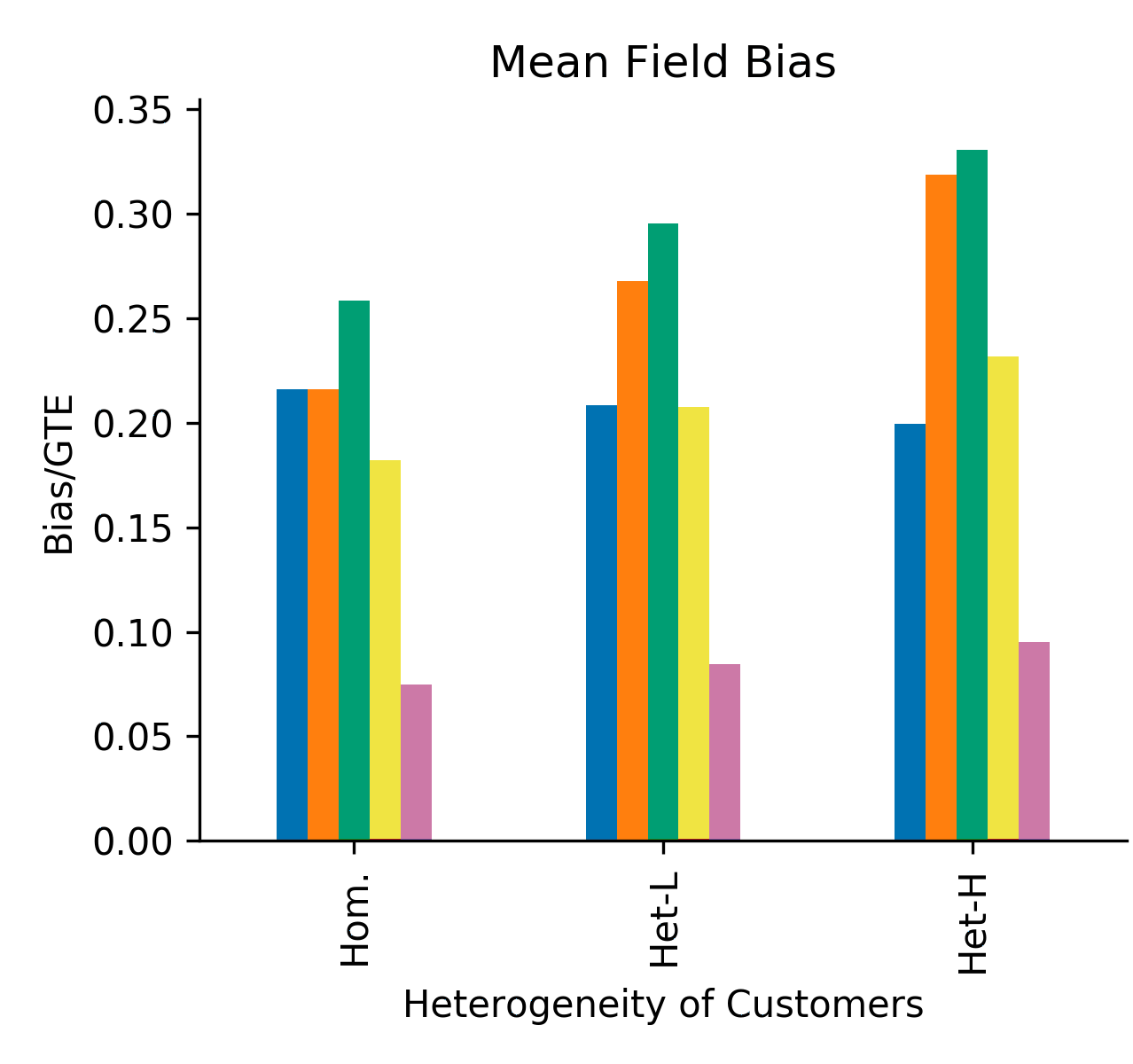}
        &
        \includegraphics[height=.26\textwidth]{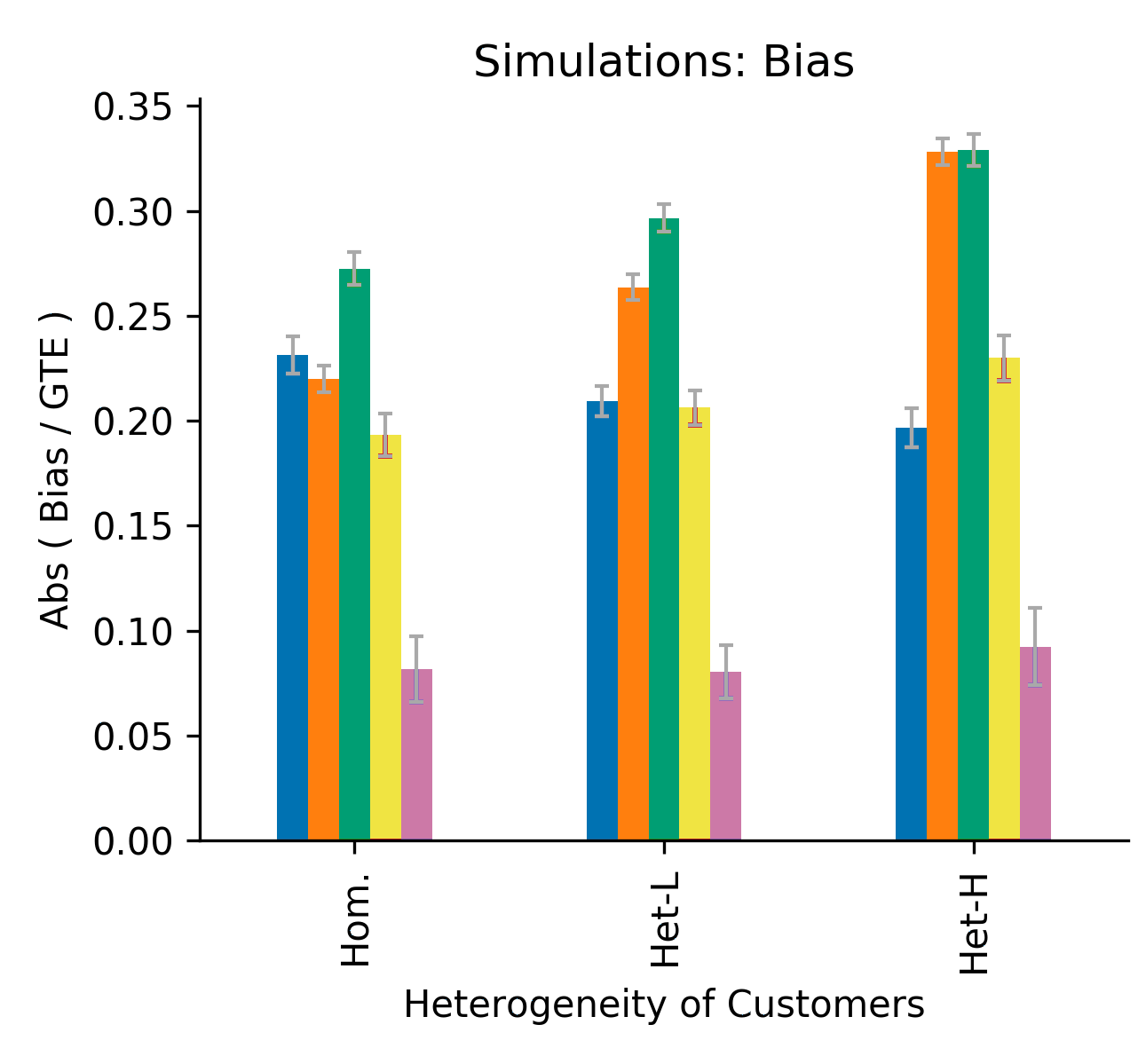}
        \\
        \includegraphics[height=.26\textwidth]{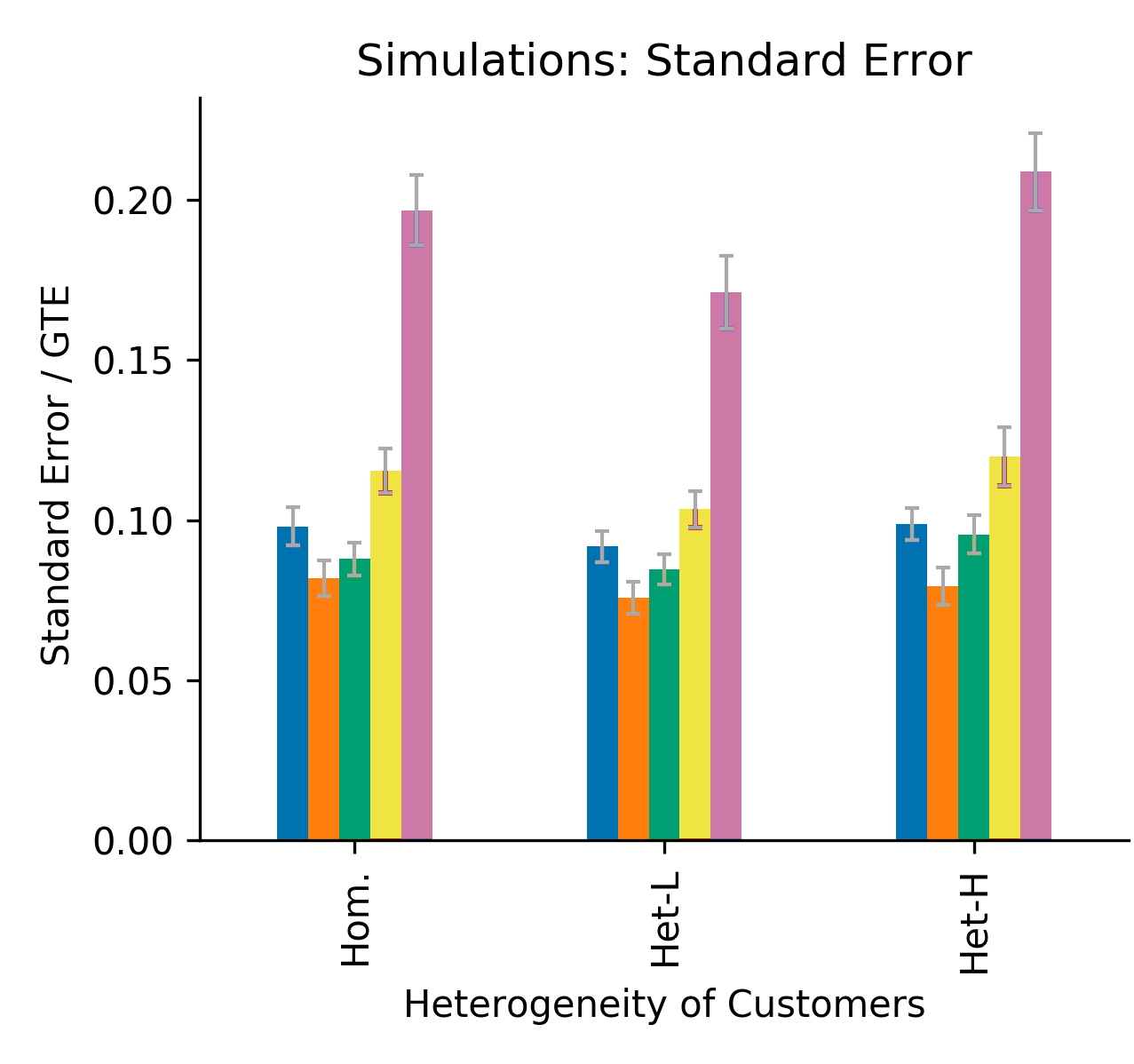}
        & 
        \includegraphics[height=.26\textwidth]{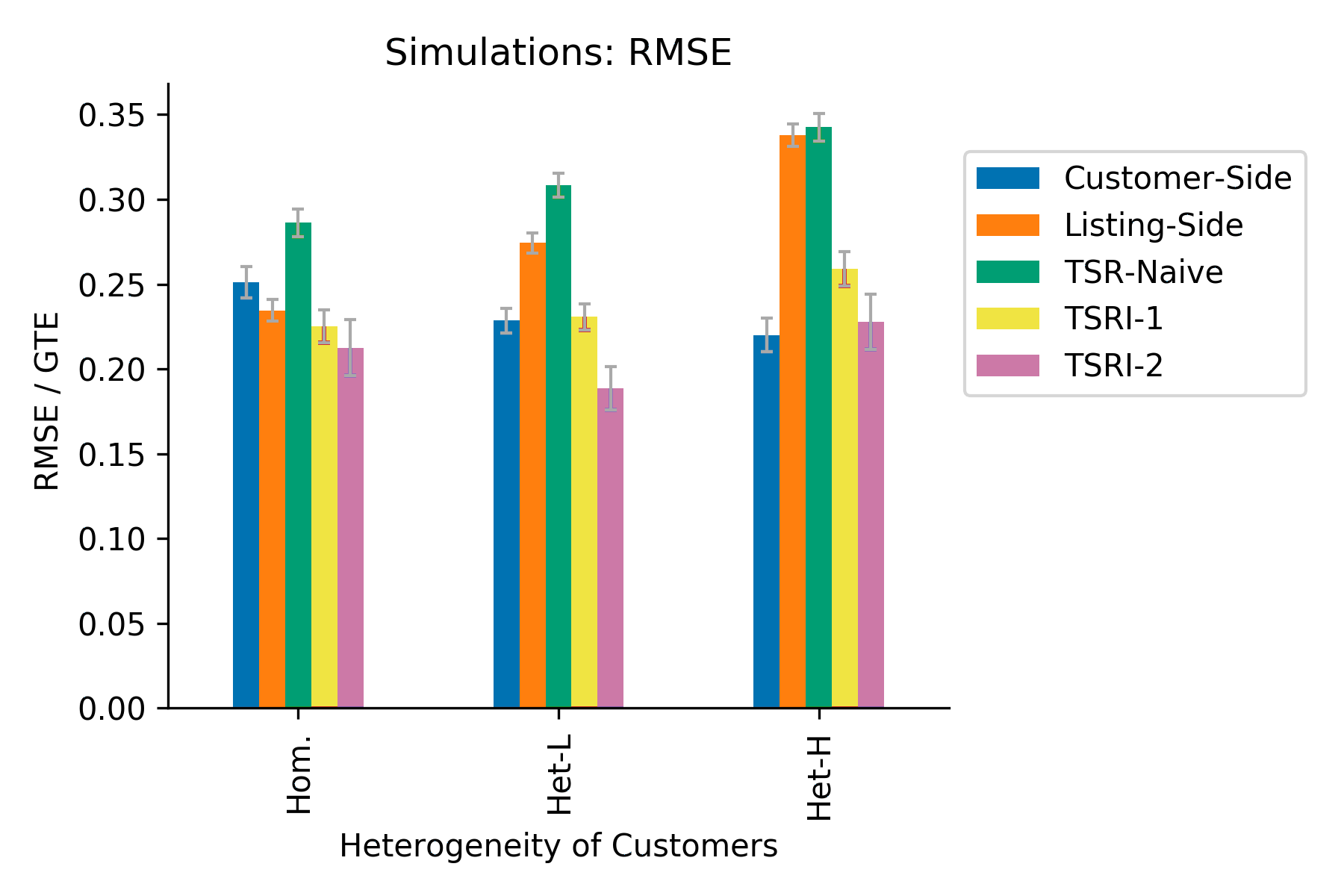}
    \end{tabular}
    \caption{(Varying heterogeneity of customers.) Results in balanced market with $\lambda=\tau=1$.
    Top left: Bias of each estimator in the mean field model (normalized by GTE). 
    Top right: Bias of each estimator in simulations, averaged across 500 runs (normalized by GTE). 
    Bottom left: Standard error of estimates, calculated across 500 runs (normalized by GTE). 
    Bottom right: $\RMSE$ of the estimates, calculated across $500$ runs (normalized by GTE). 
    }
    \label{fig:robustness_vary_customer_het}
\end{figure}

\textbf{Heterogeneity of listings.} There are two listing types $\theta_1$ and $\theta_1$ and one customer type $\gamma$. We fix the treatment effect so that $\tilde{v}(\theta) = 1.25 \cdot v(\theta)$. We additionally fix the steady state booking probabilities in global control and global treatment to be 20 percent and 23 percent, respectively.

We vary the heterogeneity of the listings by letting $v(\theta_2)>v(\theta_1)$ and varying the ratio $v(\theta_2) / v(\theta_1)$. 
\begin{itemize}
    \item Homogeneous: $v(\theta_2) / v(\theta_1) = 1$, with $v(\theta_1) = v(\theta_2) = 0.315$.
    \item Low level of heterogeneity (Het-L): $v(\theta_2) / v(\theta_1) = 1.6$, with $v(\theta_1) = 0.25, v(\theta_2) = 0.4$.
    \item High level of heterogeneity (Het-H): $v(\theta_2) / v(\theta_1)=6$, with $v(\theta_1) = 0.1, v(\theta_2) = 0.6$. 
\end{itemize}

Results are presented in Figure \ref{fig:robustness_vary_list_het}. As heterogeneity of the listings increases, the bias of the $\CR$ estimator increases significantly, the $\TSRN$ estimator increases to a lesser extent, and the $\LR$, $\TSRIo$, and $\TSRIt$ estimators remain roughly consistent. In all cases, $\TSRIt$ has the lowest bias and highest standard error. In these simulations, $\TSRIt$ is the estimator that minimizes $\RMSE$, although this can change depending on the size of the market and the relative sizes of the bias and the standard error. 

\begin{figure}
    \centering
    \begin{tabular}{l l }
        \includegraphics[height=.26\textwidth]{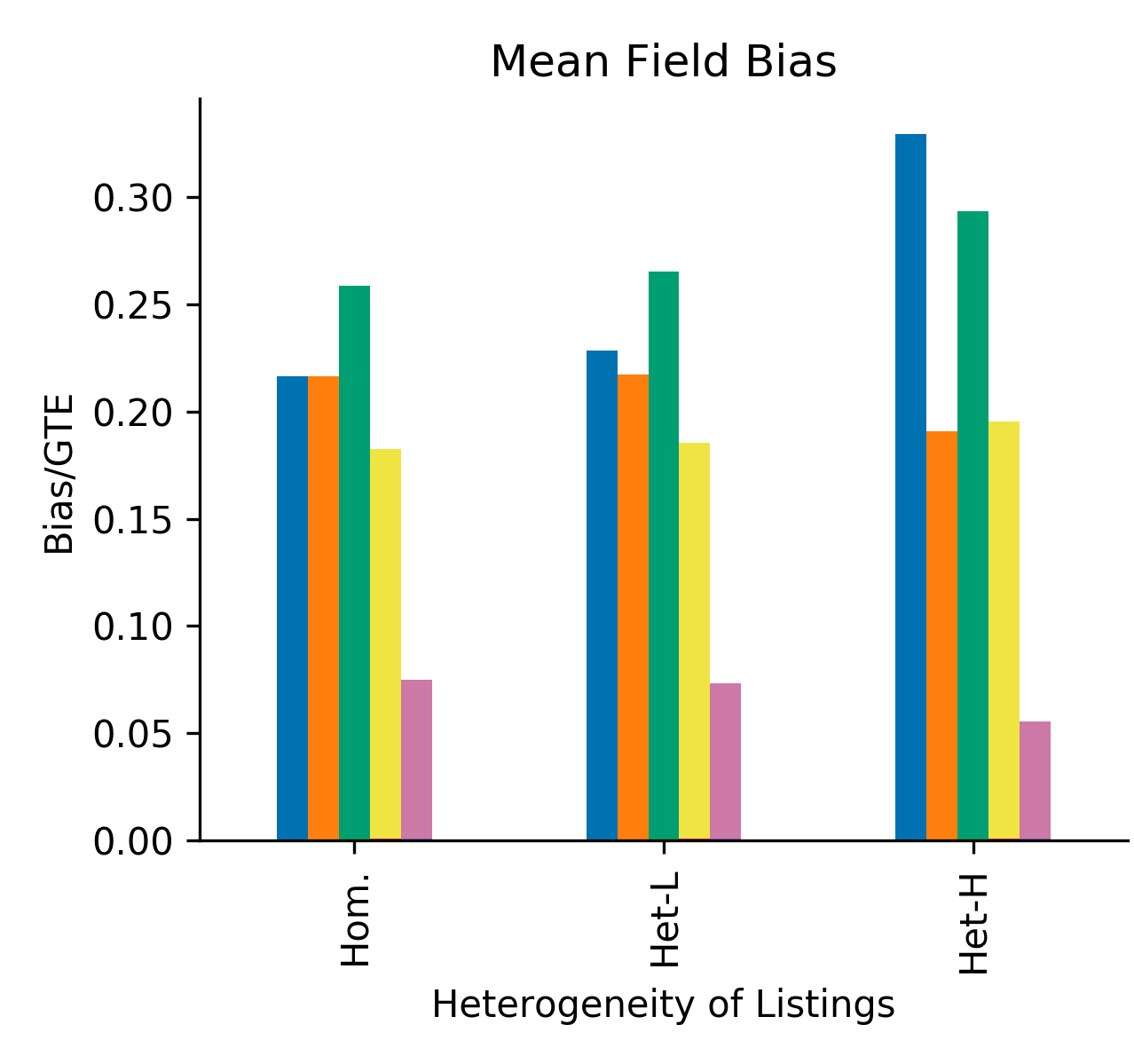}
        &
        \includegraphics[height=.26\textwidth]{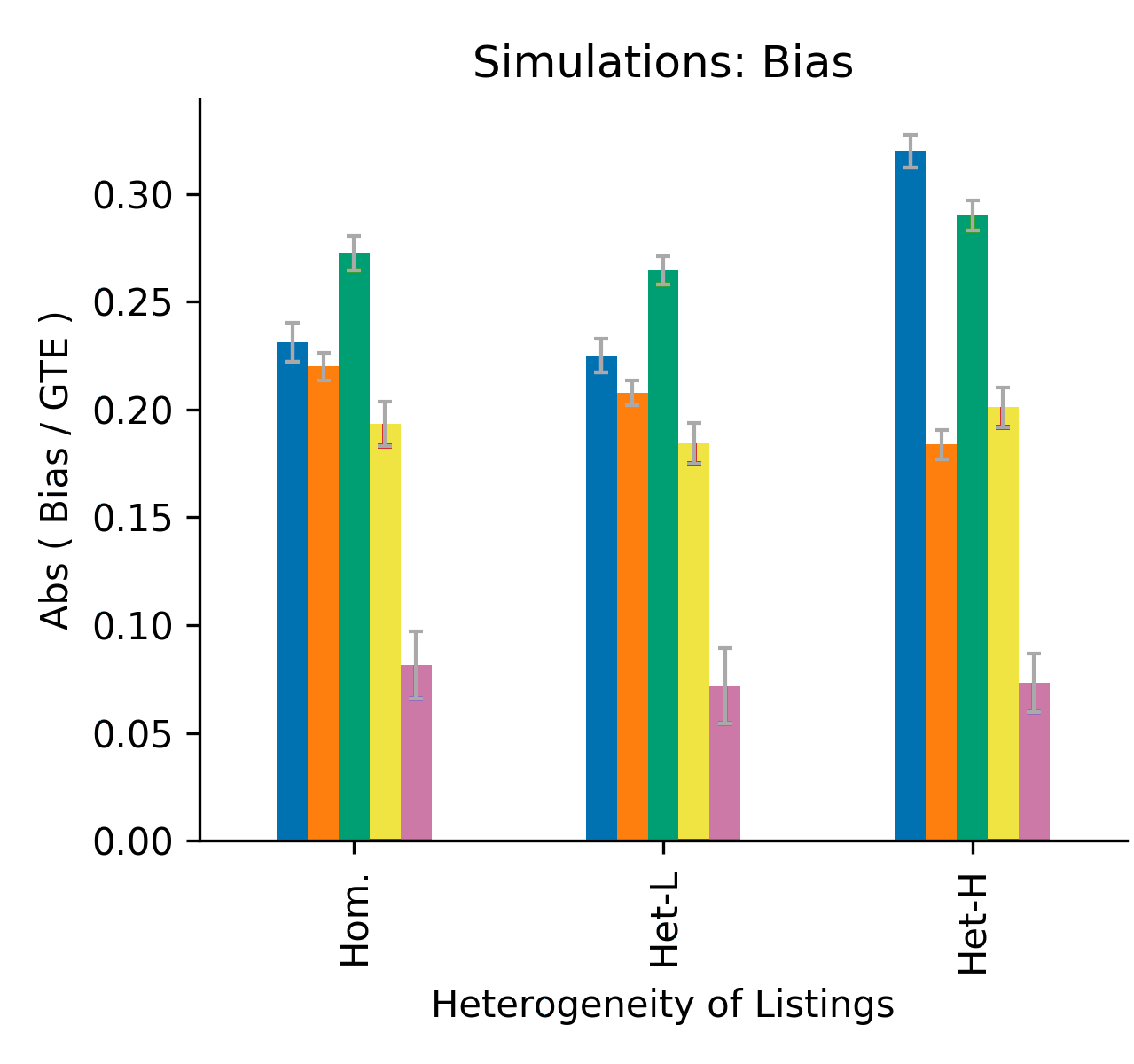}
        \\
        \includegraphics[height=.26\textwidth]{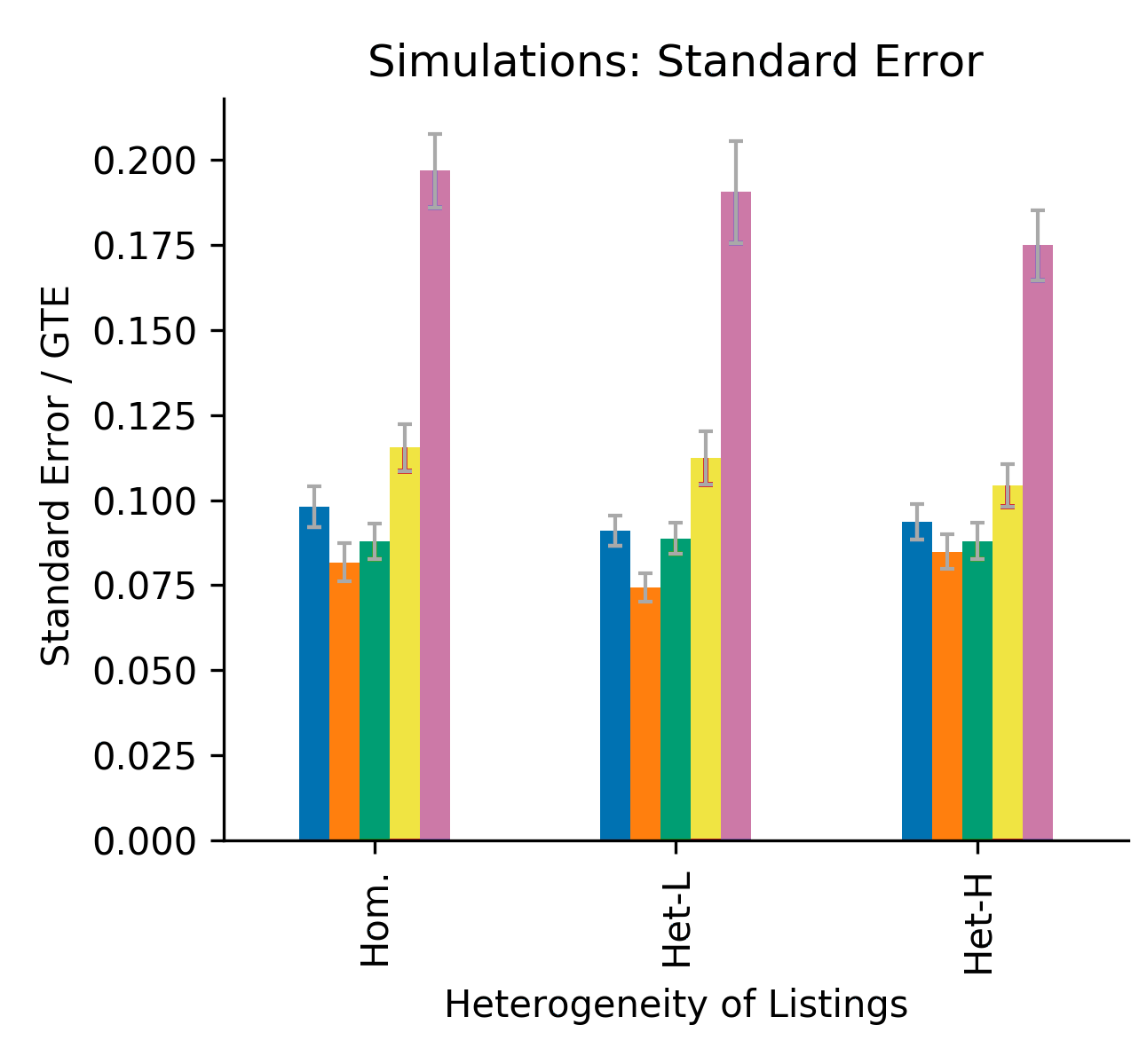}
        & 
        \includegraphics[height=.26\textwidth]{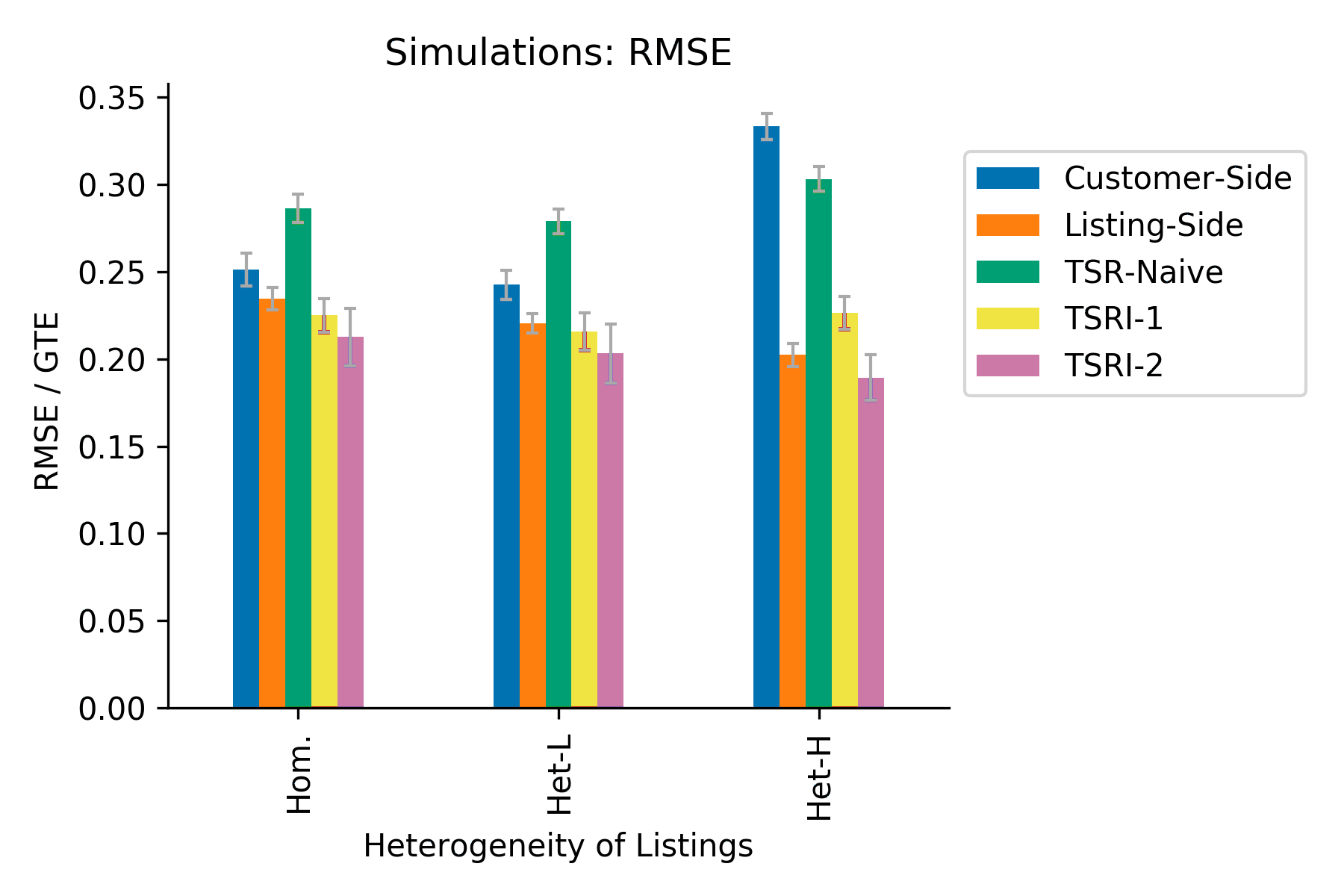}
    \end{tabular}
    \caption{(Varying heterogeneity of listings.) Results in balanced market with $\lambda=\tau=1$.
    Top left: Bias of each estimator in the mean field model (normalized by GTE). 
    Top right: Bias of each estimator in simulations, averaged across 500 runs (normalized by GTE). 
    Bottom left: Standard error of estimates, calculated across 500 runs (normalized by GTE). 
    Bottom right: $\RMSE$ of the estimates, calculated across $500$ runs (normalized by GTE). 
    }
    \label{fig:robustness_vary_list_het}
\end{figure}

\textbf{Heterogeneity of treatment effect.}
We consider the impact of heterogeneous treatment effects on the performance of the estimators. We restrict attention to where the treatment increases the utility, but the size of the increase can differ across the market. 

There are two listing types $\theta_1$ and $\theta_2$ and one customer type $\gamma$. We fix the customer's preferences pre-treatment such that the customer prefers $\theta_2$ to $\theta_1$ pre-treatment, with $v_\gamma(\theta_2)/v_\gamma(\theta_1) = 1.3$.  We compare three settings, one where the treatment effect has the same multiplicative lift on both listing types, one where the treatment amplifies the existing preference order, and one where the treatment reverses the existing preference order. 

We fix the steady state booking probabilities and global control and global treatment to be 20 percent and 23 percent, respectively. In all of the scenarios, we have $v(\theta_1) = 0.27, v(\theta_2)=0.351$.

\begin{itemize}
    \item Multiplicative: $\tilde{v}(\theta_1)=0.3375, \tilde{v}(\theta_2)=0.4388$.
    \item Heterogeneous treatment effects - amplify (HTE-amp): $\tilde{v}(\theta_1)=0.2727, \tilde{v}(\theta_2)=0.5265$.
    \item Heterogeneous treatment effects - reverse (HTE-rev): $\tilde{v}(\theta_1)=0.432, \tilde{v}(\theta_2)=0.355$.
\end{itemize}

Results are presented in Figure \ref{fig:robustness_vary_htes}. We find that when the treatment effect amplifies the existing preferences, the $\CR$ estimator has a larger bias. This is likely an artifact of the increase in listing heterogeneity after treatment (see discussion on varying listing heterogeneity and Figure \ref{fig:robustness_vary_list_het}). The bias is similar in the two settings with a multiplicative treatment effect and heterogeneous treatment effects that reverse the control preference order. In all cases, $\TSRIo$ has the lowest bias, highest standard error, and lowest $\RMSE$.

\begin{figure}
    \centering
    \begin{tabular}{l l}
        \includegraphics[height=.26\textwidth]{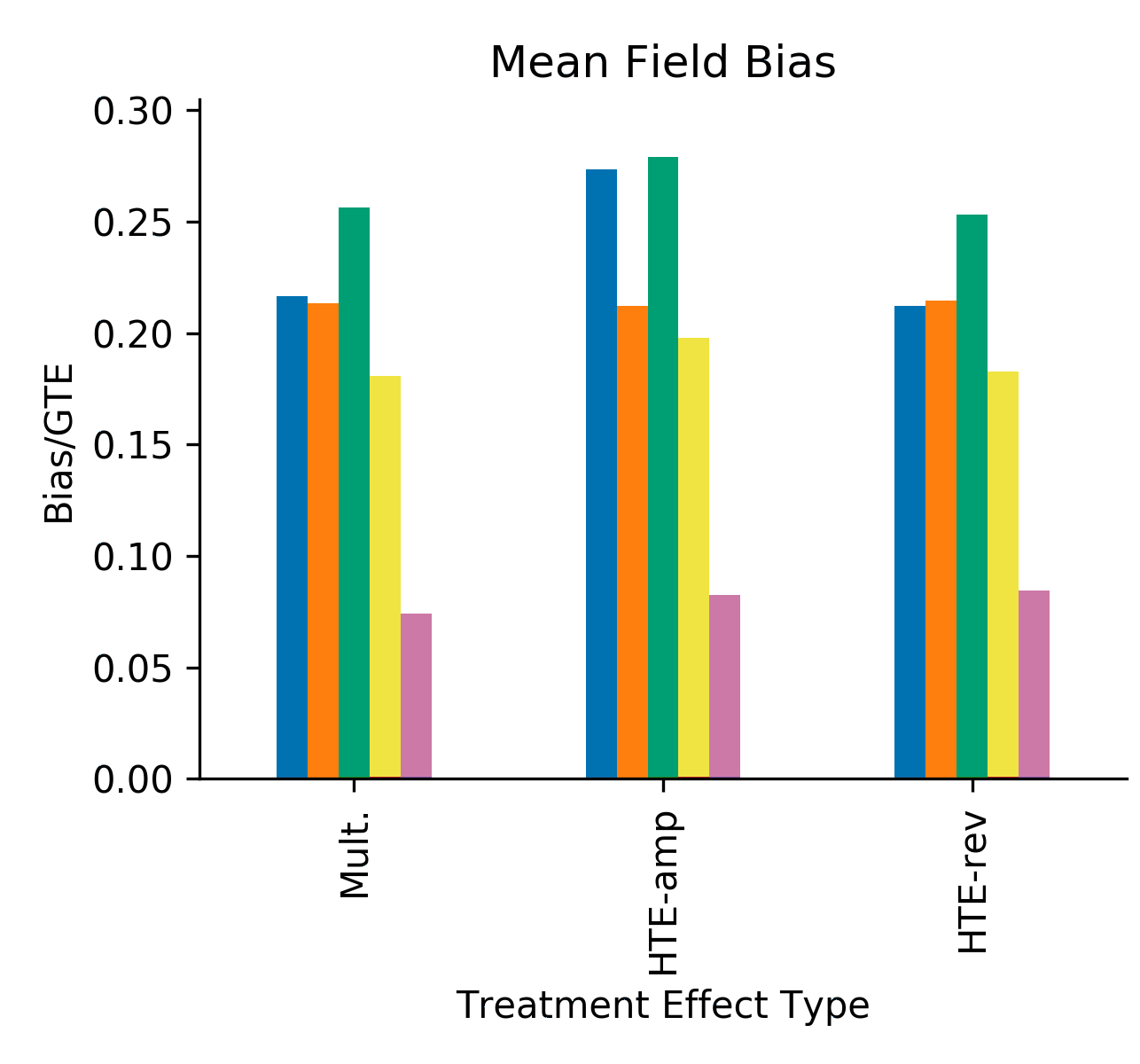}
        &
        \includegraphics[height=.26\textwidth]{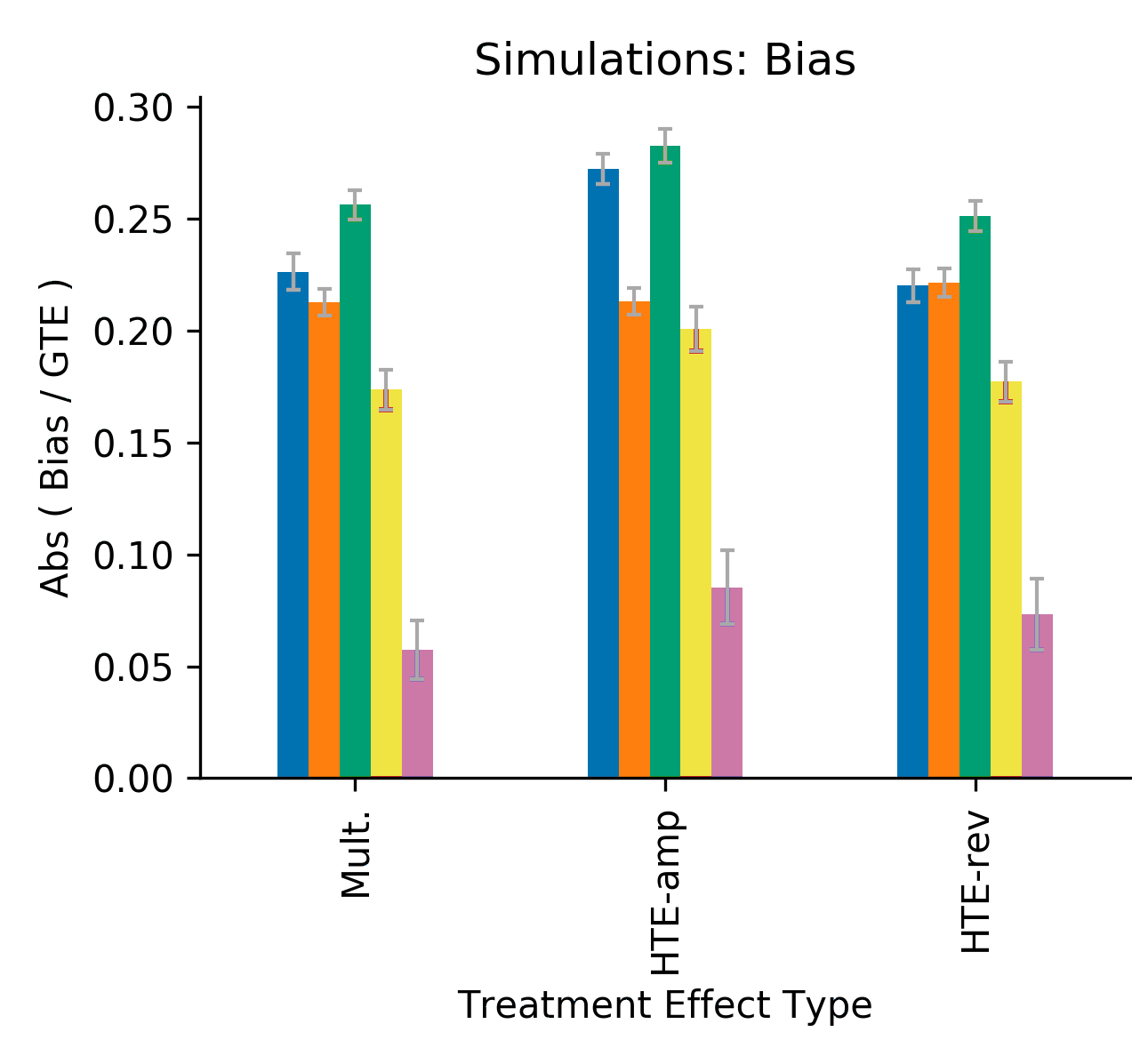}
        \\
        \includegraphics[height=.26\textwidth]{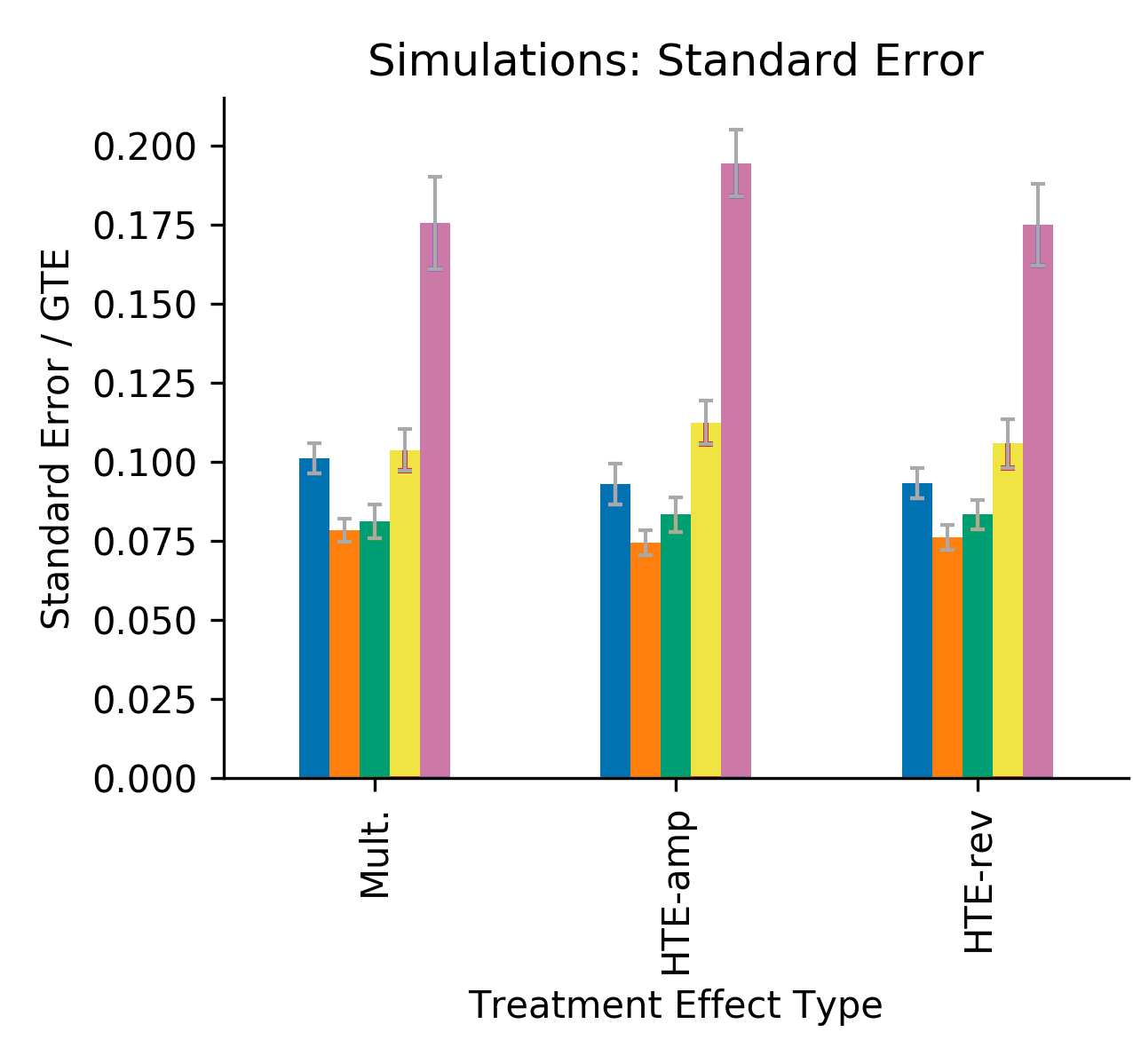}
        & 
        \includegraphics[height=.26\textwidth]{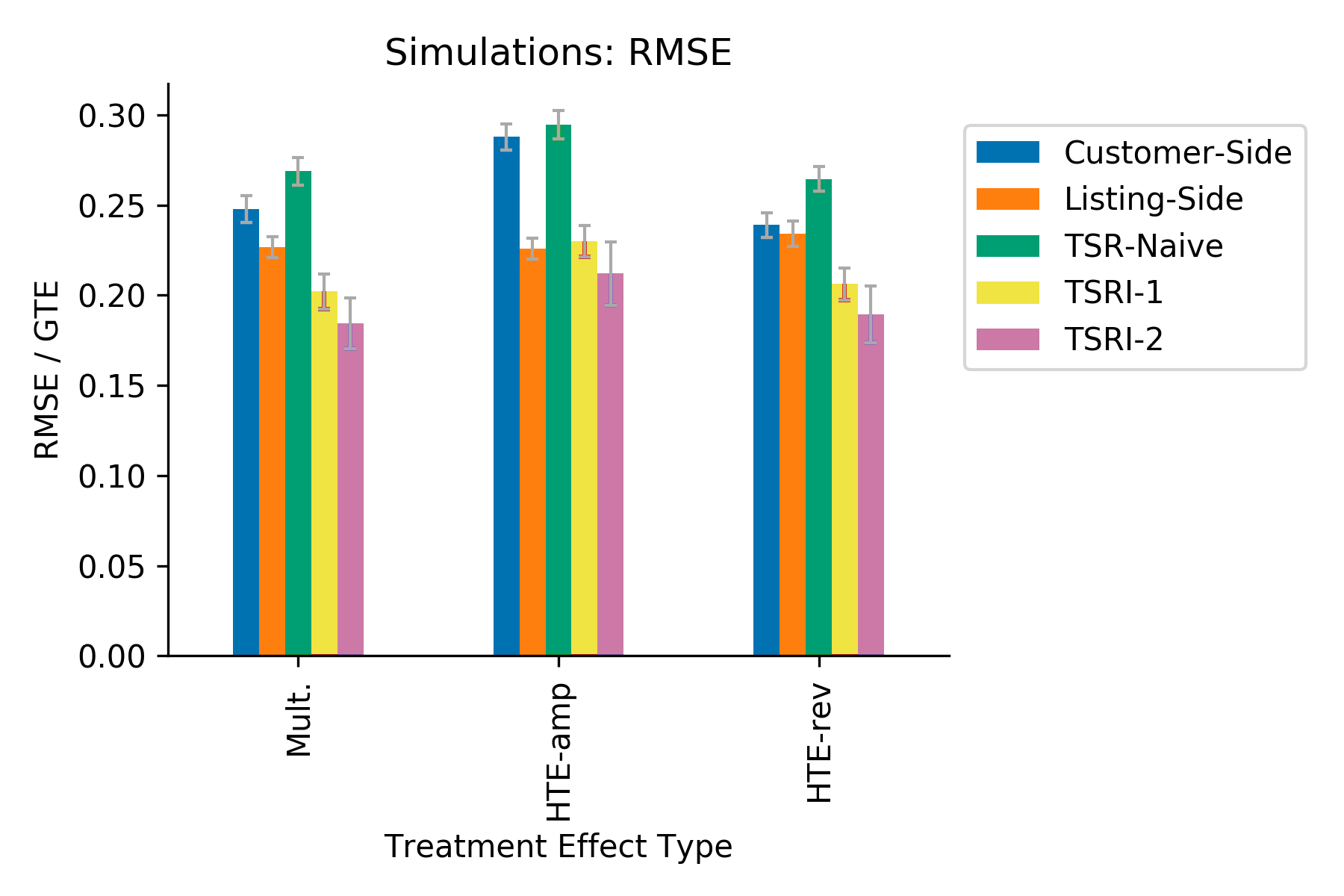}
    \end{tabular}
    \caption{(Varying heterogeneity of treatment lift on utilities.) Results in balanced market with $\lambda=\tau=1$. 
    Top left: Bias of each estimator in the mean field model (normalized by GTE). 
    Top right: Bias of each estimator in simulations, averaged across 500 runs (normalized by GTE). 
    Bottom left: Standard error of estimates, calculated across 500 runs (normalized by GTE). 
    Bottom right: $\RMSE$ of the estimates, calculated across $500$ runs (normalized by GTE).
    }
    \label{fig:robustness_vary_htes}
\end{figure}

\subsection{Robustness with varying market balance}
\label{ssec:robustness_vary_lambda}

We now replicate Figures \ref{fig:numerics_homogeneous} and \ref{fig:simulations_hom} for different market settings, verifying the results we present in the main text regarding the dependence of the estimators on market balance. In Section \ref{ssec:robustness_market_scenarios}, we consider four scenarios at a fixed market balance, by varying average utility, customer heterogeneity, listing heterogeneity, and treatment effect heterogeneity. In this section, we choose one representative set of parameters from each scenario and show how the performance of the estimators change as we change market balance. In particular, we show that the behavior of the estimators in a large market setting is still similar to the behavior obtained in the mean field limit, even in the presence of heterogeneity in the market. 

The representative settings shown are low utility (varying average utility), high level of heterogeneity (heterogeneity of customers), high level of heterogeneity (heterogeneity of listings), and heterogeneous treatment effects amplifying existing preference (heterogeneity of treatment effects). The parameters are as defined in Section \ref{ssec:robustness_market_scenarios}. One might suspect that these settings, either with a low booking probability or high levels of heterogeneity, are the ones most likely to differ from the mean field model. We show, however, that the findings in the main text are indeed robust.

Across our range of simulations, we see qualitatively similar behavior to our findings in the main text: the naive $\CR$ estimator has lower bias for small $\lambda/\tau$ and the naive $\LR$ estimator has lower bias for large $\lambda/\tau$, while the $\TSR$ estimators interpolate between the two.  The $\TSR$ estimators offer these bias reductions at the cost of higher variance.   Further, the bias of $\CR$ is higher at larger $\lambda/\tau$, and the bias of $\LR$ is higher at smaller $\lambda/\tau$.

We note that although the qualitative findings are similar, the bias in the simulations is not identical to the bias in the mean field model, especially at the smaller and larger values of $\lambda$.  We conjecture that this is due to the fact that stochastic effects matter in these extremes.  When $\lambda$ is small, since $T$ is fixed, we see a (relatively) smaller number of customer arrivals in the same time horizon; indeed, this is why standard errors are higher overall in this setting.  On the other hand, when $\lambda$ is large, then (relatively) few listings will be available.  In this setting, if the steady state number of available listings is very small relative to $N$, then our mean field approximation will begin to be less accurate, even though the qualitative behavior is similar. The discrepancies may be larger with the $\TSR$ estimators, since there are even fewer interactions happening in each of the four cells.

\begin{figure}
    \centering
    \begin{tabular}{l l }
        \includegraphics[height=.26\textwidth]{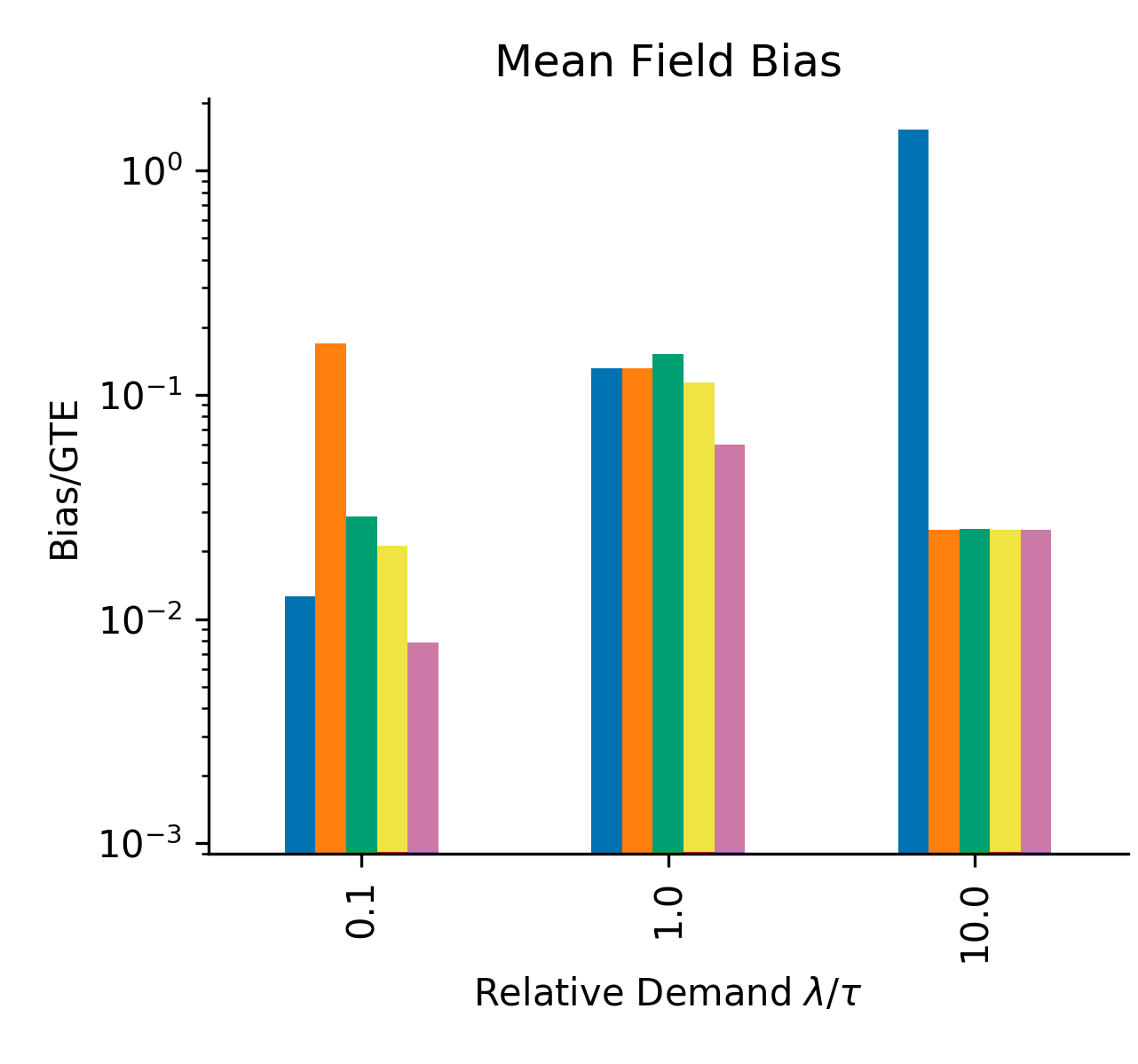}
        &
        \includegraphics[height=.26\textwidth]{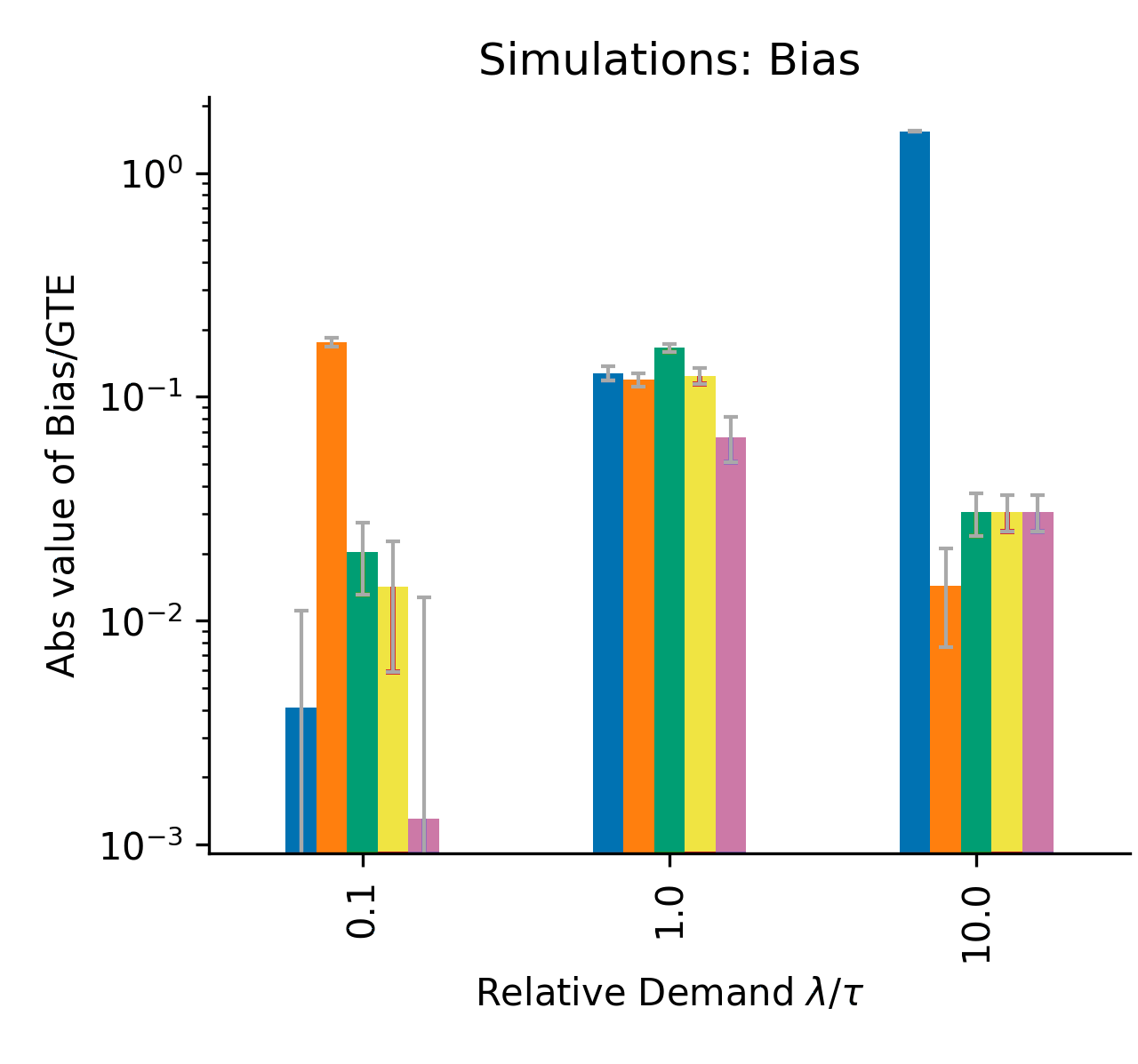}
        \\
        \includegraphics[height=.26\textwidth]{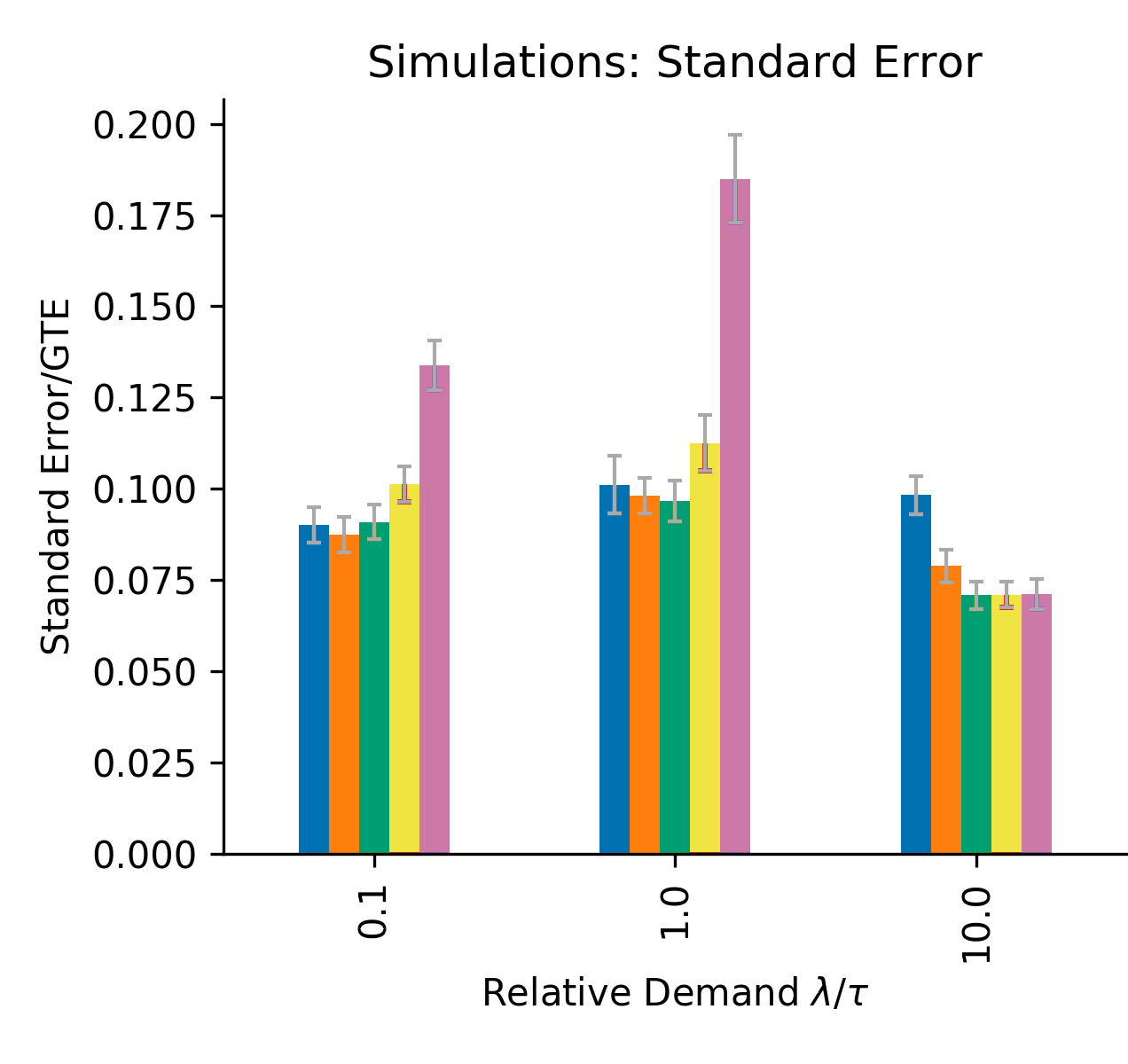}
        & 
        \includegraphics[height=.26\textwidth]{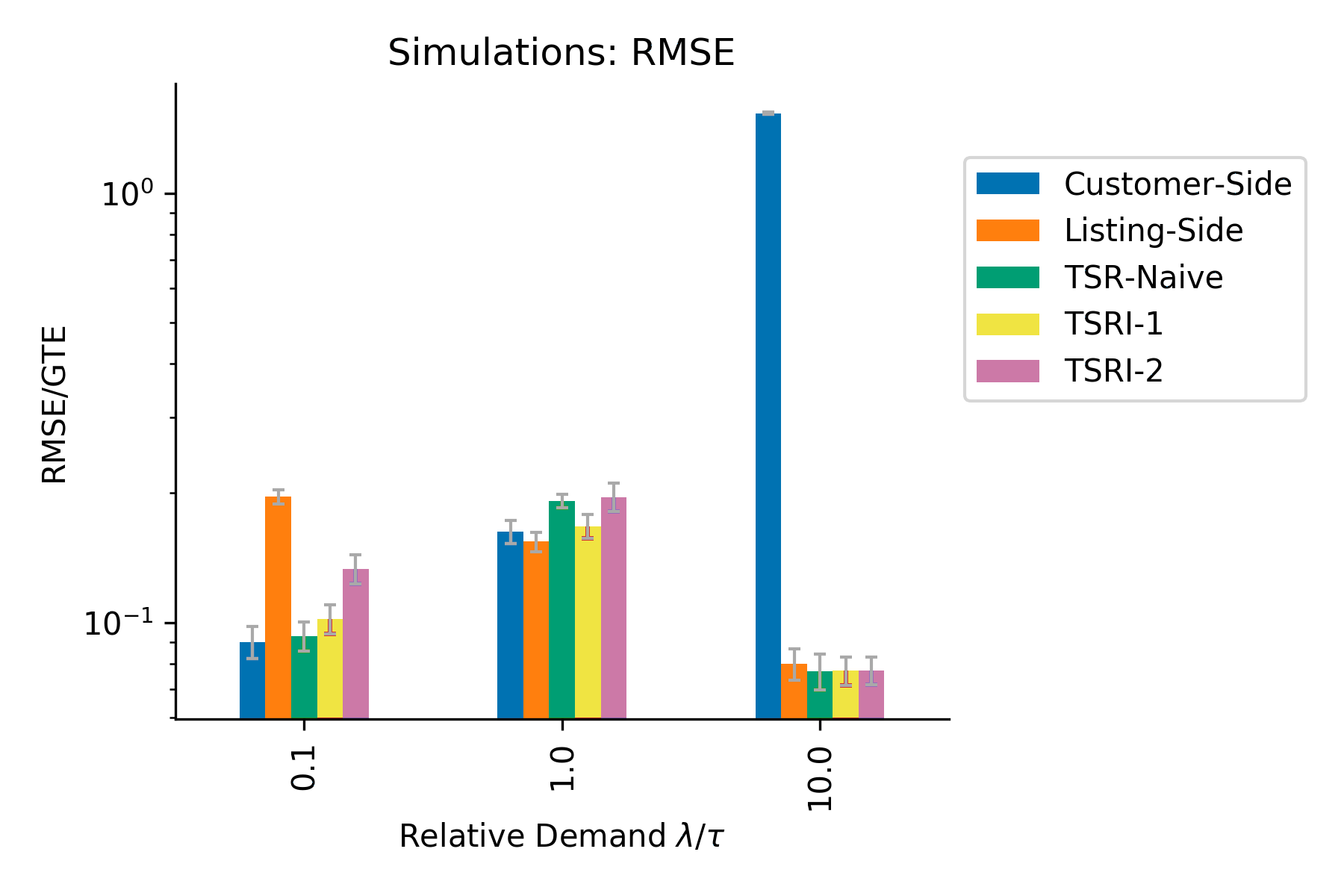}
    \end{tabular}
    \caption{(Varying market balance - small average utility.)
    Top left: Bias of each estimator in the mean field model (normalized by GTE). 
    Top right: Bias of each estimator in simulations, averaged across 500 runs (normalized by GTE). 
    Bottom left: Standard error of estimates, calculated across 500 runs (normalized by GTE). 
    Bottom right: $\RMSE$ of the estimates, calculated across $500$ runs (normalized by GTE). 
    }
    \label{fig:robustness_vary_lambda_small_utility}
\end{figure}

\begin{figure}
    \centering
    \begin{tabular}{l l }
        \includegraphics[height=.26\textwidth]{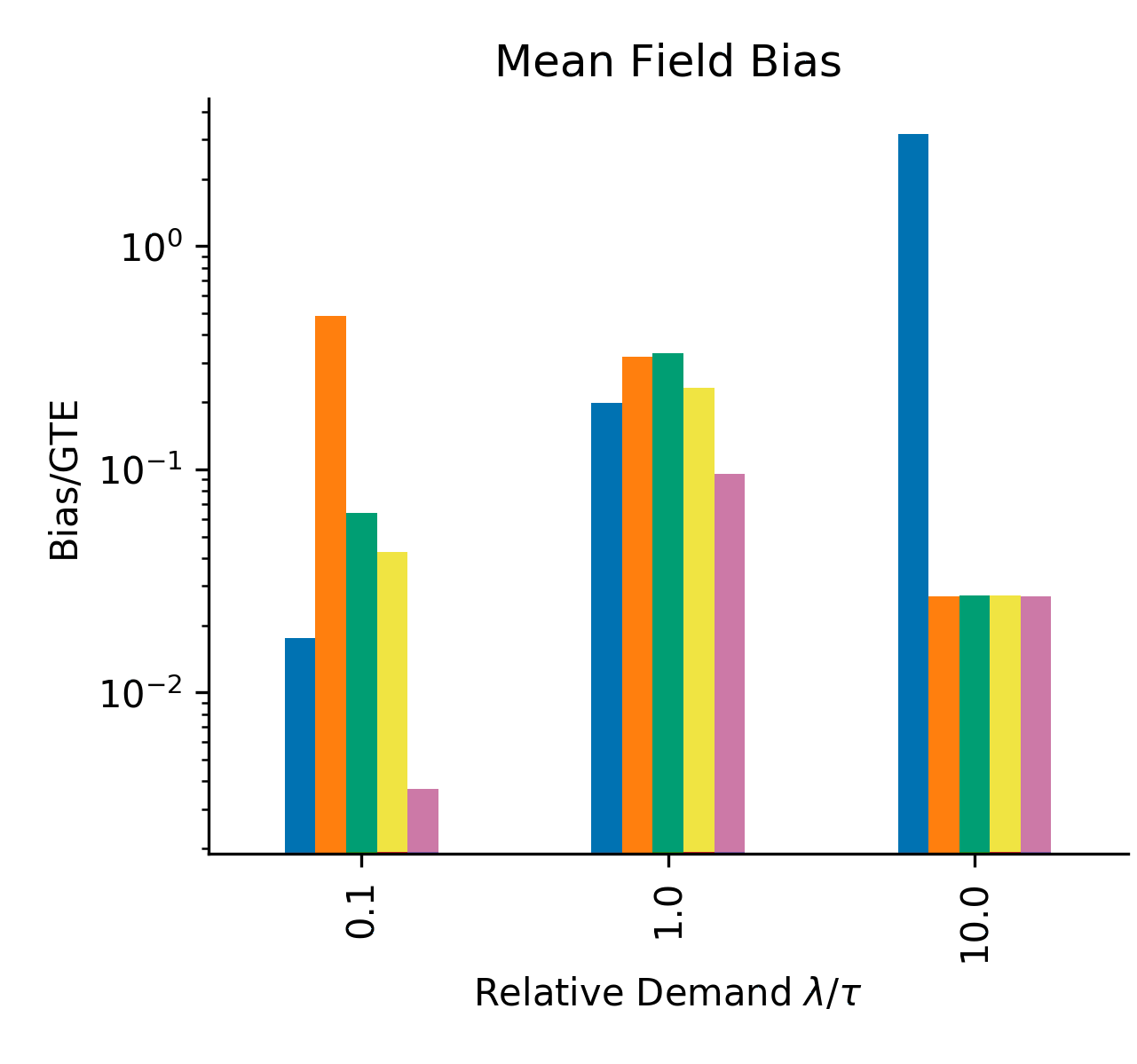}
        &
        \includegraphics[height=.26\textwidth]{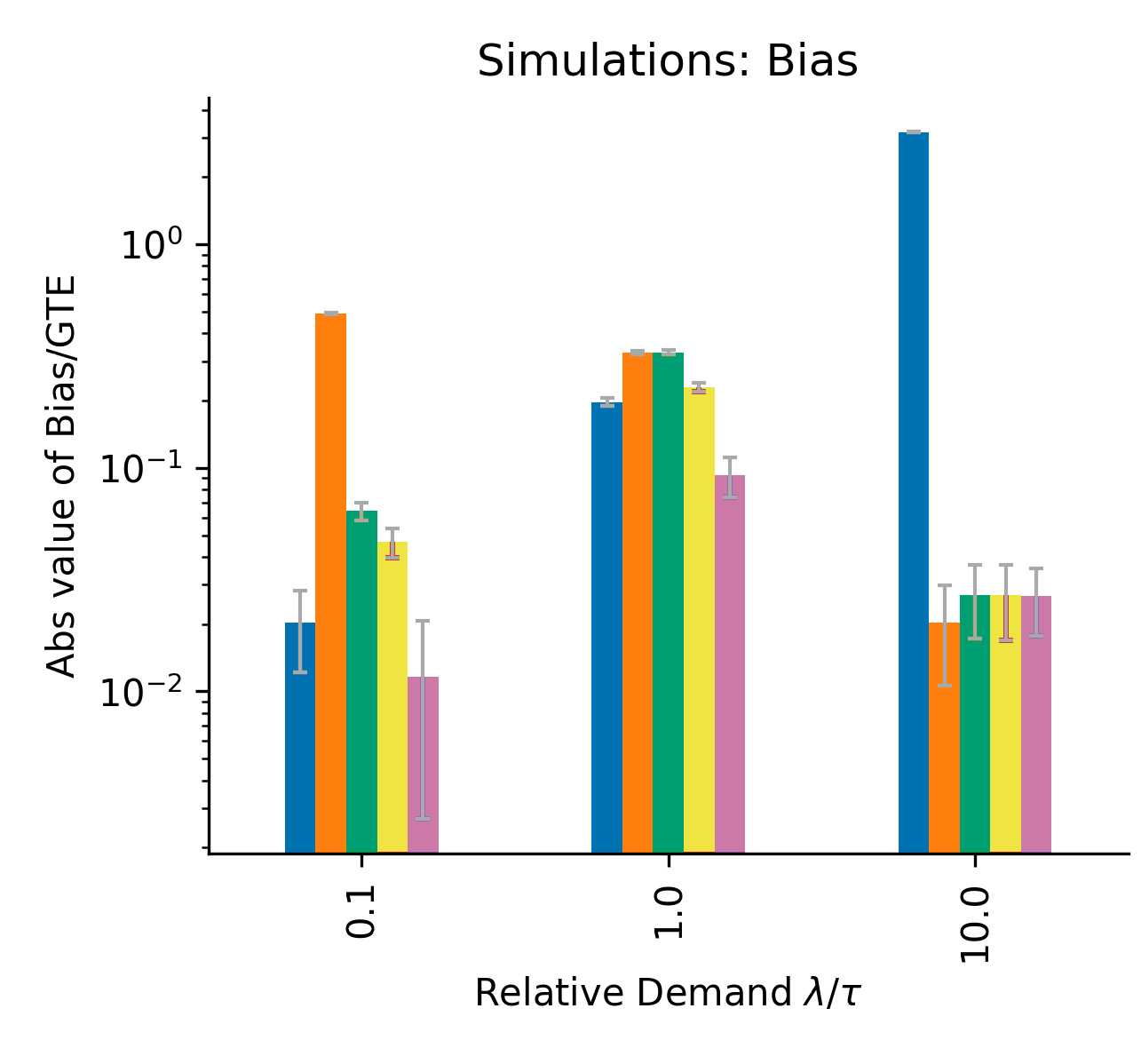}
        \\
        \includegraphics[height=.26\textwidth]{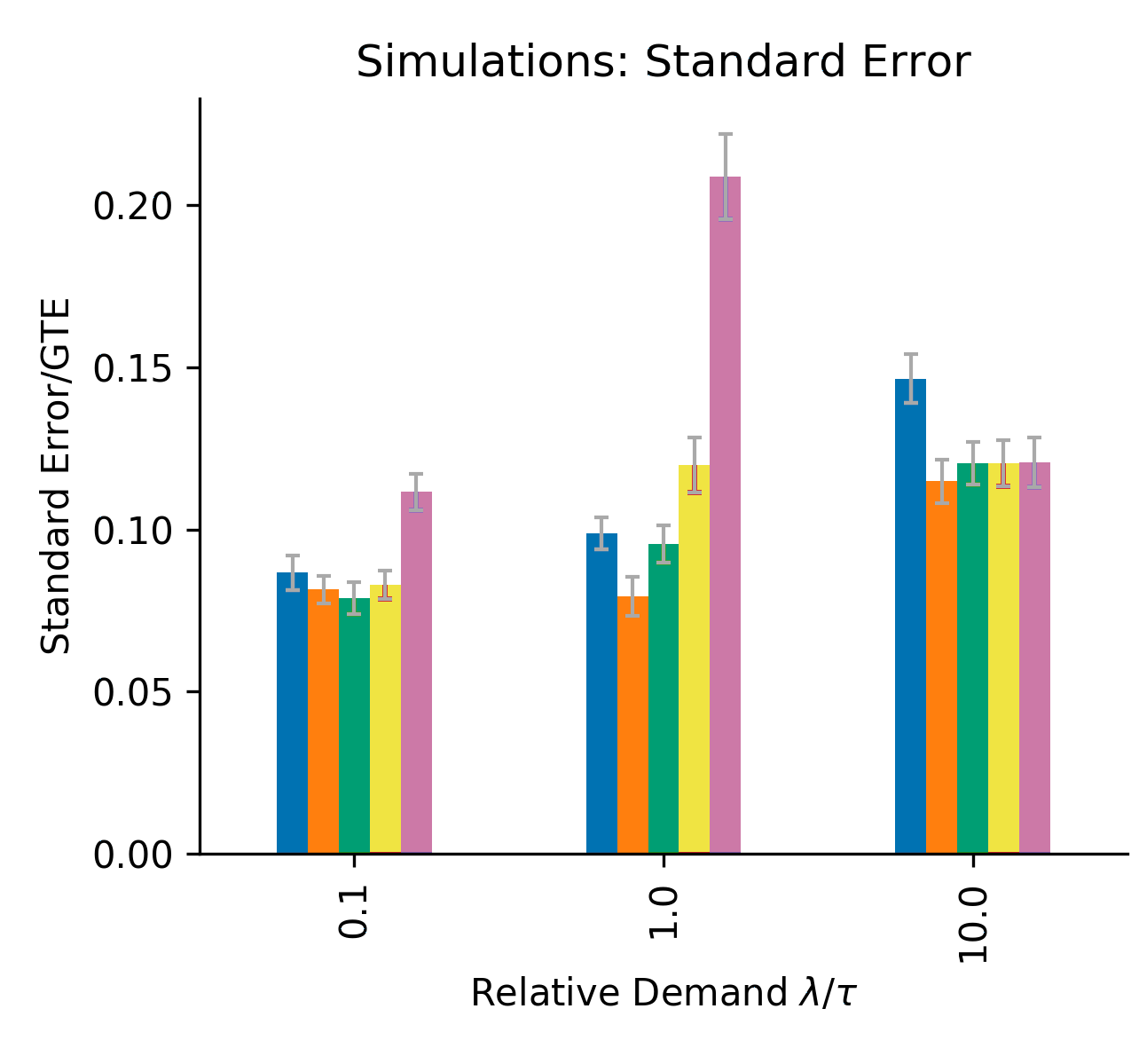}
        & 
        \includegraphics[height=.26\textwidth]{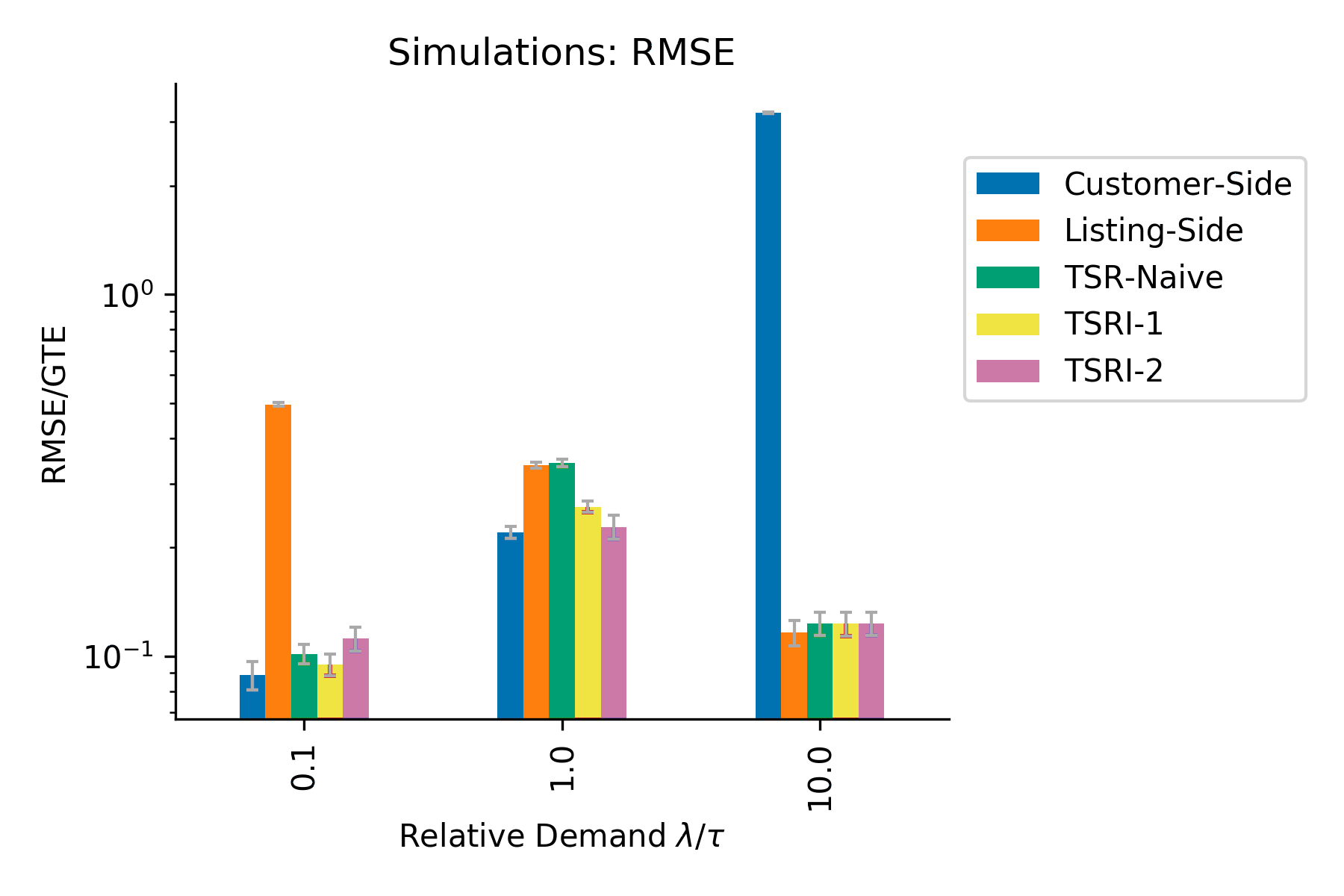}
    \end{tabular}
    \caption{(Varying market balance - large heterogeneity between customers.)
    Top left: Bias of each estimator in the mean field model (normalized by GTE). 
    Top right: Bias of each estimator in simulations, averaged across 200 runs (normalized by GTE). 
    Bottom left: Standard error of estimates, calculated across 200 runs (normalized by GTE). 
    Bottom right: $\RMSE$ of the estimates, calculated across $200$ runs (normalized by GTE). 
    }
    \label{fig:robustness_vary_lambda_heterogeneous_customers}
\end{figure}

\begin{figure}
    \centering
    \begin{tabular}{l l }
        \includegraphics[height=.26\textwidth]{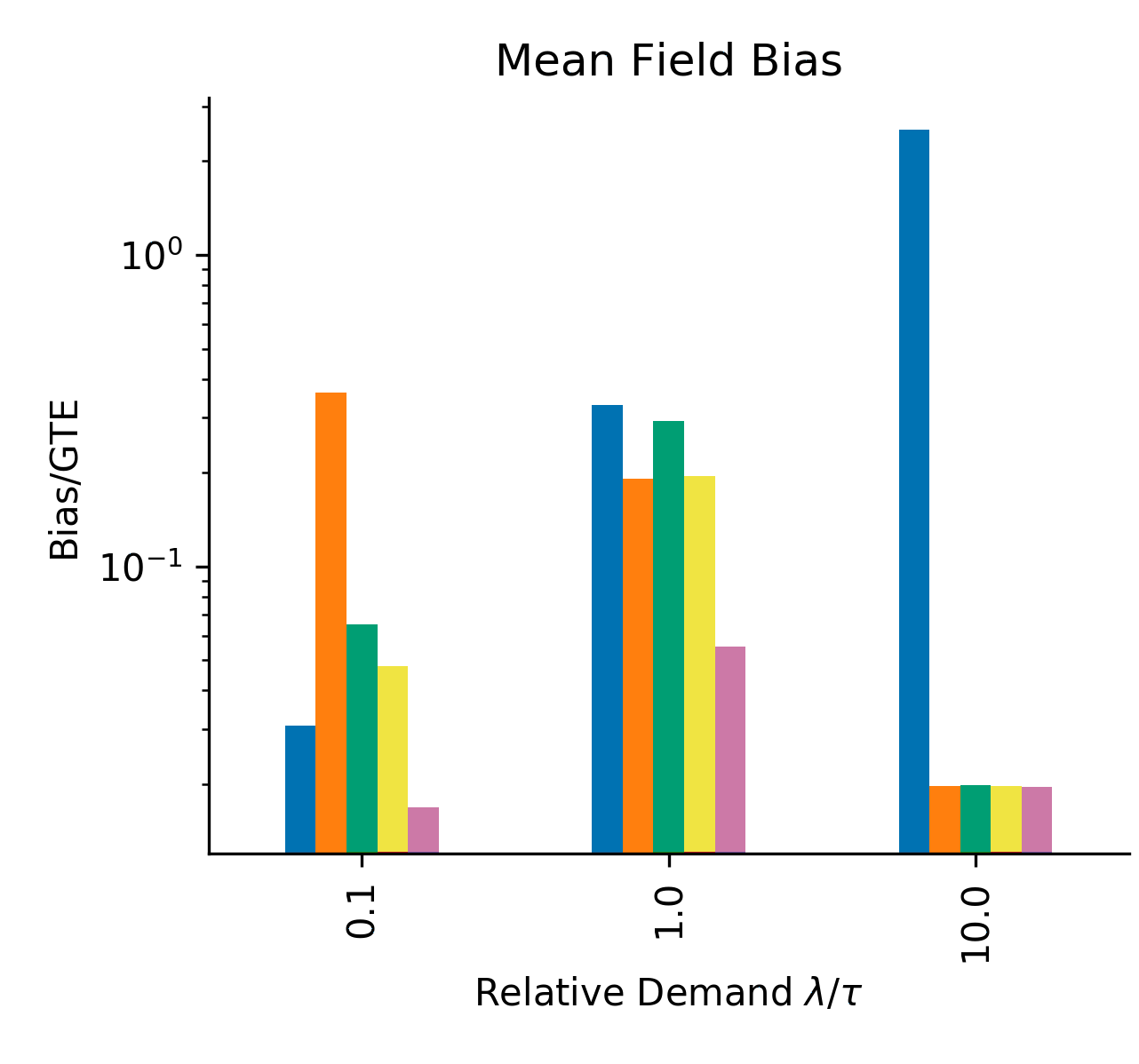}
        &
        \includegraphics[height=.26\textwidth]{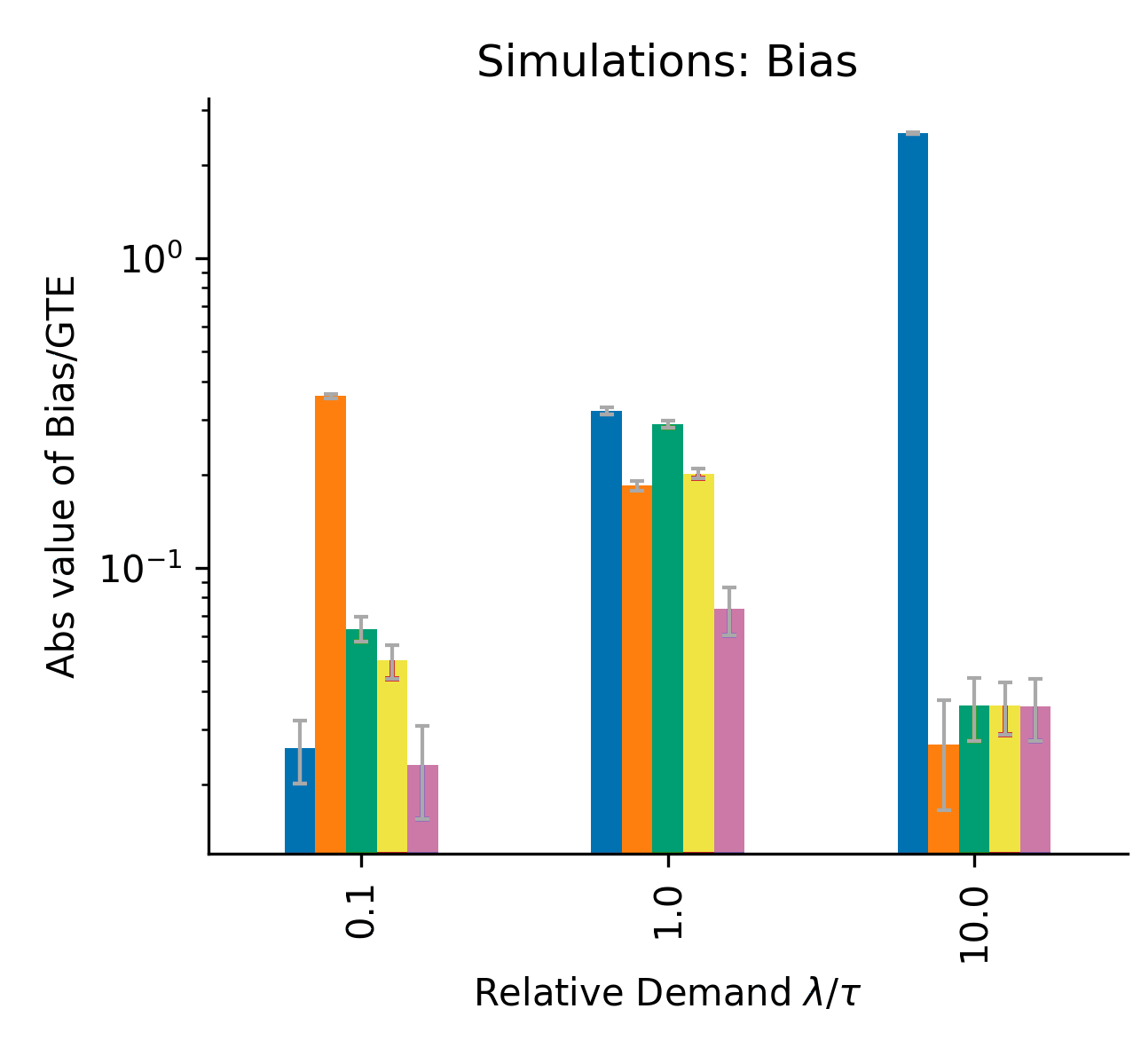}
        \\
        \includegraphics[height=.26\textwidth]{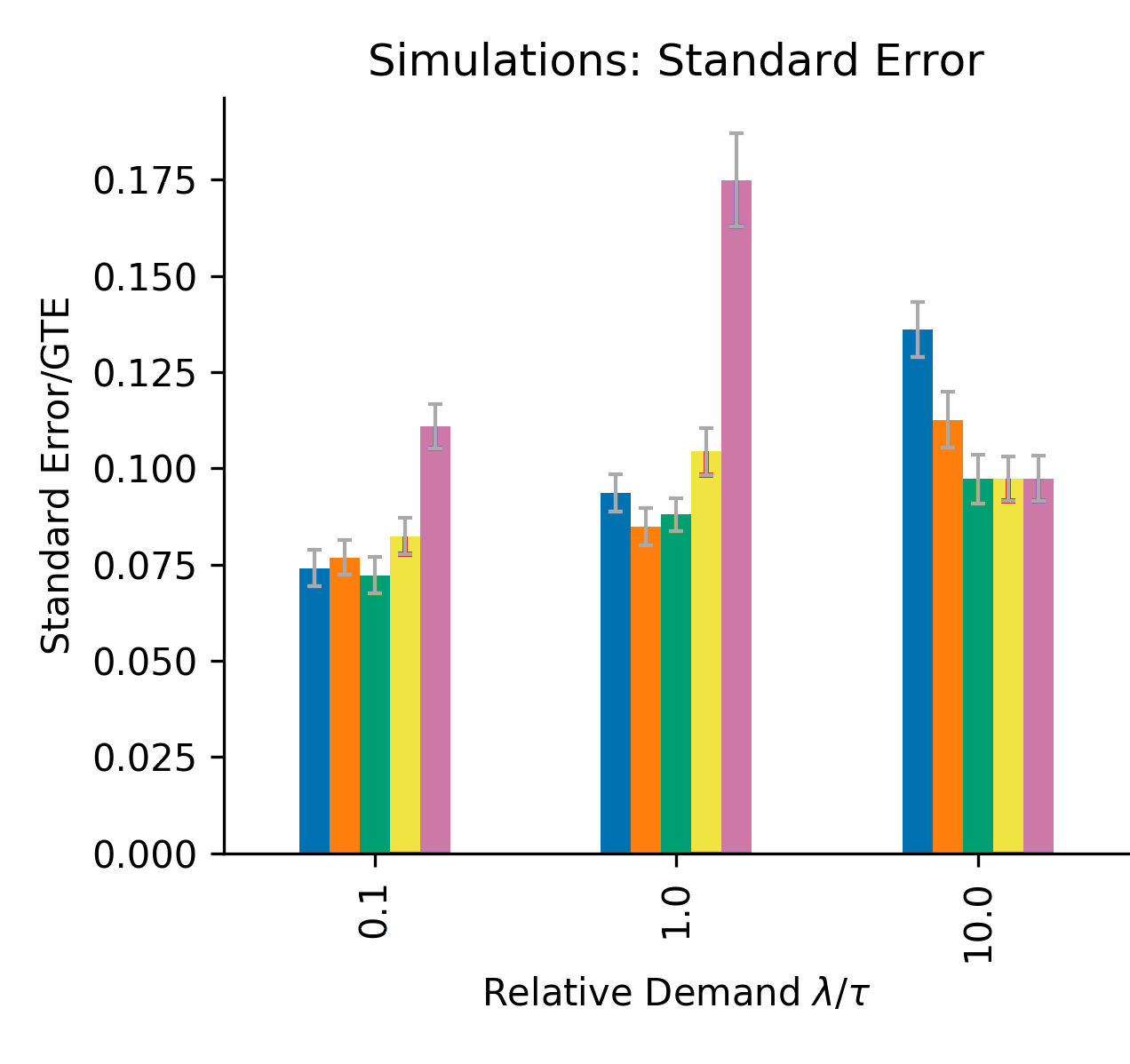}
        & 
        \includegraphics[height=.26\textwidth]{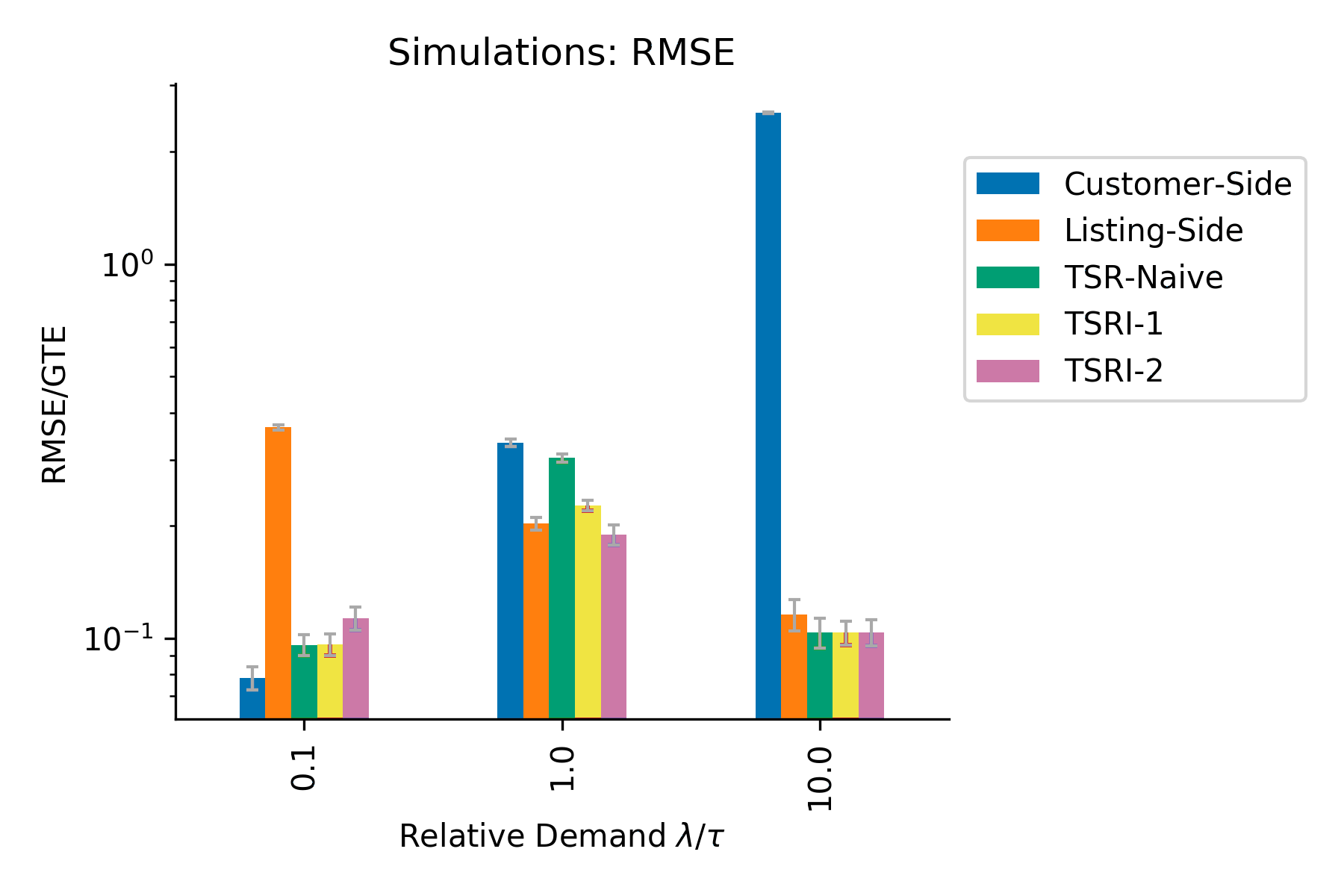}
    \end{tabular}
    \caption{(Varying market balance - large heterogeneity between listings.)
    Top left: Bias of each estimator in the mean field model (normalized by GTE). 
    Top right: Bias of each estimator in simulations, averaged across 500 runs (normalized by GTE). 
    Bottom left: Standard error of estimates, calculated across 500 runs (normalized by GTE). 
    Bottom right: $\RMSE$ of the estimates, calculated across $500$ runs (normalized by GTE). 
    }
    \label{fig:robustness_vary_lambda_heterogeneous_listings}
\end{figure}

\begin{figure}
    \centering
    \begin{tabular}{l l }
        \includegraphics[height=.26\textwidth]{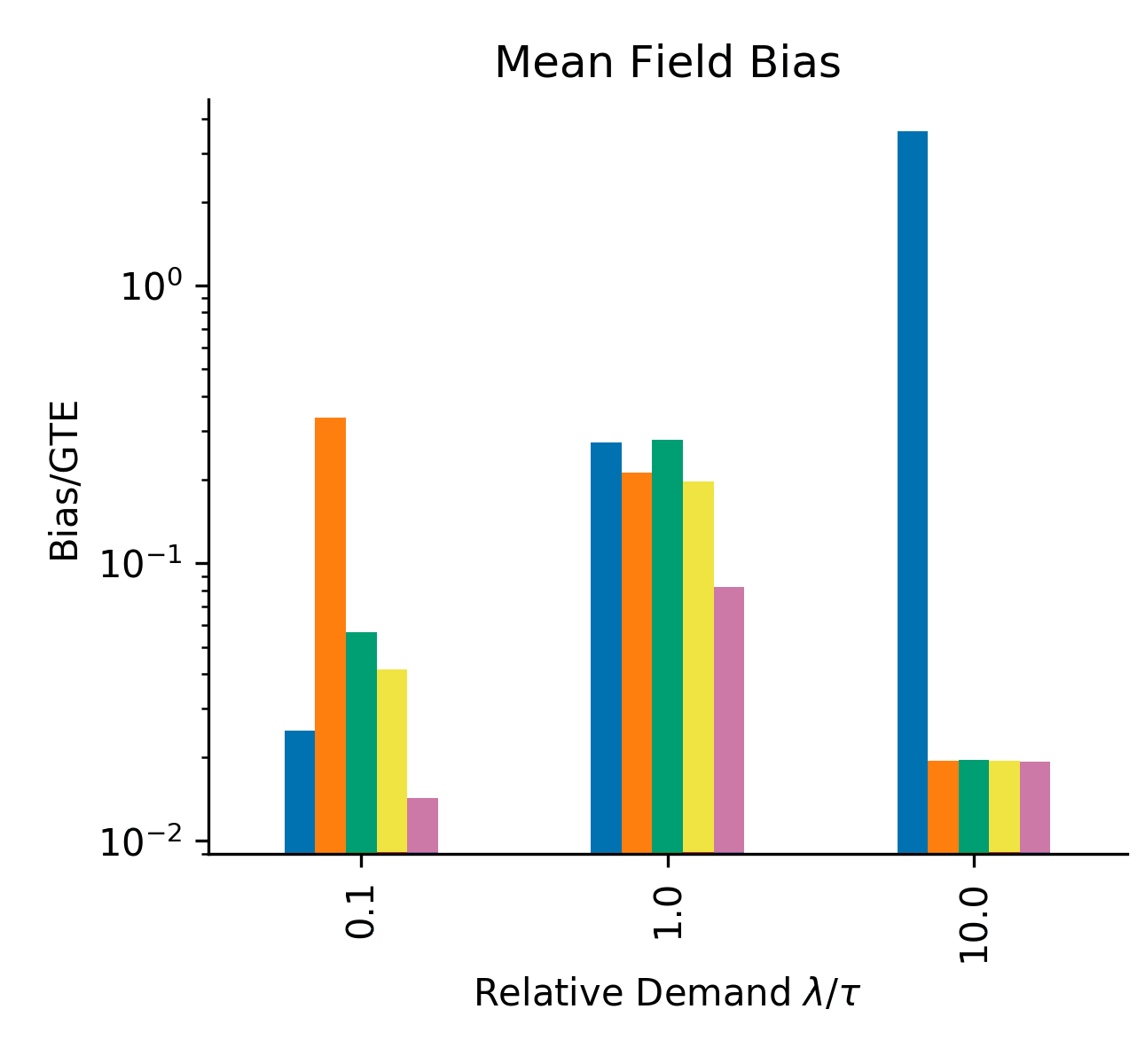}
        &
        \includegraphics[height=.26\textwidth]{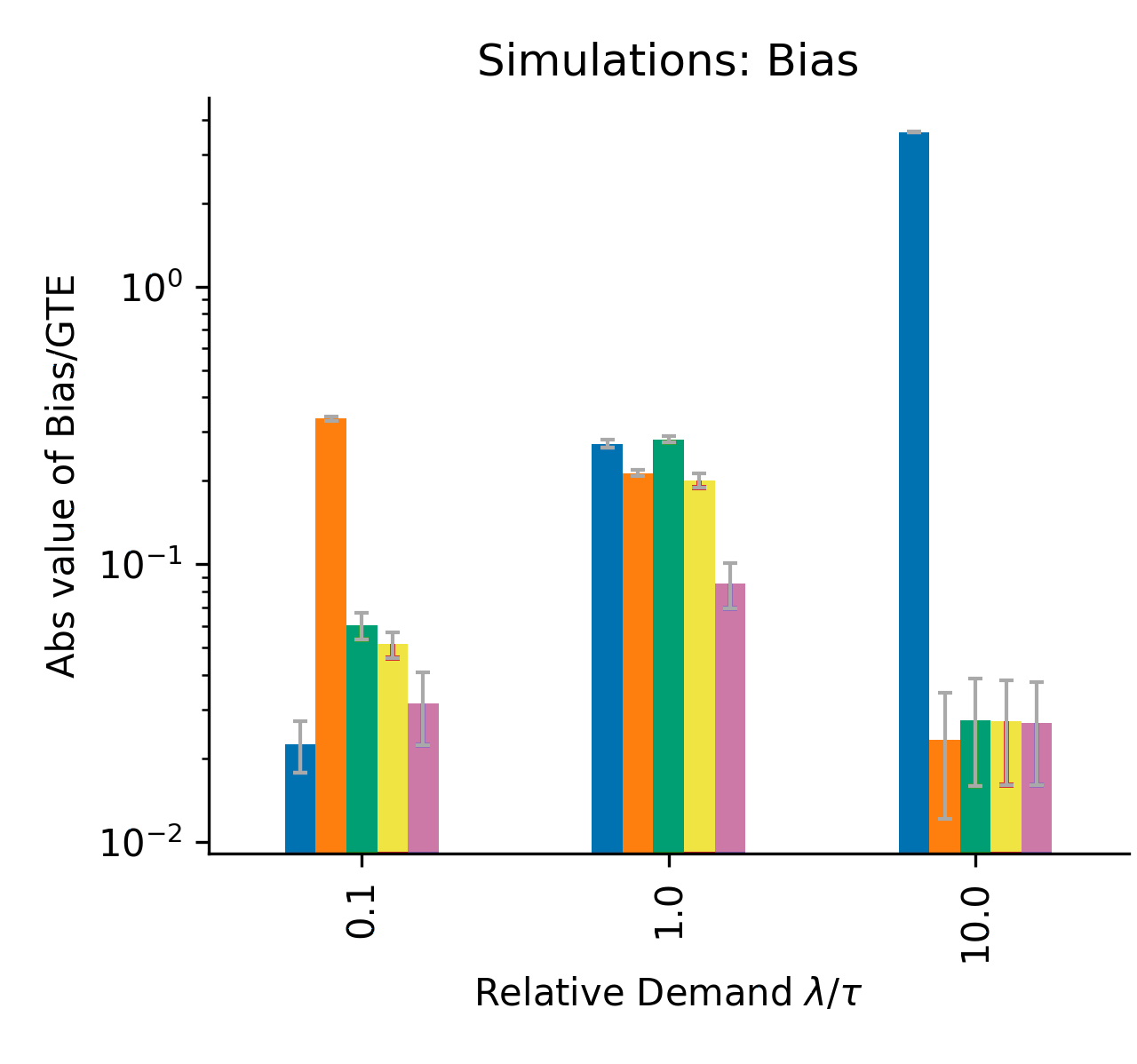}
        \\
        \includegraphics[height=.26\textwidth]{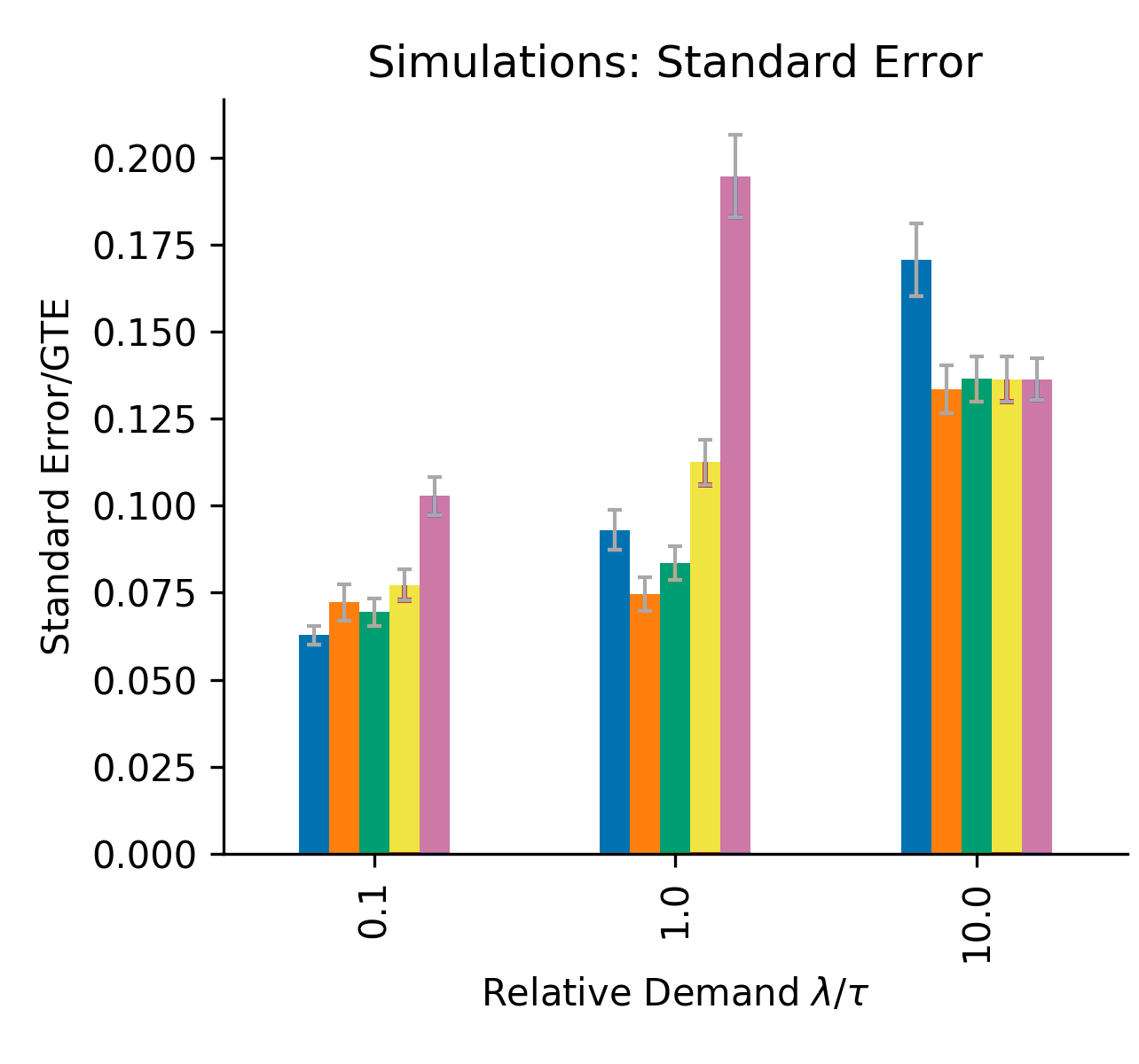}
        & 
        \includegraphics[height=.26\textwidth]{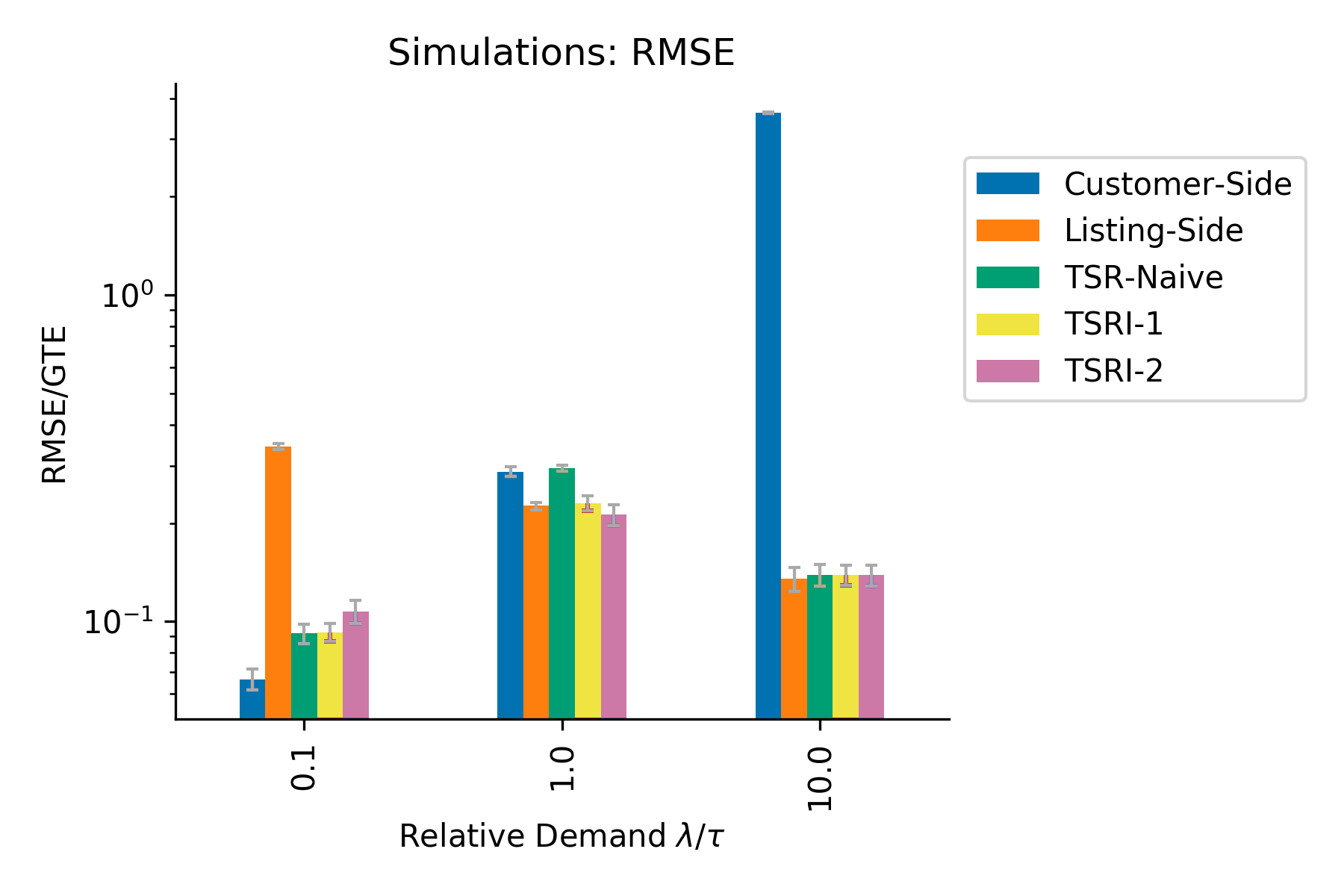}
    \end{tabular}
    \caption{(Varying market balance - heterogeneous treatment effects amplifying global control preferences.)
    Top left: Bias of each estimator in the mean field model (normalized by GTE). 
    Top right: Bias of each estimator in simulations, averaged across 200 runs (normalized by GTE). 
    Bottom left: Standard error of estimates, calculated across 200 runs (normalized by GTE). 
    Bottom right: $\RMSE$ of the estimates, calculated across $200$ runs (normalized by GTE). 
    }
    \label{fig:robustness_vary_lambda_heterogeneous_customers}
\end{figure}

\medskip
\subsection{Modifications of consideration sets.}

Here we consider a modification to the consideration set formation process described in Section \ref{sec:model}. Instead of a customer sampling each listing independently with probability $\alpha$, we now consider a situation where a customer samples a fixed set of $K = 50$ listings into their consideration set, drawn uniformly at random, as long as there are at least $50$ listings available. When there are fewer than $50$ listings, the customer samples all listings in to their consideration set. We note that another scenario to investigate is the one in which listings are drawn proportionally to their utility to the customer, to capture search and recommendation algorithms that recommend listings that the customer is likely to book. We leave deeper investigation of the formation of consideration sets (including sensitivity to the value of $K$) for future work. 

We calibrate the utilities such that the mean field model where customers sample listings into their choice set with probability $K/N = 50/N$ has a steady state booking probability of 20 percent in global control and 23 percent in global treatment, when $\lambda=\tau$. Note that with this change in consideration set formation, we can no longer guarantee that the finite system converges to the mean field model defined in Section \ref{sec:meanfield}. Thus we study these scenarios through simulations in the finite model and present the bias and standard error of the estimators in the simulations (see Figure \ref{fig:choice_set_k_50}).  As noted above, we find qualitatively similar behavior as before.

\begin{figure}[H]
    \centering
    \includegraphics[height=.27\textwidth]{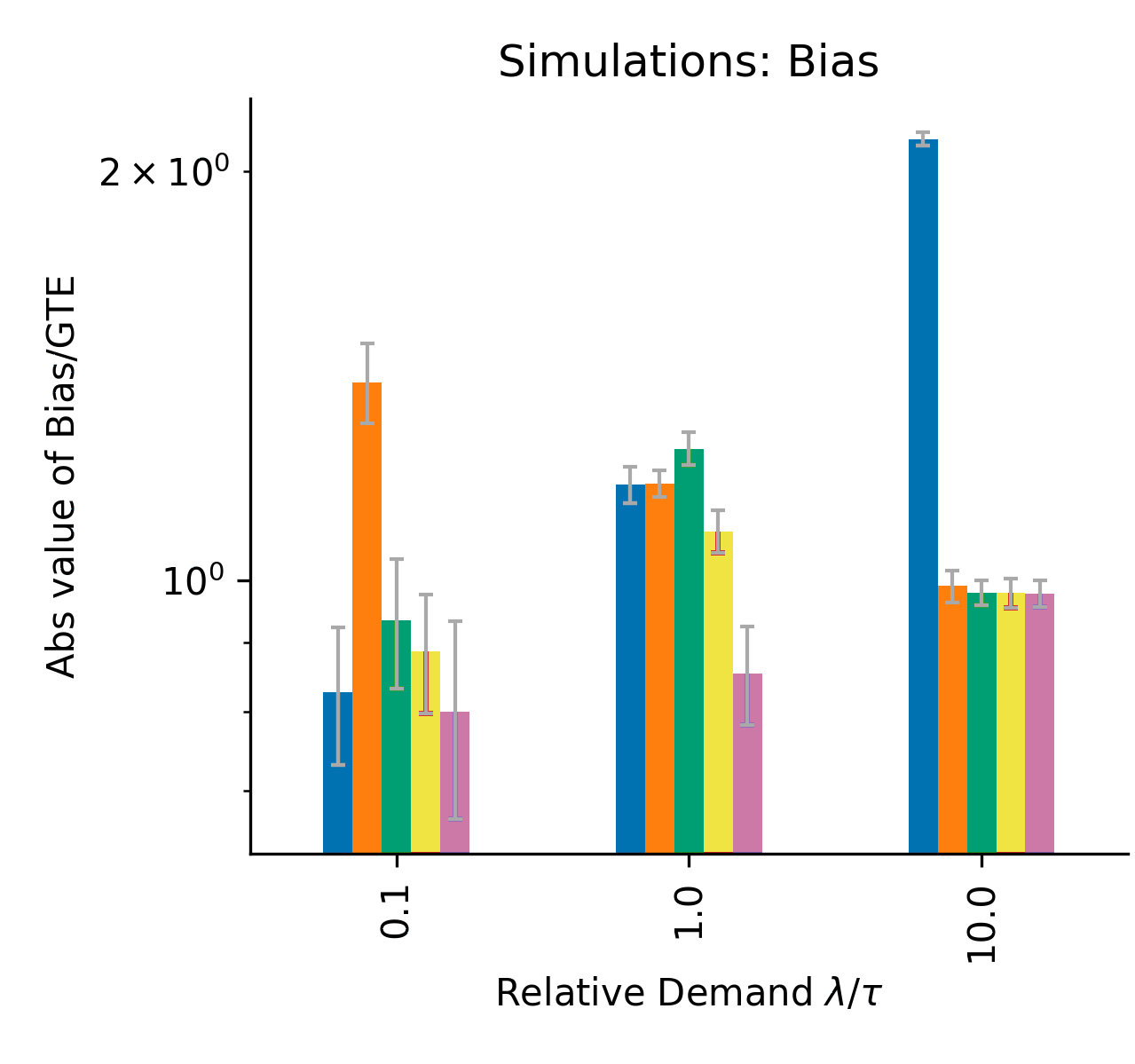}
    \includegraphics[height=.27\textwidth]{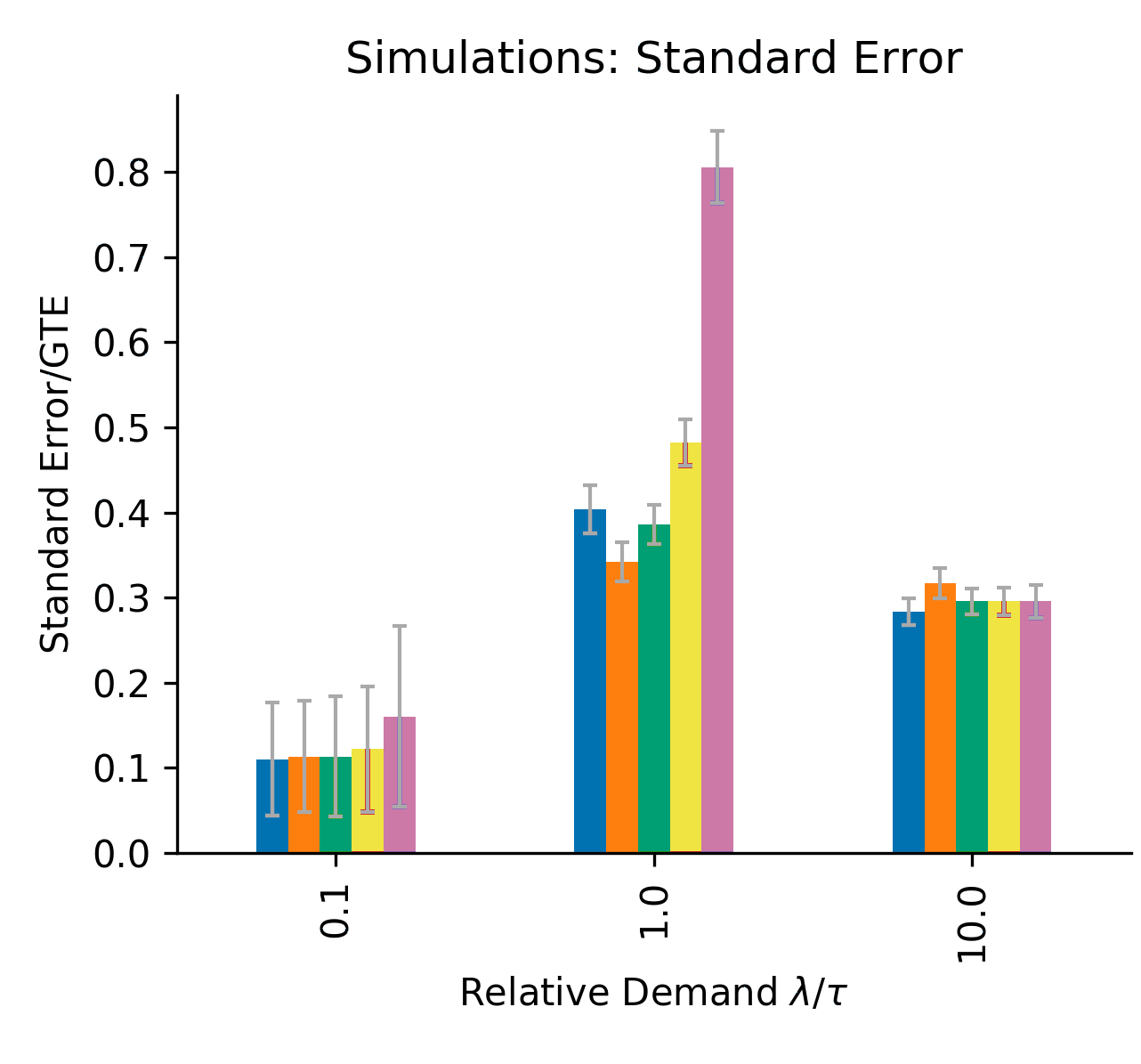} \\
    \includegraphics[height=.27\textwidth]{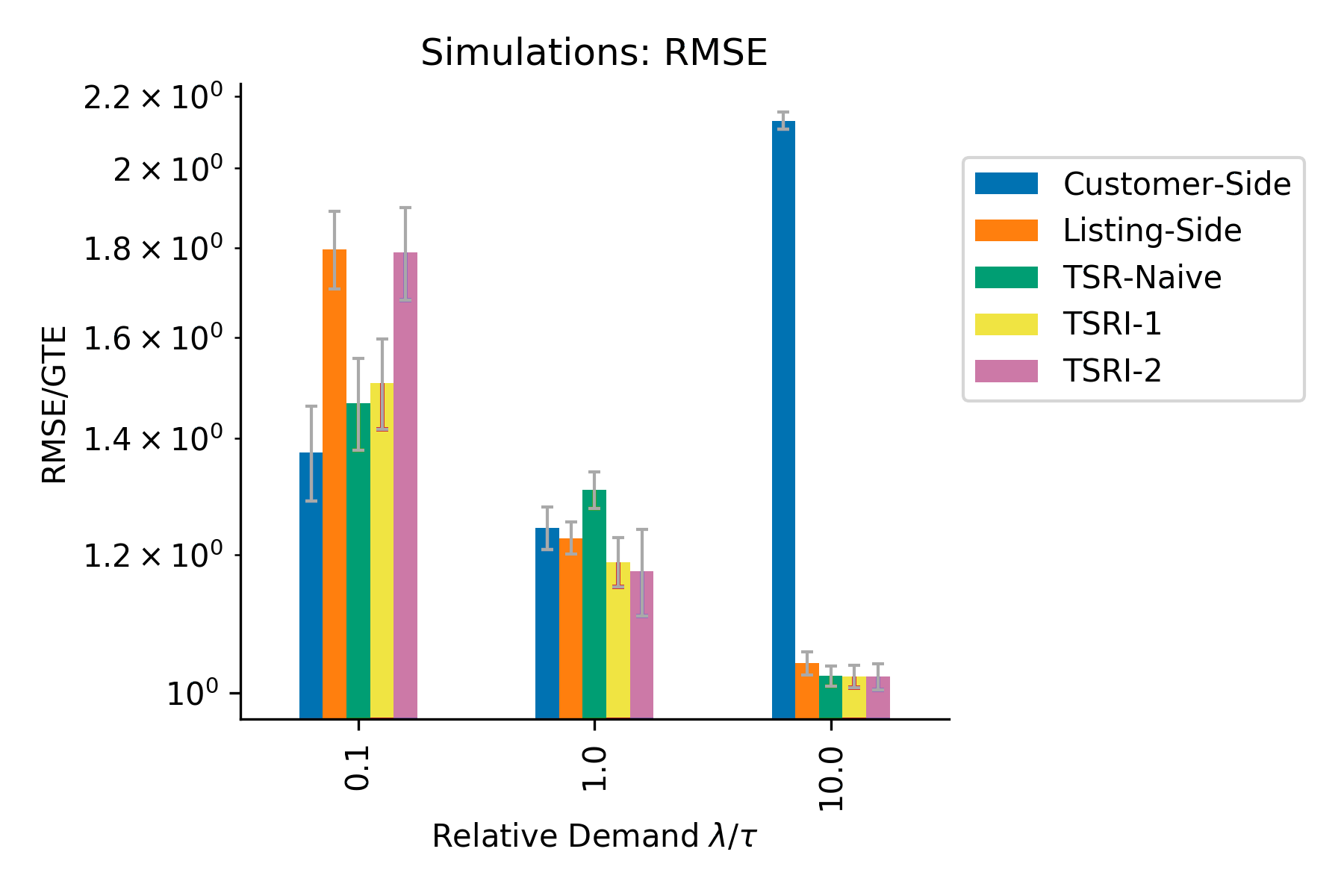}

    \caption{(Fixed size $K = 50$ consideration set.) Top left: Average bias (normalized by $\GTE$) of each estimator across 500 runs. 
    Top right: Standard error of estimates, calculated across 500 runs (normalized by $\GTE$). 
    Bottom: RMSE of estimates, calculated across 500 runs (normalized by $\GTE$).
    There is one customer type and one listing type. Customers have utility $v=6$ for control listings and $\tilde{v}=7.5$ for treatment listings.  
    }
    \label{fig:choice_set_k_50}
\end{figure}

\section{Numerics for transient dynamics}
\label{app:transient}

For practical implementation, it is important to consider the relative bias in the candidate estimators in the transient system, since experiments are typically run for relatively short time horizons. 
Theoretically, we can provide some insight when $\tau \to \infty$: in this case, the dominant term in the right hand side of \eqref{eq:ODE} is $(\rho(\theta) - s_t(\theta))\tau$.  Using this fact, it can be shown that as $\tau \to \infty$, for each $t > 0$, there holds $\v{s}_t^*(a_C, a_L) \to \v{\rho}(a_L)$ (where we define $\v{\rho}(a_L)$ as in Section \ref{ssec:ss_theory}).  In other words, the state {\em remains} at $\v{\rho}(a_L)$ at all times.   As a result in this limit the transient estimators $\widehat{\GTE}^{\CR}(T | a_C)$ and $\widehat{\GTE}^{\LR}(T | a_L)$ are equivalent to $\widehat{\GTE}^{\CR}(\infty | a_C)$ and $\widehat{\GTE}^{\LR}(\infty | a_L)$, respectively.  In particular, asymptotically as $\tau \to \infty$, the transient estimator $\widehat{\GTE}^{\CR}(T | a_C, a_L)$ will be an {\em unbiased} estimate of $\GTE$ at all times $T > 0$.  (The same is true if $\lambda \to 0$, provided the initial state is $\v{s}_0 = \v{\rho}(a_L)$.)

More generally, Figure \ref{fig:transient_dynamics} numerically investigates how the time horizon of the experiment affects the performance of the estimators in the mean field model. The system starts out in the steady state of the global control condition and evolves over the time horizon of the experiment. The relative performance of estimators depends on the time horizon of interest as well as market balance conditions. 

\begin{figure}[H]
    \centering
    \begin{tabular}{c c c }
        \includegraphics[height=0.26\textwidth]{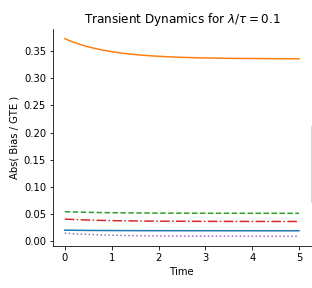}
        &
        \includegraphics[height=0.26\textwidth]{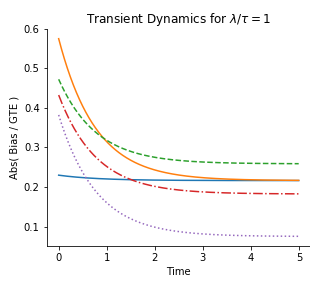}
        & 
        \includegraphics[height=0.26\textwidth]{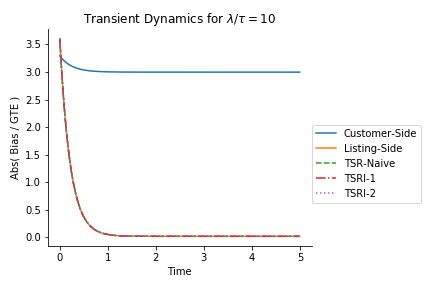}
    \end{tabular}
    \caption{Transient dynamics of estimators with $\lambda/\tau = 0.1, 1$, and $10$. Homogeneous system with one customer type and one listing type with parameters as defined in Figure \ref{fig:numerics_homogeneous}. System starts out in the global control steady state.}
    \label{fig:transient_dynamics}
\end{figure}

\end{APPENDICES}

\end{document}